\documentclass[aps,prb,twocolumn,groupedaddress,showpacs,footinbib]{revtex4}
\pdfoutput=1
\usepackage{amssymb}
\usepackage{amsmath}

\usepackage{graphicx}

\begin{document}

\title{Quantum Abacus for counting and factorizing numbers}

\author{M.V.\ Suslov$^{a,b}$, G.B.\ Lesovik$^{c}$, and G.\ Blatter$^{d}$}

\affiliation{$^{a}$Moscow Institute of Physics and Technology,
   Institutskii per.\ 9, 141700 Dolgoprudny, Moscow District, Russia}

\affiliation{$^{b}$NIX Computer Company, R\&D Department, Zvezdniy
   boulevard 19, 129085 Moscow, Russia}

\affiliation{$^{c}$L.D.\ Landau Institute for Theoretical Physics RAS,
   117940 Moscow, Russia}

\affiliation{$^{d}$Theoretische Physik, ETH-Zurich, CH-8093
   Z\"urich, Switzerland}

\date{\today}

\begin{abstract}
  We generalize the binary quantum counting algorithm of Lesovik, Suslov, and
  Blatter [Phys.\ Rev.\ A {\bf 82}, 012316 (2010)] to higher counting bases.
  The algorithm makes use of qubits, qutrits, and qudits to count numbers in a
  base 2, base 3, or base $d$ representation. In operating the algorithm, the
  number $n<N=d^K$ is read into a $K$-qudit register through its interaction
  with a stream of $n$ particles passing in a nearby wire; this step
  corresponds to a quantum Fourier transformation from the Hilbert space of
  particles to the Hilbert space of qudit states. An inverse quantum Fourier
  transformation provides the number $n$ in the base $d$ representation; the
  inverse transformation is fully quantum at the level of individual qudits,
  while a simpler semi-classical version can be used on the level of qudit
  registers.  Combining registers of qubits, qutrits, and qudits, where $d$ is
  a prime number, with a simpler single-shot measurement allows to find the
  powers of 2, 3, and other primes $d$ in the number $n$.  We show, that the
  counting task naturally leads to the shift operation and an algorithm based
  on the quantum Fourier transformation. We discuss possible implementations
  of the algorithm using quantum spin-$d$ systems, $d$-well systems, and their
  emulation with spin-1/2 or double-well systems.  We establish the 
  analogy between our counting algorithm and the phase estimation algorithm
  and make use of the latter's performance analysis in stabilizing our scheme.
  Applications embrace a quantum metrological scheme to measure a voltage
  (analog to digital converter) and a simple procedure to entangle
  multi-particle states.
\end{abstract}

\pacs{03.67.Ac 
      03.67.Bg 
      73.23.-b 
}

\maketitle

\section{Introduction}\label{sec:intro}

The representation of an integer number $n$ and its decomposition into prime
factors are basic mathematical operations. Quantum mechanics offers a new
perspective on these tasks, as well known since the seminal work of Peter Shor
\cite{Shor_94} on the efficient factorization of large numbers, with a drastic
impact on the security of codes. But even the counting of small numbers and
their factorization may prove useful, e.g., in the manipulation of physical
number states (and superpositions thereof) or in the entanglement of flying
qubits \cite{BoseHome,Beenakker,LSB_09}. A quantum algorithm to count $n <
N=2^K$ particles propagating in a wire using an array of qubits (a $K$-qubit
register) has been proposed recently \cite{LSB_09}; a very similar scheme has
been proposed by D'Helon and Milburn \cite{Milburn_96} in order to find the
number state distribution of a vibrational excitation in a system of trapped
laser-cooled ions. Besides providing the number $n$ in a binary form (base 2
counting), a simplified version of this algorithm tests for the divisibility
of $n$ by $2^k$ for a given $k \leq K$ and thus provides the power of 2 in the
factorization of $n$. In the present article, we generalize this algorithm to
perform a base $d$ counting and a test for the factor $d^k$ in the
decomposition of $n$. 

In order to accomplish this goal, we make use of a minimal formulation of the
counting task in terms of the problem of distinguishing between different
known quantum states in a single-shot measurement. This reduction to a few
very basic elements naturally connects the counting task with the quantum
Fourier transformation and provides us with a constructive scheme for the
setup of a (non-demolition) quantum counting algorithm.  We study various
possible (hardware) implementations of this algorithm, paying special
attention to the case of a ternary (base 3) counting system involving qutrits
as elementary counting devices. We establish the relation between our quantum
counting algorithm and the phase estimation algorithm and discuss several
applications.
\begin{figure}[ht]
  \includegraphics[width=8.0cm]{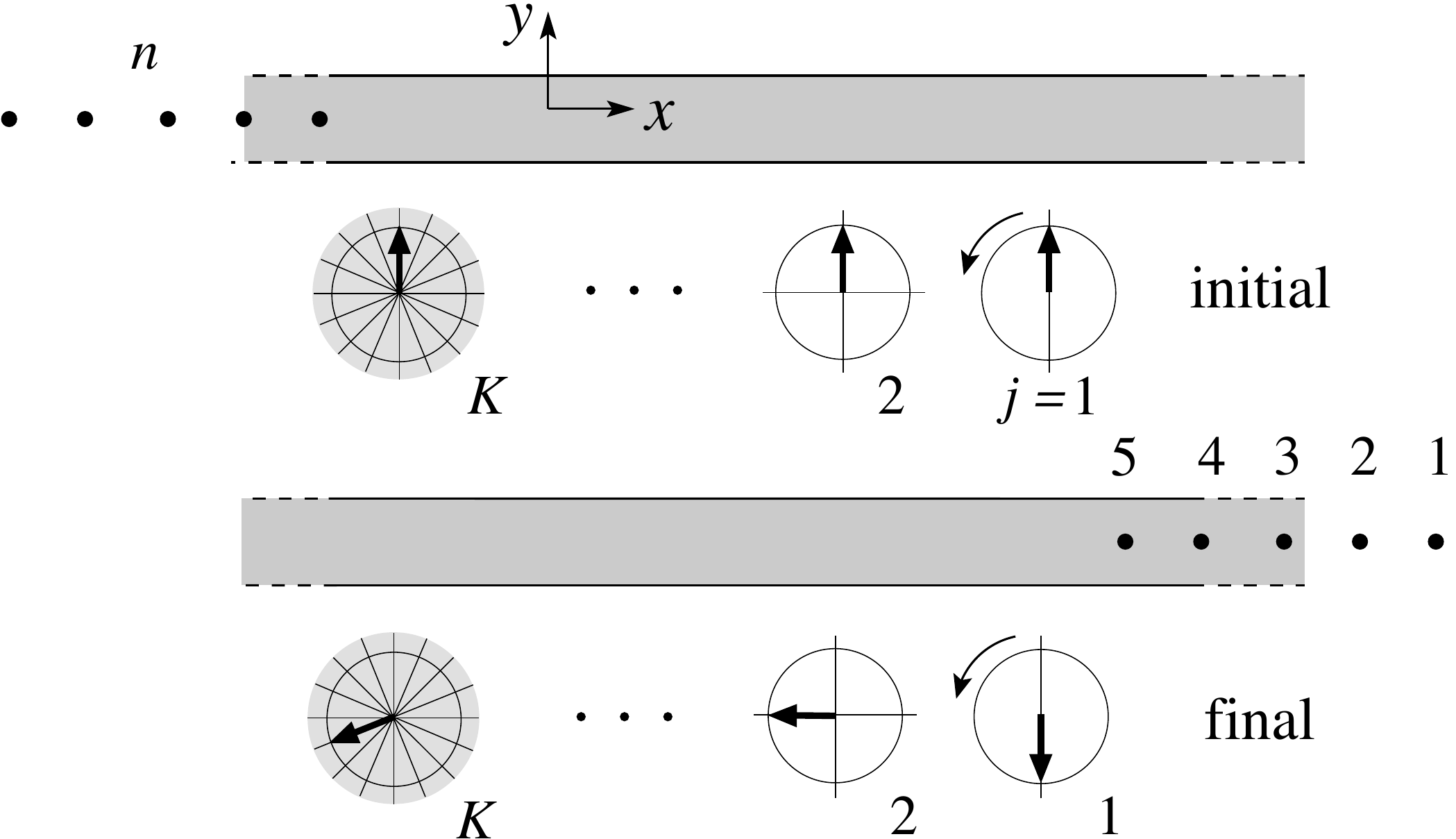}
  \caption[]{\label{fig:counting} Schematic representation of the quantum
  counting algorithm (shown is the case of base-two counting with qubits). A
  particle number state $|n\rangle_{\scriptscriptstyle \Phi}$ is fed into a
  quantum wire to undergo quantum counting. The interaction between the
  charged particles and the qubits rotates the spin/qubit states, thereby
  generating the Fourier transformation $\mathsf{F}$ taking the initial state
  $\mathsf{F} (|0\rangle_{\scriptscriptstyle Q})$ into the state
  $\mathsf{F}(|n\rangle_{\scriptscriptstyle Q})$.  A suitable manipulation and
  readout either provides the number state $|n\rangle_{\scriptscriptstyle Q}$
  (this is an inverse quantum Fourier transformation with a subsequent single
  shot measurement in the computational basis or a semi-classical quantum
  Fourier transformation involving a sequential measurement) or the maximal
  factor $2^{K}$ in $n$ (this readout involves `qubit-rotations' followed by a
  single shot measurement).}
\end{figure}

Our counting algorithm is inspired by the problem of counting in mesoscopic
systems\cite{LLL,Ensslin} and in quantum optics\cite{Brune}.  A
straightforward setup counting particles in a non-invasive manner requires of
the order of $N^2$ individual counting elements, see below. A more
sophisticated setup providing a unary counting scheme reduces this effort to
an order-$N$ process, and a drastic further reduction to a $(\log_d N)^2$
scaling can be achieved when going over to a counting base $d$.

The base-2 algorithm proposed in Ref.\ \onlinecite{LSB_09} involves two steps,
a first (analog) one where a finite train of $n$ charged particles traversing
a quantum wire, see Fig.\ \ref{fig:counting}, is coupled to an array of $K$
nearby qubits, thereby rotating the states of the qubits in a prescribed
manner. In more abstract terms, this step corresponds to reading a number
state $|n\rangle_{\scriptscriptstyle \Phi} \in {\cal H}_{\scriptscriptstyle
\Phi}$ in the Hilbert space of number states of particles into a state
$|\Psi_n\rangle_{\scriptscriptstyle Q}$ of a $K$-qubit register in the Hilbert
space ${\cal H}_{\scriptscriptstyle Q}$ of the $K$ qubits. More generally, a
superposition of number states gets entangled with the $K$-qubit register (not
copied, as demanded by the no-cloning theorem) during the counting process. In
a second step, the qubit state in ${\cal H}_{\scriptscriptstyle Q}$ is
manipulated and read out, providing either the maximal power of 2 contained in
$n$ or the number $n$ in binary form, depending on the readout algorithm.

The two steps correspond to two quantum Fourier transformations of special
kind: expressed in the usual (cf.\ Ref.\ \onlinecite{NielsonChuang})
computational basis $|0\rangle_{\scriptscriptstyle Q}=|00\dots
0\rangle_{\scriptscriptstyle Q}, |1\rangle_{\scriptscriptstyle Q}=|00\dots
1\rangle_{\scriptscriptstyle Q}, \dots, |2^K-1\rangle_{\scriptscriptstyle Q}
= |11\dots 1\rangle_{\scriptscriptstyle Q}\in {\cal H}_{\scriptscriptstyle Q}$
of the $K$-qubit register, the passage of the $n$ particles in the quantum
wire transforms the initial state
\begin{equation}\label{eq:in}
   |\Psi_0\rangle_{\scriptscriptstyle Q}
   = \mathsf{F}(|0\rangle_{\scriptscriptstyle Q}) = \frac{1}{\sqrt{N}} 
    \sum_{k=0}^{2^K-1} |k\rangle_{\scriptscriptstyle Q},
\end{equation}
i.e., the lowest Fourier harmonic of the $K$-qubit register, into the $n$-th
harmonic
\begin{equation}\label{eq:fi}
   |\Psi_n\rangle_{\scriptscriptstyle Q}
   = \mathsf{F}(|n\rangle_{\scriptscriptstyle Q}) = \frac{1}{\sqrt{N}} 
    \sum_{k=0}^{2^K-1} \exp(2\pi i \,n k/2^K) |k\rangle_{\scriptscriptstyle Q},
\end{equation}
the quantum Fourier transform of the state $|n\rangle_{\scriptscriptstyle Q} =
|n_1\,n_2\,\dots n_K\rangle_{\scriptscriptstyle Q}$, where $n$ is written in
binary form, $n = n_1\,n_2 \dots n_K = n_1 2^{K-1} + n_2 2^{K-2} +\dots + n_K
2^0$, cf.\ Ref.\ \onlinecite{NielsonChuang}.

Depending on the desired information (precise counting or divisibility) and
the available hardware (an operating quantum computer or a set of qubits), the
readout of the qubit register can be done in various ways.  The most efficient
version for this second step of the algorithm is the application of a second
(inverse) quantum Fourier transformation $\mathsf{F}^{-1}$ on the qubit state
$\mathsf{F}(|n\rangle_{\scriptscriptstyle Q})$, taking it back into the state
$|n\rangle_{\scriptscriptstyle Q}$.  A simultaneous measurement of the $K$
qubits then provides the particle number $n$ in binary form; in case of a
superposition of particle number states, the measurement will execute a
projection onto one of them.  Alternatively, the number state may be used in a
further computation.

However, performing an inverse quantum Fourier transformation on the state
$\mathsf{F}(|n\rangle_{\scriptscriptstyle Q})$ requires a quantum computer (or
at least those qubit operations required in the execution of a quantum Fourier
transformation).  Instead, we can make use of a procedure which is basically
identical to the semi-classical Fourier transform suggested by Griffiths and
Niu \cite{GriffithsNiu_96}, a conditional measurement algorithm involving a
sequential readout, where the reading of the $j$-th qubit depends on the
results of the previous $j-1$ measurements. This measurement still provides
the full information on a pure number state; if the initial state is a
superposition of number states, the semi-classical algorithm will project the
state (upon sequential measurements of the qubits) to one of its components.

A specially efficient readout is available if we are interested in the power
of 2 contained in $n$ rather than $n$ itself. In this case, a simultaneous
(rather then conditional) readout algorithm can be applied directly to the
state $\mathsf{F}(|n\rangle_{\scriptscriptstyle Q})$; this algorithm then
provides a divisibility check of $n$ by $2^k$, $k<K$.

The present article deals with the generalization of this algorithm. The most
obvious task to generalize is the determination of other powers of primes in
$n$, i.e., to find the factorization of $n$, $n=2^{k_2} 3^{k_3} 5^{k_5}
\dots$. This can be achieved by going over to generalized qubits, three-level
systems or qutrits, $d$-level systems or qudits, etc. Equivalently, this
corresponds to changing the representation of the number $n$ from binary (base
2) to ternary (base 3), quinary (base 5), etc. Again, the two-step algorithm
first transfers the information from the physical number state
$|n\rangle_{\scriptscriptstyle \Phi}$ into the computational $K$-qudit
register ($\to \mathsf{F}(|n \rangle_{\scriptscriptstyle Q})$) through a
particle-qudit interaction and then extracts the information in the qudit
register via an inverse Fourier transform ($\to |n \rangle_{\scriptscriptstyle
Q}$). This readout step involves a full quantum transform on the level of each
single qudit, while a semi-classical transform \cite{GriffithsNiu_96} suffices
to extract the information on the level of the $K$-qudit register.

In order to carry out the above program, it is very helpful to have an
abstract understanding of the counting process. Indeed, a minimal abstract
formulation of quantum counting in an $N$-dimensional Hilbert space (allowing
to distinguish or count at most $N$ objects) naturally leads us to two types
of basis states, the computational basis (corresponding to the states
$|n\rangle_{\scriptscriptstyle Q}$), in which the result of the counting
process is measured, and the counting basis $|\psi_n
\rangle_{\scriptscriptstyle Q}$ where the actual counting process is done---it
turns out that just these two basis-sets are naturally related by the quantum
Fourier transformation, $|\psi_n\rangle_{\scriptscriptstyle Q} = \mathsf{F}(|n
\rangle_{\scriptscriptstyle Q})$.  Furthermore, the abstract analysis of the
counting process provides us with a recipe how the algorithm can be physically
implemented.

In the following, we compare the efficiency of various quantum counting
algorithms (Sec.\ \ref{sec:efficiency}) and then briefly repeat our previous
base 2 quantum algorithm with qubits, including the sequential and single-shot
readout schemes, see section \ref{sec:qubits}. We then present in Sec.\
\ref{sec:fourier} our basic analysis of the quantum counting process,
providing a natural link between quantum counting and the quantum Fourier
transformation as well as a constructive scheme helping us to generalize
counting to a base $d$ system.  We then proceed with the simplest
generalization to qutrits in section \ref{sec:qutrits}, the problem of
counting in a base-3 representation or power counting of 3, and discuss its
further generalization to qudits.  To be specific, we consider an
implementation with a three-level system in the form of three quantum dots and
also discuss various ideas for other hardware implementations of the base-3
algorithm in section \ref{sec:implementations}, a spin-1 system, serving
rather as a Gedanken experiment for illustration, and two practical versions
of emulating a qutrit with qubits. In Sec.\ \ref{sec:PEA}, we discuss an
interesting correspondence between our counting algorithm and the phase
estimation algorithm \cite{NielsonChuang,Kitaev,Cleve} (no such correspondence
is yet known for our divisibility check) and apply this insight in the
proposal for a quantum voltage-detector (an analog-digital converter). Another
application, a scheme to create multi-particle entangled states in a
Mach-Zehnder interferometer, is discussed in Sec.\ \ref{sec:mpentanglement}.
We summarize and conclude in Sec.\ \ref{sec:conclusion}.

\section{Efficiency of quantum counting}\label{sec:efficiency}

In a broader context, the efficiency of our quantum counting algorithm has to
be compared with other schemes. E.g., the most straightforward non-invasive
way of counting the number of (charged) particles flowing in a wire (directed
along $x$) is to use a spin-counter polarized in the $xy$-plane and rotating
the state by an incremental angle $\phi = \pi/N$ (around the $z$-axis) upon
passage of a particle \cite{LLL}.  This spin rotation is achieved through the
magnetic field pulse generated by the passing charge \cite{Baz,Rybachenko};
see also Refs.\ \onlinecite{Brune,Guerlin} for a related experiment in a
quantum optical setup. The precise correspondence of the angle with the number
of passed particles then requires an accurate measurement, to a precision of
$\phi = \pi/N$, of the spin-counter's final state polarization---this either
necessitates a large number $M> N^2/\pi^2$ of repetitions of the counting
experiment, or a single-shot readout of $M$ identical counters all measuring
the passage of the particles: Measuring the spin along the $y$-axis, the
(theoretical) probability to find it pointing upwards is given by $P^\uparrow
= \langle m^\uparrow \rangle_k/M = \cos^2 (n\phi/2)$, where $\langle
m^\uparrow \rangle_k$ denotes the average of finding $m^\uparrow$ of the $M$
spins pointing up in a sequence of $k \to \infty$ realizations of the entire
experiment. On the other hand, the one-time measurement $m_m^\uparrow$
provides the experimental result $P_m^\uparrow = m_m^\uparrow/M$, from which
we can find the number $n = (2/\phi) \arccos[(P_m^\uparrow)^{1/2}]$.  As this
procedure is a statistical one, we have to determine how many spins
(measurements) $M$ are needed to predict the particle number $n$ with
certainty. We then require that the difference in probability (we assume $N >
n \gg 1$) $\delta P^\uparrow = |P^\uparrow(n+1)-P^\uparrow(n)| \approx
|\partial_n P^\uparrow| = (\phi/2) \sin(n\phi)$ has to be much larger than the
uncertainty $[\langle (\delta m^\uparrow)^2\rangle_k]^{1/2} \equiv [\langle
(m^\uparrow-\langle m^\uparrow\rangle_k)^2\rangle_k]^{1/2}$ in the
measurement, $\delta P^\uparrow \gg [\langle(\delta m^\uparrow)^2
\rangle_k]^{1/2}/M$.  Given the binomial statistics of the measurement process
[the values $\uparrow$ and $\downarrow$ are measured with probabilities
$P^\uparrow$ and $(1-P^\uparrow)$], we obtain $\langle (\delta
m^\uparrow)^2\rangle_k = P^\uparrow (1-P^\uparrow) \, M$ and combining these
results, we find that
\begin{equation} \label{eq:M}
   M \gg 1/\phi^2 > N^2/\pi^2 \gg 1
\end{equation}
spins are needed in order to accurately measure the particle number $n < N$.

\section{Qubits: counting powers of 2}\label{sec:qubits}

We first provide a more detailed discussion of the base-2 counting algorithm
with the setup in Fig.\ \ref{fig:counting}, where the $n < N=2^K$ particles to
be counted flow in a quantum wire along $x$. Single electron pulses can be
generated by appropriate voltage pulses \cite{LLL,Keeling} or through
injection from a quantum dot \cite{Feve} and the counters are conveniently
thought of as individual spins, cf.\ Refs.\ \onlinecite{Baz,Rybachenko,LLL}.
We use spin states polarized along the $z$-axis as our computational basis,
$|\!\uparrow\rangle \leftrightarrow |0\rangle$ and $|\!\downarrow\rangle
\leftrightarrow -i|1\rangle$.  

{\it Preparation}: Initially, the $K$ spins or qubits (we use these terms
synonymously, cf.\ Ref.\ \onlinecite{Hassler}) are polarized along the
positive $y$-axis, i.e., the initial states read $|+y\rangle_j =
[|\!\uparrow\rangle_j + i|\!\downarrow \rangle_j]/\sqrt{2}$, $j=1,\dots, K$.
Identifying $|\!\uparrow\rangle_j \leftrightarrow |0\rangle_j$ and
$|\!\downarrow\rangle_j \leftrightarrow -i|1\rangle_j$, the product state
$|\Psi_0 \rangle_{\scriptscriptstyle Q} = \prod_{j=1}^K
[(|0\rangle_j+|1\rangle_j)/ \sqrt{2}]$ is identical with the equally weighted
sum of $K$-qubit register states $|0\rangle_{\scriptscriptstyle Q},
|1\rangle_{\scriptscriptstyle Q}, \dots, |2^K-1\rangle_{\scriptscriptstyle
Q}$, $|\Psi_0 \rangle_{\scriptscriptstyle Q} = (1/\sqrt{2^K}) \sum_k
|k\rangle_{\scriptscriptstyle Q}$.  This state then coincides with the lowest
harmonic in the Fourier transformed computational basis: indeed, the quantum
Fourier transform takes a state
\begin{equation}\label{eq:X}
   |X\rangle_{\scriptscriptstyle Q} 
   \equiv \sum_l x_l|l\rangle_{\scriptscriptstyle Q}
\end{equation}
into the state 
\begin{equation}\label{eq:Y}
   \mathsf{F}(|X\rangle_{\scriptscriptstyle Q}) 
   \equiv \sum_k y_k|k\rangle_{\scriptscriptstyle Q} 
   = |Y\rangle_{\scriptscriptstyle Q}
\end{equation}
with
\begin{equation}\label{eq:yx}
   y_k=(1/\sqrt{N})\sum_l x_l \exp(2\pi i\, lk/N).
\end{equation}
The initial state $|\Psi_0\rangle_{\scriptscriptstyle Q}$ of the counting
setup then is given by $x_l = \delta_{l0}$ and hence $|\Psi_0
\rangle_{\scriptscriptstyle Q} = \mathsf{F}(|0 \rangle_{\scriptscriptstyle
Q})$.

{\it Counting and Fourier transformation}: Assuming a transverse coupling
between the charged particle and the spin, the passage of a particle rotates
the spins in the $x$-$y$ plane. The couplings of the spin counters to the wire
are chosen such that the $j$-th spin is rotated (anti-clockwise) by the amount
$\phi_j = 2\pi/2^j$ (a rotation by $\mathsf{U}_z (\phi_j) = \exp(-i\phi_j
\sigma_z/2)$ with $\sigma_z$ a Pauli matrix).  The passage of $n$ particles
then rotates the $j$-th spin by the amount $n \phi_j$ and thus it ends up in
the state $[|\!\uparrow\rangle_j +i\exp(2\pi i \, n/2^j)|\!\downarrow
\rangle_j] /\sqrt{2}$, where we have dropped the overall phase $\exp(-\pi i\,
n /2^j)$.  Again, we identify $|\!\uparrow\rangle_j \leftrightarrow
|0\rangle_j$ and $|\!\downarrow\rangle_j \leftrightarrow -i|1\rangle_j$ and
use the binary representation of integers $k = k_1 2^{K-1} + \dots + k_K 2^0 =
k_1 \,k_2 \dots k_K $ to rewrite the product $|\Psi_n\rangle_Q = \prod_{j=1}^K
[(|0\rangle_j+\exp(2\pi i \, n/2^j)|1\rangle_j/2]$ over qubit states as a sum
over register states,
\begin{eqnarray}\label{eq:qFtps}
  |\Psi_n\rangle_{\scriptscriptstyle Q} \!&=&\!
       \prod_{j=1}^K \frac{|0\rangle_j+\exp(2\pi i \,n \,2^{K-j}/2^K)
       |1\rangle_j}{\sqrt{2}}
    \\  \nonumber
    \!&=&\! \frac{1}{\sqrt{2^K}}\!\! \sum_{k_1, \dots,k_K = 0,1} 
    \!\!\!\!\!\!\!\!\!
    e^{2\pi i \,n (\sum_{l=1}^K\! k_l 2^{K-l})/2^K}
    |k_1 \dots k_K\rangle_Q
   \\  \nonumber
    \!&=&\! \frac{1}{\sqrt{2^K}} 
    \sum_{k=1}^{2^K} e^{2\pi i \, n \, k/2^K} |k\rangle_Q.
\end{eqnarray}
From the  comparison with the Fourier transform Eqs.\ (\ref{eq:Y}) and
(\ref{eq:yx}) we find that $x_l = \delta_{ln}$, and hence the passage of the
$n$ particles (the counting process) takes the state $\mathsf{F}
(|0\rangle_{\scriptscriptstyle Q})$ into the $n$-th harmonic
$\mathsf{F}(|n\rangle_{\scriptscriptstyle Q})$.  We call the states
$|n\rangle_{\scriptscriptstyle Q}$ the computational basis and the transforms
$|\Psi_n\rangle_{\scriptscriptstyle Q} = \mathsf{F}
(|n\rangle_{\scriptscriptstyle Q})$ define the counting basis. The fact that
the qubits in the $K$-qubit register reside in a product state and hence
remain unentangled \cite{GriffithsNiu_96,Cleve} is a crucial element of our
algorithm and in fact decisive for the next step, the readout of the result
with the help of a semi-classical quantum Fourier (back) transformation.
\begin{figure}[ht]
  \includegraphics[width=7.0cm]{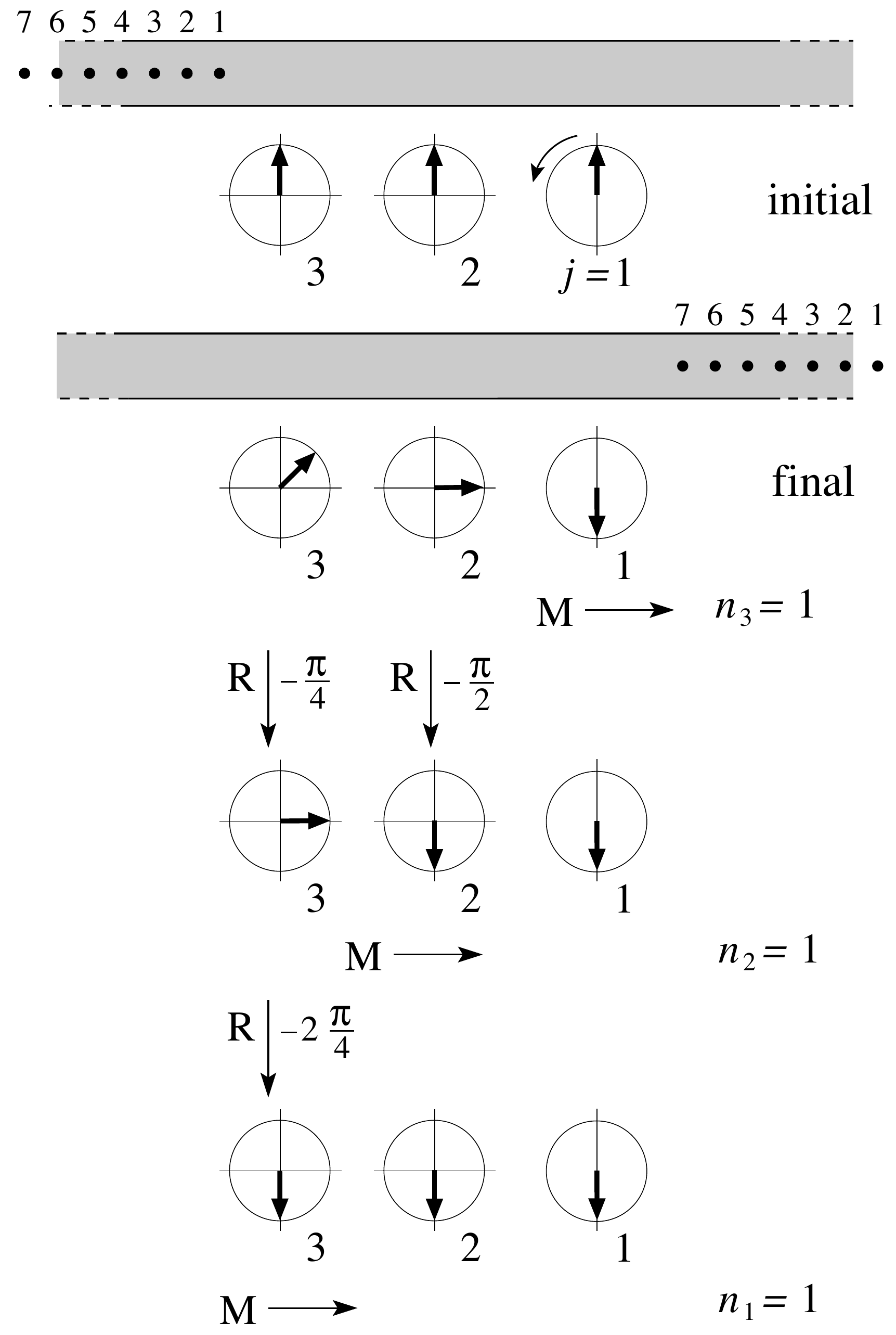}
  \caption[]{Counting of $n=7$ particles by  $K=3$ qubits. After passage of
  the particles, the initial state with all spins pointing along the $y$-axis
  and corresponding to the lowest Fourier harmonic is transformed to the final
  state, the $n$-th Fourier harmonic with properly rotated spins. The
  indivdiual counter qubits are then measured (M) and the binary digits $n_j$,
  $j = 3,2,1$ are determined in reverse order.  After each measurement, the
  remaining qubits are rotated (R) to undo the counting of the corresponding
  moduli: Following the measurement of the first qubit, the rotations by the
  last (odd numbered) particle are undone. After measurement of the second
  qubit the rotation by the last odd-numbered pair is undone.}
  \label{fig:rotation}
\end{figure}

{\it Readout and inverse Fourier transformation}: Let us then turn to the
second step, the semi-classical quantum Fourier (back) transformation which
provides us with the desired result, the binary representation of $n$. We
start with the measurement of the first spin: since this has been rotated by
the angle $n\pi$, we measure it along the $y$-axis.  If we find it pointing
upward, the number's parity is even and we store a `0' in the first position
$n_K$ of the binary number; in case we find it pointing downward, the parity
is odd and we store the digit `1'. Besides providing the number's first binary
digit, the parity, the outcome of the measurement also tells us whether the
second spin (rotated by $n\pi/2$) is directed along the $y$-axis (even parity)
or along the $x$-axis. This information allows us to measure the second spin
along the correct axis; for an even $n$, we measure the spin along the
$y$-axis (and store a 0 (1) in $n_{K-1}$ if the spin is pointing up (down)),
while for an odd-parity $n$, we measure the spin along the $-x$-axis.  The
iteration of the readout algorithm is straightforward: the $j$-th spin is
measured along the direction $m_{j-1} \phi_j$ with the integer $m_{j-1}=
n_{K-j+2} \dots n_{K-1}n_K$ corresponding to the binary number encoded in the
$j-1$ previous measurements.  The $j$-th position in the binary register then
assumes a value $n_{K-j+1} = 0$ or $n_{K-j+1}=1$, depending on the measurement
result, 0 for a spin pointing along the axis and 1 for a spin pointing
opposite. This sequential measurement algorithm provides us with the binary
representation of $n$.

In order to render the algorithm more efficient, rather than rotating the axis
of measurement, the spins are rotated backwards by the corresponding angles.
These rotations by $-m_{j-1} \phi_j$ are conveniently done incrementally:
after measurement of the $j$-th spin with outcome `0' or `1', all spins $J >j$
are rotated by $-n_{K-j+1} 2^{j-1} \phi_J$. These rotations undo the action of
odd-numbered groups of particles: The first rotation by $-2\pi n_K\, 1/2^J$
acting on the qubits $J>1$ compensates for the last odd-numbered particle. The
second rotation by $-2\pi n_{K-1}\,2/2^J$ acting on the qubits $J>2$
compensates for the last odd-numbered pair. The third rotation by $-2\pi
n_{K-2} 4/2^J$ acting on the qubits $J>3$ compensates for the last
odd-numbered quartet, etc. These rotations make the next spin to be measured
point up or down, since the action of those particles giving an intermediate
result has been subtracted.  The entire algorithm is illustrated in Fig.\
\ref{fig:rotation} for the case of $n=7$ particles counted by $K=3$ qubits.

Formally, the availability of a sequential readout algorithm can be derived
from a suitable representation of the product state Eq.\ (\ref{eq:qFtps}): the
fraction $n/2^j$ in the phase $\exp(2\pi i \, n/2^j)$ of the $j$-th qubit has
to be known only modulo 1 and making use of the relation
\begin{eqnarray}\label{eq:n_rel}
  \frac{n}{2^j} \Big|_{\mathrm{mod}(1)} &=&
  0.n_{K-j+1}\dots n_K \\
  &=& \frac{n_K2^0+n_{K-1} 2^1+ \dots+n_{K-j+1}2^{j-1}}{2^j},
  \nonumber
\end{eqnarray}
the $j$-th qubit state can be written in the form $[|0\rangle_j+\exp(2\pi i \,
0.n_{K-j+1}\dots n_K) |1\rangle_j]/\sqrt{2}$, where we make use of the binary
representation of fractions $0.n_1 n_2 \dots n_K = n_1/2+n_2/4+\dots +
n_K/2^K$, see Ref.\ \onlinecite{NielsonChuang}.  The final state after passage
of the particles can then be written in the form
\begin{equation} \label{eq:qFt}
  |\Psi_n\rangle_{\scriptscriptstyle Q} =
       \prod_{j=1}^K \frac{|0\rangle_j+\exp(2\pi i \> 0.n_{K-j+1}\dots n_K)
       |1\rangle_j}{\sqrt{2}}.
\end{equation}
This representation demonstrates that the state of the first qubit $j=1$
involves only the smallest digit $n_K$ of the seeked number $n$, the second
one involves the fraction $0.n_{K-1}n_K$, and so on. Hence each qubit state
requires knowledge of the states of previous qubits and its measurement adds
one digit more to the binary representation of $n$.

{\it Divisibility by $2^k$}: A variant of the above counting algorithm
provides a test for the divisibility of $n$ by powers of two. Consider the
state of the first $j=1,2,\dots, k \leq K$ spins after the passage of $n$
particles. If the number $n$ contains the factor $2^k$, then the $k$ qubits
will all point along the positive $y$-axis (and, for $k<K$, the $k+1$-th qubit
will point down; alternatively, if $n=2^K$ then all spins in the register have
returned to their initial state pointing along $y$).  A single-shot
measurement of the $K$-qubit register along the $y$-axis thus provides the
(maximal) factor $2^k$ in $n$.  The formal proof of this statement is given in
Ref.\ \onlinecite{LSB_09}.

\section{Distinguishability and quantum counting}\label{sec:fourier}

We now reduce the problem of quantum counting to the task of distinguishing
between quantum states. This reduction will quite naturally lead us to the
definition of two basis sets, one serving the counting process itself by
admitting a simple manipulation of phases during the counting step (the
computational basis) and the other keeping track of the counting (counting
basis); the two are related by the operation of quantum Fourier
transformation.

We start from the premise that quantum counting corresponds to the process of
associating distinct states of an auxiliary quantum system (the counter) to
the size (cardinality) of a physical state. We assume that we want to count at
most $N$ objects, hence our auxiliary quantum system shall count the objects
modulo $N$. During the counting process, the initial state $|\Psi_0\rangle$ of
the counter is, upon passage of $n$ objects, transformed to the final state
$|\Psi_n\rangle$ (we can safely drop the index $Q$ in this section). We define
the unitary operation $\mathsf{C}_1$ to describe the passage of one particle,
\begin{equation}\label{eq:C1}
   |\Psi_{1}\rangle = \mathsf{C}_1 |\Psi_0\rangle,
\end{equation}
and a simple iteration produces the state
\begin{equation}\label{eq:Cj}
   |\Psi_n\rangle = \mathsf{C}_1^n |\Psi_0\rangle 
   \equiv \mathsf{C}_n |\Psi_0\rangle
\end{equation}
upon passage of $n$ particles. We now require that we can distinguish between
the states $|\Psi_n\rangle$ and $|\Psi_0\rangle$ in a single-shot measurement,
implying that the states should be orthogonal, $\langle \Psi_n |\Psi_0\rangle
= 0$. So far, we only require that we can distinguish between `no particles'
associated with the state $|\Psi_0\rangle$ and a state with `some particles
$n$' with $0 < n < N$ and associated with the state $|\Psi_n\rangle$, without
being able to decide between different number states with different $n$'s.  It
turns out that a setup solving this reduced task is also able to distinguish
between the different particle number states $|\Psi_n\rangle$. Indeed, using 
Eq.\ (\ref{eq:Cj}) and the fact that $\langle \Psi_n |\Psi_0\rangle
= 0$ for all $0<n<N$, we find that (we choose $0 < l < n < N$)
\begin{eqnarray}\label{eq:orth}
   \langle \Psi_l |\Psi_n\rangle &=& \langle \Psi_0 | \mathsf{C}_l^\dagger 
   \mathsf{C}_n |\Psi_0\rangle = \langle \Psi_0| \mathsf{C}_{n-l}
   |\Psi_0\rangle \\
   \nonumber
   &=&  \langle \Psi_0|\Psi_{n-l}\rangle = 0
\end{eqnarray}
and hence the states $|\Psi_n\rangle$ are all orthogonal and distinguishable.
Finally, our wish to count modulo $N$ requires cyclicity, i.e.,
\begin{equation}\label{eq:cyclicity}
   \mathsf{C}_N|\Psi_0\rangle = \exp{i\Theta} |\Psi_0\rangle;
\end{equation}
if the dimension of our auxiliary counter system is given by $N$, then the
cyclicity Eq.\ (\ref{eq:cyclicity}) follows automatically: Applying
$\mathsf{C_1}$ iteratively to the counting states $|\Psi_n\rangle$, one finds
that all states $|\Psi_{n+1}\rangle$ are orthogonal to the previous states
$|\Psi_l\rangle$, $\langle \Psi_l|\Psi_{n+1}\rangle$, $0\leq l \leq n < N-1$.
Once we arrive at $n=N-1$, the last state completing the Hilbert space, the
further application of $\mathsf{C_1}$ produces a state $|\Psi_N\rangle =
\mathsf{C_1} |\Psi_{N-1}\rangle$ which has to be a superposition of the
previous states, $|\Psi_N\rangle = \sum_{n=0}^{N-1}
\langle\Psi_n|\Psi_N\rangle |\Psi_n\rangle$. However, since all matrix
elements $\langle \Psi_n|\Psi_N\rangle = \langle
\Psi_n|\mathsf{C}_1\Psi_{N-1}\rangle = \langle \Psi_{n-1}|\Psi_{N-1}\rangle$
vanish for $0<n\leq N-1$, we must have $|\Psi_N\rangle \propto |\Psi_0\rangle$
and since $\mathsf{C}_1$ is unitary, we arrive at the result Eq.\
(\ref{eq:cyclicity}).  Otherwise, for a larger dimensionality of the auxiliary
system, the condition Eq.\ (\ref{eq:cyclicity}) has to be imposed as a
separate requirement.

In the end, the (minimal) auxiliary counter system is described by an
$N$-dimensional Hilbert space ${\cal H}$ with orthonormal counting basis
$|\Psi_n\rangle \in {\cal H}$, $n = 0, \dots, N-1$ and $\langle
\Psi_l|\Psi_n\rangle = \delta_{ln}$, and a unitary (shift or counting)
operator $\mathsf{C}_1$ taking one counting state to the next, $\mathsf{C}_1
|\Psi_n\rangle = |\Psi_{n+1}\rangle$, and the property of cyclicity,
$\mathsf{C}_1^N = \mathsf{C}_N = \exp (i\Theta)$.  In a specific physical
implementation, the phase $\Theta$ is determined by the dynamical evolution of
the system during counting.  The states $|\Psi_n\rangle$ keep track of the
numbers in the counting process, i.e., due to their orthogonality they
uniquely identify the cardinality of the counted set.

In order to further characterize the properties of our auxiliary counting
system, we determine the eigenvalues and eigenvectors of the counting operator
$\mathsf{C}_1$.  Expressed in the basis $\{|\Psi_n\rangle\}_{n=0}^{N-1}$, the
latter assumes the form
\begin{equation}\label{eq:QF_S}
   \mathsf{C}_1 
          = \left(
          \begin{array}{ccccc}
          0 & 0& \dots & 0 & e^{i\Theta}\\
          1 & 0& \dots & 0 & 0\\
          0 & 1& \dots & 0 & 0\\
          0 & 0& \dots & 0 & 0\\
          0 & 0& \dots & 1 & 0
          \end{array}
          \right).
\end{equation}
Its eigenvalues and eigenvectors are easily found: the determinant of
$\mathsf{C}_1 - \lambda \openone$ is given by $(-\lambda)^N +
(-1)^{N-1}e^{i\Theta}$ and hence the eigenvalues of $\mathsf{C}_1$ are the $N$
roots of 1 on the unit circle in the complex plane shifted by $\Theta/N$,
$\lambda_k = \exp(2\pi i k/N +i\Theta/N)$, $k=0,1,\dots,N-1$.  The associated
eigenvector $|k\rangle$ is given by $\langle \Psi_n | k\rangle = \exp(-2\pi i\,
k n/N-i\,n\Theta/N)/\sqrt{N}$. Note that a phase $\Theta = 2\pi l$ simply
renumbers the eigenvectors and eigenvalues by $l$.  

The eigenstates $|k\rangle$ of the counting operator $\mathsf{C}_1$ show a
particularly simple behavior in the counting process---they merely pick up a
phase, and these phases are distributed homogeneously over the unit circle.
Hence, expressing the (unknown) counting states $|\Psi_n\rangle$ through the
eigenstates $|k\rangle$ of the counting operator $\mathsf{C}_1$, we obtain
\begin{eqnarray}\label{eq:QF_evv}
  |\Psi_n\rangle &=& \frac{1}{\sqrt{N}} \sum_{k=0}^{N-1} e^{2\pi i\, k
  n/N+i\,n\Theta/N} |k\rangle
  \\ \nonumber 
  &=&e^{i\,n\Theta/N} \mathsf{F}(|n\rangle).
\end{eqnarray}
Making use of the eigenstates $|k\rangle$, the counting step can be
implemented in an extremely simple and minimally invasive manner: upon passage
of a particle, each state $|k\rangle$ shall pick up the additional phase
$\exp[i(2\pi\, k+\Theta)/N]$ and $|\Psi_n\rangle$ goes over to $|\Psi_{n+1}
\rangle$, which is just the action of $\mathsf{C}_1$ or of counting.  Hence,
if we choose as our computational basis the set of orthonormal eigenstates $\{
|k\rangle \}_{k=0}^{N-1}$ of the shift operator $\mathsf{C}_1$, then the
counting process can be implemented in a `soft' way, adding only phases to the
computational states (note that these are the states in which our final
projective measurement will be done). At the same time, the counting basis,
which is interconnected by the shift operator $\mathsf{C}_1$, is made from the
states $|\Psi_n\rangle= \exp(i\,n\Theta/N) \mathsf{F}(|n\rangle)$, which is
nothing but the quantum Fourier transform (up to a phase) of the eigenstates
$|n\rangle$. In the further general discussion below we set $\Theta = 0$; the
required transformation eliminating a finite $\Theta$ will be discussed later.

\begin{figure}[ht]
 \includegraphics[width=8.0cm]{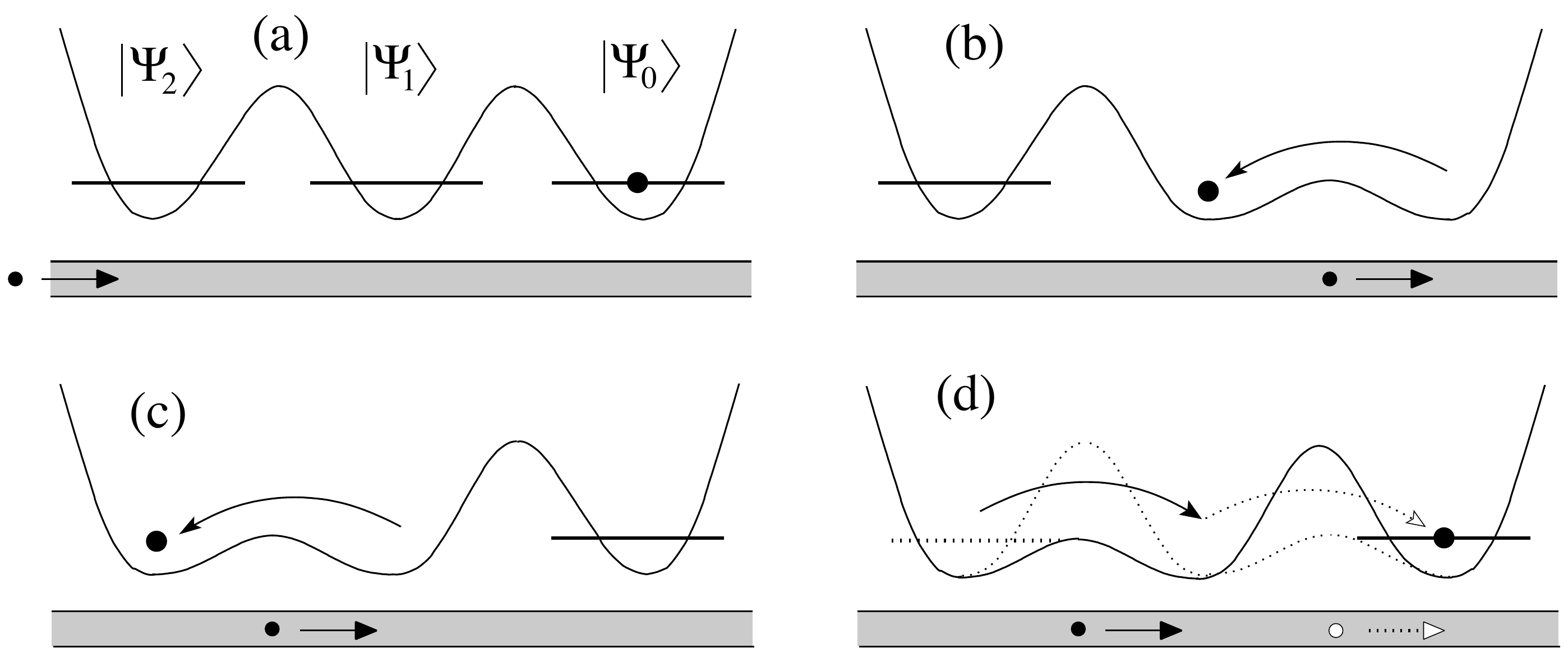}
  \caption[]{\label{fig:real_space_counter} Sketch for a real space counter in
  the form of a 3-level qutrit operating in the amplitude mode; the interaction
  between the counter particle and the particle in the quantum wire is
  assumed to be attractive (a slightly modified version can be found for a
  repulsive interaction). Subsequent lowering of the barriers separating
  the quasi-classical states $|0\rangle$, $|1\rangle$, and $|2\rangle$
  upon passage of a particle moves the counter particle to the left, cf.\
  (a) $\to$ (b) $\to$ (c). The third particle passing drags the counter
  particle all the way back to the initial state, cf.\ (d).}
\end{figure}

At this point one may ask if there are other counting algorithms which do not
exploit the quantum Fourier transformation---the answer is yes, but such
alternative schemes do not provide the `soft' counting involving only the
addition of phases.  As an example, consider the setup shown in Fig.\
\ref{fig:real_space_counter}, a multi-well system (one counter particle in a
$d$-well potential landscape) operating in the amplitude mode.  Assuming an
attractive interaction \cite{repulsive}, the passing particle lowers or
removes the barrier between adjacent semi-classical states, allowing the
counter particle to move between two wells. Here, the semi-classical states
localized in the individual wells play the role of the counting states
$|\Psi_n\rangle$: initializing the counter in the (right-most) state
$|\Psi_0\rangle$, cf.\ Fig.\ \ref{fig:real_space_counter}, the first $d-1$
particles passing will push the counter particle to the left in steps of one,
$|\Psi_n\rangle \to |\Psi_{n+1}\rangle$, while the $d$-th particle will drag
the counter particle back to the right until it ends up in the initial state
$|\Psi_0\rangle$. The difference of this device with the `soft' counting
device discussed above is in the choice of the computational basis: rather
than selecting the Fourier transformed states of $|\Psi_n\rangle$, which are
eigenstates of the shift operator $\mathsf{C}_1$ and only pick up phases
during the counting process, here we choose as a computational basis the
counting basis itself, hence, $|n\rangle = |\Psi_n\rangle$. As a result,
rather than adding phases during counting, we shift the counter particle in
real space.  Obviouly, this setup is difficult to realize as quite some fine
tuning is required to generate a clean shift operation; furthermore, the shift
operation in real space will generate an appreciable back action on the
passing particles. On the other hand, such a counter is not supposed to evolve
coherently between counting steps, hence the requirements on the coherence
time are reduced.

\begin{figure}[ht]
 \includegraphics[width=6.0cm]{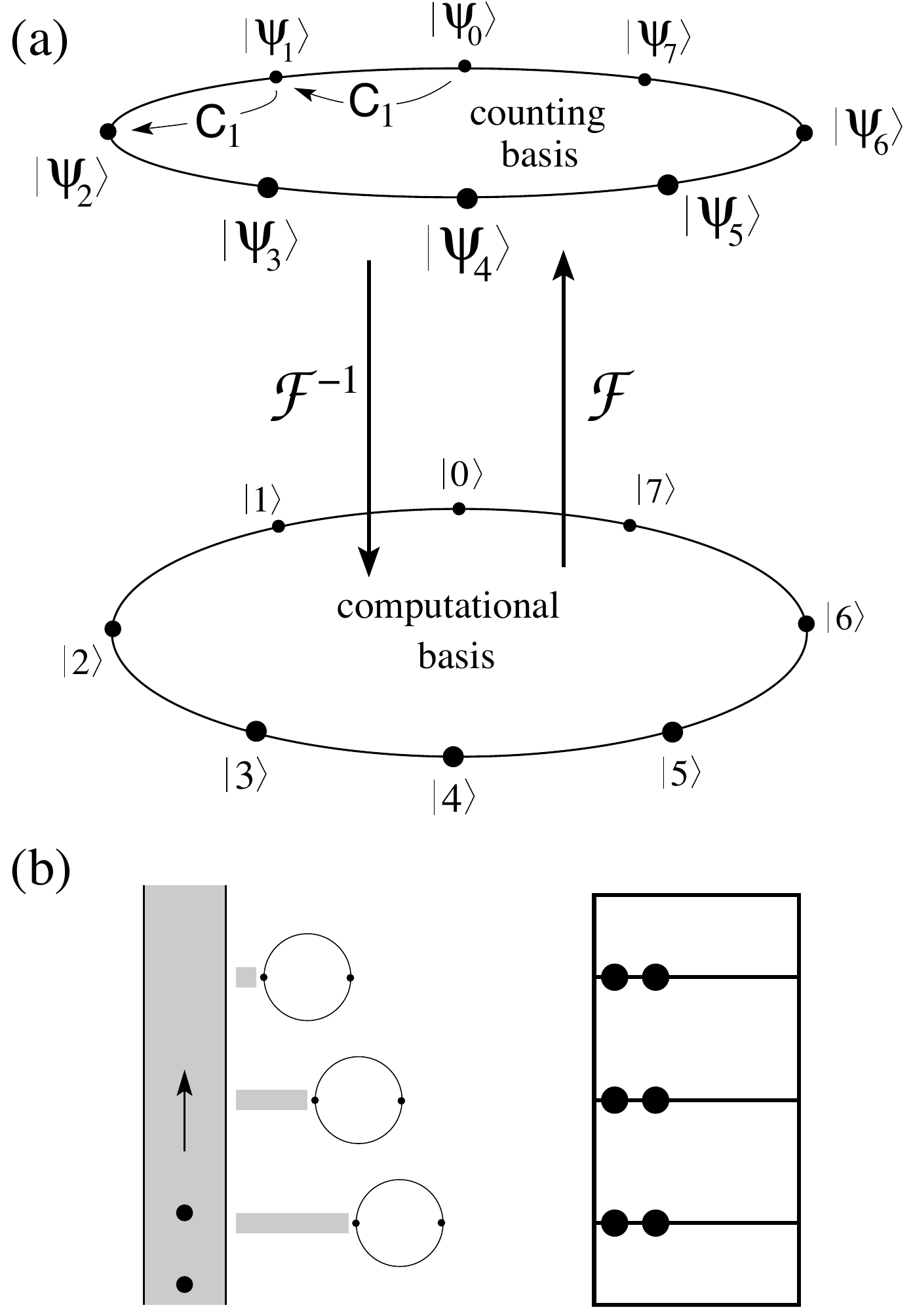}
  \caption[]{\label{fig:circle} (a) Counting device for $N=8$ (corresponding to
  a qudit with $d=N=8$). The (semi-classical) states $|k\rangle$, $k=0,\dots,
  N-1$, form the computational basis; their Fourier transforms
  $|\Psi_n\rangle$ define the counting basis. In the elementary counting step,
  the shift operator $\mathsf{C}_1$ transforms one Fourier mode
  $|\Psi_n\rangle$ into the next $|\Psi_{n+1}\rangle$. (b) Emulation of the
  $N=8$ qudit through three qubits, implementing quantum counting with $d=2$
  and $K=3$ (left) and corresponding classical Abacus (right).}
\end{figure}

In abstract terms, the counting process can be illustrated through the
counting states $|\Psi_n\rangle$ arranged in a circle with the shift operator
$\mathsf{C}_1$ transforming one state to the next. The goal then is to have
for the Fourier transformed basis states $|k\rangle$ a set of (measureable)
states which merely pick up phases when interacting with the passing
particles; these states then shall form our computational states. In this
basis, the counting process (the shift operator $\mathsf{C}_1$) transforms one
Fourier mode into the next, cf.\ Fig.\ \ref{fig:circle}(a).

So far, the abstract computational states $|n\rangle$ have been chosen in a
trivial way, without any additional structure; correspondingly, our counting
process is still a simple (and inefficient) one, requiring as many different
states as we have objects to count. The great reduction in the (hardware)
complexity of the counting process appears with the introduction of a counting
base.  Introducing a counting base $d$ ($d=2~(3)$ for binary (ternary)
counting), we have $N=d^K$ and can reduce the hardware requirement from $N$ to
$\log_d N = K$; correspondingly, the efficiency of the algorithm increases
dramatically from a linear in $N$ complexity to $\log_d N$.  To illustrate the
case, we consider a simple example $N=8$, $d=2$, and $K=3$, counting up to 8
with the help of 3 qubits. The primitive version involves a loop with 8 states
$|0\rangle, |1\rangle, \dots, |7\rangle$. This can be reduced to three loops
with two states each, $|0\rangle_j, |1\rangle_j$, $j=1,2,3$, cf.\ Fig.\
\ref{fig:circle}.  The preparation of the large loop generates the state
$|\Psi_0^{(8)}\rangle = \sum_{k=0}^7|k\rangle/\sqrt{8}$, the properly prepared
reduced system is in the state $|\Psi_0^{(2^3)}\rangle = \prod_{j=1}^3
[|0\rangle_j + |1\rangle_j]/\sqrt{2}$. The counting step on the large loop
adds the phase $\exp(2\pi i k/8)$, $|k\rangle \to \exp(2\pi i k/8)|k\rangle$,
to the state $|k\rangle$ in $|\Psi_0^{(8)}\rangle$. On the other hand, the
counting step on the small loops adds sequentially smaller phases:
$|\nu\rangle_1 \to \exp(2\pi i \nu/2)|\nu\rangle_1$, $|\nu\rangle_2 \to
\exp(2\pi i \nu/4)|\nu\rangle_2$, $|\nu\rangle_3 \to \exp(2\pi i
\nu/8)|\nu\rangle_3$, $\nu = 0,1$. One easily checks that the state
\begin{eqnarray}\nonumber
  |\Psi_0^{(2^3)}\rangle \!&=&\! \frac{
   [|0\rangle_1 + |1\rangle_1]\otimes[|0\rangle_2 + |1\rangle_2]
   \otimes [|0\rangle_3 + |1\rangle_3]}{\sqrt{8}}\\
      \label{eq:Psi_aba2}
      &=& [|000\rangle + |001\rangle + |010\rangle + |011\rangle \\
      \nonumber
      &+& |100\rangle + |101\rangle + |110\rangle + |111\rangle]/\sqrt{8}
\end{eqnarray} 
(the three entries refer to the qubits 1, 2, and 3, e.g., $|011\rangle \equiv
|0_11_21_3\rangle \equiv |0\rangle_1|1\rangle_2|1\rangle_3$) transforms to the
state
\begin{eqnarray}\label{eq:Psi_aba3}
  |\Psi_1^{(2^3)}\rangle \!&=&\! [|000\rangle + e^{i\pi /4} |001\rangle
                                              + e^{i\pi /2} |010\rangle
     \\ \nonumber
     &+& e^{3i \pi /4} |011\rangle + e^{i\pi} |100\rangle 
       + e^{5i\pi/4} |101\rangle
     \\ \nonumber
     &+& e^{3i\pi/2} |110\rangle + e^{7i\pi/4} |111\rangle]/\sqrt{8}
\end{eqnarray}
upon passage of one particle; this is identical to $|\Psi_1^{(8)}\rangle$ upon
identifying the state $|k\rangle$ with its binary equivalent $|k_1 k_2
k_3\rangle$.  Hence, the complexity of the hardware used in the counting task
is easily reduced by going over to a counting base; in our case, rather than
implementing an 8-level system, the same counting task can be accomplished
with the help of 3 qubits. This step in reduction of complexity is nothing but
going over to a (quantum) Abacus,  cf.\ Fig.\ \ref{fig:circle}(b).  Its
physical implementation involves qudits (playing the role of the rows in the
classical Abacus); performing one cycle in the $j$-th qudit shifts the state
of the next qudit by one unit.

The insight provided by the `soft counting procedure' and the reduction in
hardware provided by choosing a counting base (base $2, 3, \dots, d$)
described above gives us a recipe how to construct a physical implementation
of the counting process: Define a quantum mechanical system with $N = d^K$
orthogonal states $|k\rangle$, $k=0,\dots, N-1$, with an identical (trivial)
time evolution; these states form our computational basis and often appear in
the form of semi-classical (measureable) states.  The non-trivial time
evolution of these states originates from their interaction with the particles
during their passage. The coupling of the states to the particles has to be
arranged in a way such that the state $|k\rangle$ picks up a phase $\exp(2\pi
i k/N)$ upon passage of one particle.  Defining the interaction Hamiltonian
$\mathsf{H}_\mathrm{int}$, this implies an equidistant distribution of the
matrix elements $\int_0^{t_k} dt \langle k| \mathsf{H}_\mathrm{int}|k\rangle$;
assuming that all levels interact with the particles during equal time
intervals $t_k =t_c$, we can conclude that the spectrum of
$\mathsf{H}_\mathrm{int}$ is equidistant. Furthermore, returning back to the
cyclic phase $\Theta$ in the counting operator $\mathsf{C}_1$, we can
compensate for a finite value via an energy shift in the interaction
Hamiltonian $\mathsf{H}_\mathrm{int}$: adding a constant energy $\Theta
\hbar/t_c$ to $\mathsf{H}_\mathrm{int}$, we redefine the counting operator
operator $\tilde{\mathsf{C}}_1 = e^{-i\Theta/N} \mathsf{C}_1$ and the counting
basis $|\tilde\Psi_n\rangle = e^{-i\Theta\, n/N} |\Psi_n\rangle$, $0 \leq n <
N$, and find that the new condition for the cyclicity reads
$|\tilde\Psi_N\rangle = \tilde{\mathsf{C}}_1| \tilde\Psi_{N-1}\rangle =
e^{-i\Theta}|\Psi_{N}\rangle = e^{-i\Theta} e^{i\Theta}|\Psi_0\rangle =
|\tilde\Psi_0\rangle$.
\begin{figure}[ht]
 \includegraphics[width=5.0cm]{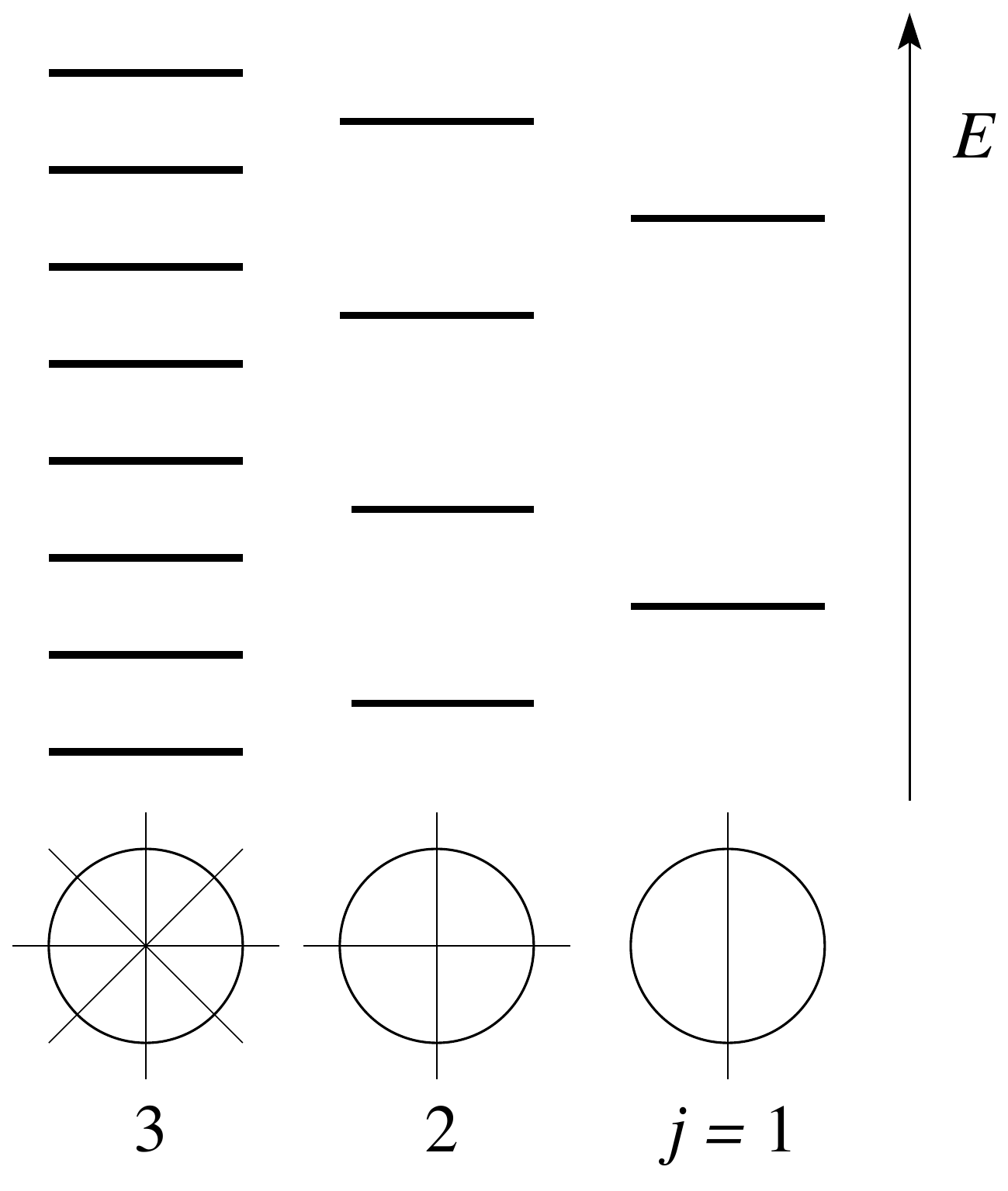}
  \caption[]{\label{fig:spectrum}
  Equidistant spectrum for the Hamilonian $\mathsf{H}_\mathrm{int}$
  describing $K=3$ qubits.}
\end{figure}

As a specific example, we can examine the situation for the base 2 counting in
Sec.\ \ref{sec:qubits}; here, the Hamiltonian (more precisely, the logarithm
of the shift operator, $(1/i) \ln \mathsf{C}_1$) describing one count can
be written in the form 
\begin{equation}\label{eq:QF_Hint}
   \frac{1}{\hbar}\int_0^{t_c} dt \mathsf{H}_\mathrm{int} =
   \sum_{j=1}^K \frac{\pi}{2^j} \sigma_z^{\rm\scriptscriptstyle (j)},
\end{equation}
where the Pauli matrix $\sigma_z^{\rm\scriptscriptstyle (j)}$ operates on the
$j$-th qubit, resulting in the expected equidistant spectrum, cf.\ Fig.\
\ref{fig:spectrum}. In this example, the energy zero is located in between 2
eigenstates and hence $\Theta = \pi$. Shifting the zero-energy point to one of
the eigenstates, we then can get rid of the cyclic phase $\Theta$; e.g.,
shifting the energy zero to the top-most level we identify the state
$|\uparrow\uparrow\uparrow\rangle$ with the computational state $|0\rangle$.

\section{Qutrits: counting powers of 3}\label{sec:qutrits}

\begin{figure}[ht]
  \includegraphics[width=6.0cm]{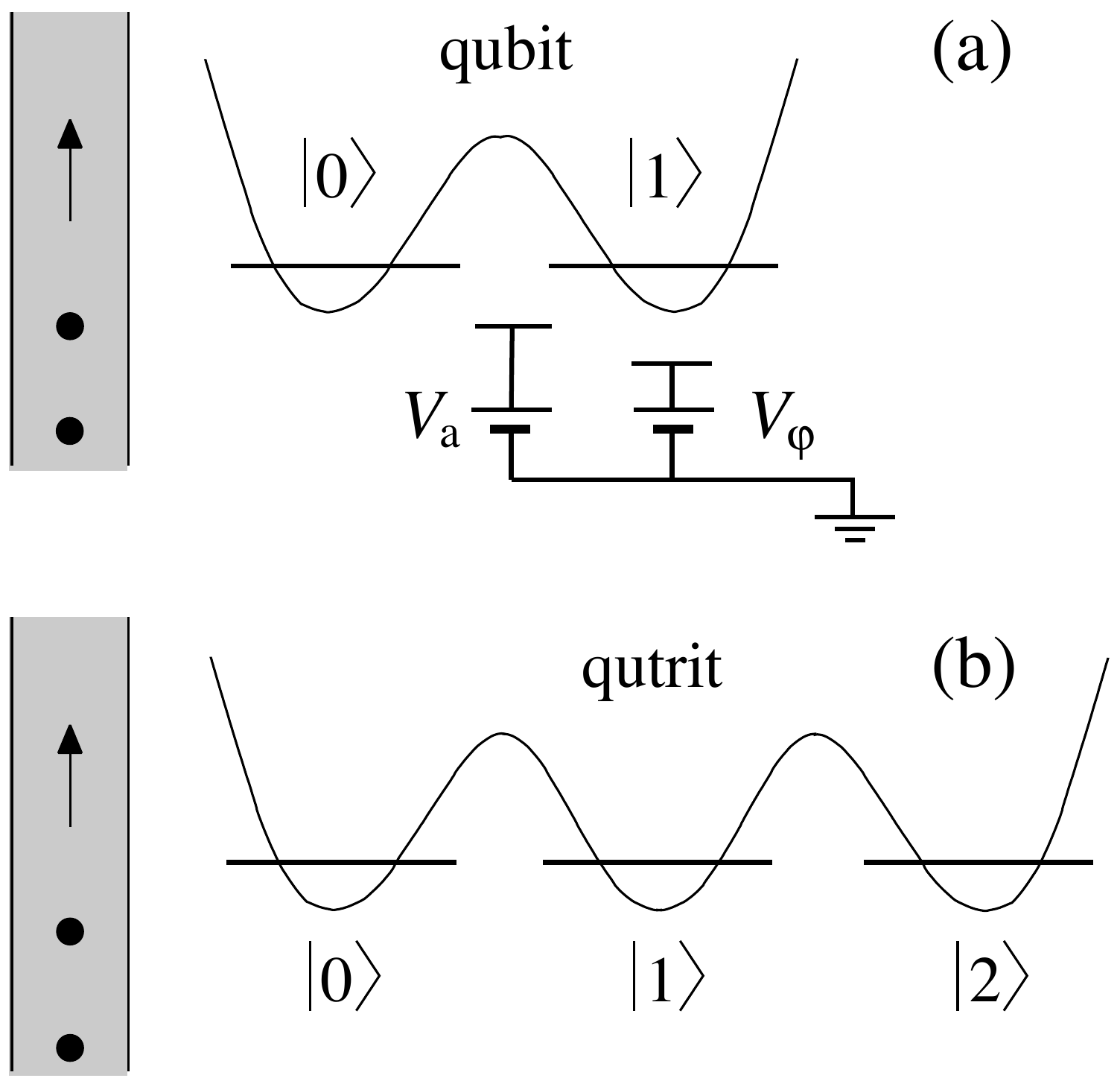}
  \caption[]{Double- and triple-well potentials defining qubits (a) and
  qutrits (b). The qubit is manipulated via voltage pulses $V_\mathrm{a}$
  lowering the barrier between the wells and changing the amplitude, or voltage
  pulses $V_\varphi$ disbalancing the wells and changing the phase of the
  qubit's state (and similar for qutrits).}
  \label{fig:d_t_well}
\end{figure}

As a first step towards the generalization to qutrits, we reformulate the
base 2 counting algorithm in terms of manipulations of a particle in a double well
potential with semi-classical (computational) states $|0\rangle$ and
$|1\rangle$, cf.\ Fig.\ \ref{fig:d_t_well}(a), as directly realizable with a
double-dot charge qubit \cite{Hayashi,Petta,Gorman}.

We assume the double-dot charge qubit to be aligned perpendicular to the wire,
such that the two wells couple differently to the charge of the passing
electrons, cf.\ Fig.\ \ref{fig:d_t_well}(a). We work with the quasi-classical
states $|0\rangle\leftrightarrow |\!\uparrow\rangle$ and $|1\rangle
\leftrightarrow |\!\downarrow\rangle$ (our computational basis) and consider a
`phase mode operation' of the counter, with a large barrier separating the
quasi-classical states, resulting in an exponentially small tunneling
amplitude $\propto \Delta$, with $2\Delta$ the gap between the two true
eigenstates.  The qubit is manipulated by applying voltage pulses, either to
lower the barrier between the two wells in order to change the amplitude
(rotation around the $x$-axis; e.g., opening a gap $2\Delta$ between the qubit
eigenstates during the time $t= \hbar \pi/2\Delta$ moves the state $|0\rangle$
to the state $|1\rangle$) or to disbalance the two wells in order to change
the phase (rotation around the $z$-axis; e.g., lifting the right well by
$\delta$ during the time $t=\hbar \varphi/\delta$ adds a relative phase
$\exp(-i\varphi)$ to the state $|1\rangle$).

The algorithm involves three steps: To {\it prepare} the qubit, we start from
the semi-classical state $|0\rangle$ and apply the unitary (amplitude-shift)
operator $\mathsf{A}_{10} = \exp(-i\mathsf{H}_{10}t/\hbar)$ with the tunneling
Hamiltonian $\mathsf{H}_{10}= - \Delta (|0\rangle \langle 1|+|1\rangle \langle
0|)$ and the time $t=\hbar \pi/4\Delta$ to produce the balanced superposition
$|\Psi_0\rangle = ((|0\rangle +i|1\rangle)/\sqrt{2}$,
\begin{eqnarray}\label{eq:qubits_U10}
   \mathsf{A}_{10} &=& \exp(-i\mathsf{H}_{10}t/\hbar)|_{t=\hbar \pi/4\Delta}\\
   &=& \frac{1}{\sqrt{2}}\left(
          \begin{array}{cc}
          1 & i \\
          i & 1 
          \end{array}
          \right);
    \nonumber
\end{eqnarray}
in a spin language this corresponds to a rotation around the $x$-axis by
$-\pi/2$ to produce the initial state $|+y\rangle$.

Next, we let the particles pass the (first) counter qubit. Upon {\it passage
of one particle}, the semi-classical qubit states $|\nu\rangle$ pick up a
phase $\exp (2\pi i \nu /2)$. The unitary operator $\mathsf{C}_1 = \exp[i\pi
(0\,|0\rangle \langle 0|+1\,|1\rangle \langle 1|)]$ then takes $|\Psi_0
\rangle$ into $|\Psi_1\rangle = \mathsf{C}_1|\Psi_0\rangle = (|0\rangle +
ie^{i\pi} |1\rangle)/\sqrt{2}$; the passage of a second particle brings the
state $|\Psi_1\rangle$ back to $|\Psi_0\rangle$, hence $\mathsf{C}_1$ is cyclic,
$\mathsf{C}_1^2 = 1$. This operation corresponds to the rotation of the spin by
$\pi$ around the $z$-axis. The states $|\Psi_0 \rangle$ and $|\Psi_1\rangle$
define the counting basis.

Finally, we translate the {\it measuring} process. The application of the
operator $\mathsf{A}_{10}^{-1}= \mathsf{A}_{10}^\dagger$ transforms
$|\Psi_0\rangle$ back to $|0\rangle$ and $|\Psi_1\rangle$ to $|1\rangle$ (up
to a phase); a simple check that the qubit is in the state $|0\rangle$ tells,
that the number $n$ of passed particles is even.

The subsequent qubits $j>1$ are prepared in the same way. Since it is the task
of the qubits $j >1$ to detect the passage of groups of particles, these
counters are more weakly coupled to the wire. In particular, the operator
$\mathsf{C}_1$ for the second qubit measures the passage of pairs, hence the
phase added to the state $|1\rangle$ is $\pi/2$, and similar for the following
qubits. Regarding the final step of measurement, there are two variants: in
order to extract the maximal factor of $2^k$ in $n$, it is sufficient to apply
the operation $\mathsf{A}_{10}^\dagger$ and then check, whether the counter
resides in the state $|0\rangle$; the first qubit residing in the state
$|1\rangle$ determines the power $k$. On the other hand, if the goal is to
find the cardinality $n$, then before applying $\mathsf{A}_{10}^\dagger$ the
qubits have to be properly prepared through a `rotation around the $z$-axis'
(rather then `rotating the axis of measurement'). These `rotations' are
conveniently done incrementally and involve a phase shift of the
quasi-classical state $|1\rangle$: after measurement of the $j$-th qubit with
outcome `0' or `1', the measured value is stored as the digit $n_{K-j+1}$
(note the reversal in the sequence) and all qubits $J >j$ are given a phase
shift $-n_{K-j+1} 2^{j-1} \phi_J$ on the state $|1\rangle$. After application of
$\mathsf{A}_{10}^\dagger$, the qubit $j+1$ is measured.

An interesting subtlety concerns the possible entanglement of the qubits with
the passing particles during counting. The spin counter discussed in section
\ref{sec:qubits} is entirely unproblematic regarding this aspect, as the gauge
interaction leaves the counted particle essentially unchanged. On the other
hand, a charge qubit acting as a counter can become strongly entangled with
the counted particle: With the qubit charge in a quantum superposition, with
1/2 probability to be closeby the wire (state $|\uparrow\rangle$) and 1/2
probability to be further away $|\downarrow\rangle$, the particle passing by
is first decelerated and then reaccelerated by the qubit charge. This process
happens with different strengths depending on the qubit state. As a result,
the particle wave function may split after passing the qubit, with one part
(the fast one, $|f\rangle$) moving ahead of the other (the slow one,
$|s\rangle$).  The calculation of the probability ${\cal P}_{y}$ to find the
counter state pointing along the $y$-direction is given by the partial trace
over the particle space; if the two states  $|f\rangle$ and $|s\rangle$ are
distinguishable, $\langle f |s\rangle = 0$, then ${\cal P}_{y} = 1/2$
independent of the phase $\phi$ picked up by the counter qubit. Hence, it is
crucial that the counter does not generate a wave function splitting when the
particle passes by, i.e., the counter only works properly if $|f\rangle
\approx |s\rangle$ and the final state is essentially non-entangled.  Using
the charge qubit as a measuring device, the requirement of weak splitting
boils down to the condition\cite{expl_wpsplitting} $\phi \ll \xi k_{\rm
\scriptscriptstyle F}$, where $\phi$ is the angle of the qubit rotation
quantifying the qubit-particle interaction, $k_{\rm \scriptscriptstyle F}$ is
the Fermi wave vector in the quantum wire, and $\xi$ is the width of the wave
packet.

In a more quantitative analysis, we can consider the evolution of a
(Lorentzian, cf.\ Ref.\ \onlinecite{Keeling}) wave packet $\Psi(x)$ with
Fourier amplitudes $f(k) = \sqrt{4 \pi \xi} \exp[-(k-k_{\rm\scriptscriptstyle
F})\xi] \Theta (k-k_{\rm \scriptscriptstyle F})$ subject to the scalar field
of a charge qubit.  For simplicity, we assume that the state $|0\rangle$ acts
with a potential $V(x)$ on the wave function, while the state $|1\rangle$,
which is further away from the wire, has no influence on the particle.
Furthermore, we choose the potential $V(x)$ such that the qubit state is
rotated by $\phi$, i.e., after the passage of the particle, the initial qubit
state $|\Psi_0\rangle = (|0\rangle+i|1\rangle)/\sqrt{2}$ shall be rotated into
the state $|\Psi_\phi\rangle = (e^{i\phi}|0\rangle + i|1\rangle)/\sqrt{2}$.
Assuming a smooth potential $V(x)$, we determine the asymptotic particle-qubit
state within a semi-classical approximation and project the wave function onto
the qubit state $|\Psi_\phi\rangle$; tracing over the particle degree of
freedom, we find the probability $P_\phi = 1 - \phi^2/8k_{\rm
\scriptscriptstyle F}^2\xi^2$ to measure the qubit in the state
$|\Psi_\phi\rangle$ and the second term provides the probability to observe a
wrong result. Note that the error probability involves the square of the small
parameter $\phi/\xi k_{\rm \scriptscriptstyle F}$.

The base 3 counting with qutrits in the form of triple wells follows the same
scheme as the base 2 counting with double-dot qubits described above. We start
out with one qutrit counter, i.e., $K=1$, which initially resides in the state
$|0\rangle$, cf.\ Fig.\ \ref{fig:d_t_well}(b).  First, the qutrit is {\it
prepared} in a balanced state with equal weights in each of the three
semi-classical ground states (the phases may be chosen arbirarily). The
(amplitude-shift) operators $\mathsf{A}_{10}$ and $\mathsf{A}_{21}$ performing
this task generalize the operator $\mathsf{A}_{10}$ above. The angle
$\chi_{10} = \Delta_{10} t/\hbar$ in the operator $\mathsf{A}_{10}$ (we write
the matrices in the semi-classical basis),
\begin{eqnarray}\label{eq:qtrit_prep_10}
   \mathsf{A}_{10} 
          &=& \exp [i\chi_{10}(|0\rangle \langle 1|+|1\rangle \langle 0|)]\\
          &=& \left(
          \begin{array}{ccc}
          \cos\chi_{10} & i \sin \chi_{10}& 0 \\
          i \sin \chi_{10} & \cos\chi_{10} &0 \\
          0 & 0& 1
          \end{array}
          \right),
\end{eqnarray}
is chosen such that 2/3 of the wave function is shifted to the state
$|1\rangle$, hence $\tan \chi_{10} = \sqrt{2}$. The matrix $\mathsf{A}_{10}$
then assumes the form
\begin{equation}\label{eq:qtrit_prep_10_ex}
   \mathsf{A}_{10}
          = \frac{1}{\sqrt{3}}\left(
          \begin{array}{ccc}
          1  & i \sqrt{2} & 0 \\
          i \sqrt{2} & 1 &0 \\
          0 & 0& \sqrt{3}
          \end{array}
          \right).
\end{equation}
The operator $\mathsf{A}_{21}$ transfers weight between states $|1\rangle$
and $|2\rangle$,
\begin{eqnarray}\label{eq:qtrit_prep_21}
   \mathsf{A}_{21}
          &=& \exp [i\chi_{21}(|1\rangle \langle 2|+|2\rangle \langle 1|)] \\
          &=& \left(
          \begin{array}{ccc}
          1 & 0& 0\\
          0 & \cos\chi_{21} & i \sin \chi_{21} \\
          0 & i \sin \chi_{21} & \cos \chi_{21}
          \end{array}
          \right).
\end{eqnarray}
Its task is to take the state $\mathsf{A}_{10} |0\rangle = (|0\rangle+i
\sqrt{2}|1\rangle)/\sqrt{3}$ to the balanced state $|\Psi_0\rangle =
(|0\rangle+e^{i\varphi_1}|1\rangle+e^{i\varphi_2}|2\rangle)/\sqrt{3}$
(with approriate phases $\varphi_{1,2}$). Choosing the time such that
$\chi_{21} = \Delta_{21} t/\hbar = \pi/4$ we obtain the operator
\begin{equation}\label{eq:qtrit_prep_21_ex}
   \mathsf{A}_{21}
          = \frac{1}{\sqrt{2}}\left(
          \begin{array}{ccc}
          \sqrt{2} & 0 & 0\\
          0 & 1 & i \\
          0 & i & 1
          \end{array}
          \right),
\end{equation}
and the balanced state $|\Psi_0\rangle = \mathsf{A}_{21} \mathsf{A}_{10}
|0\rangle$ reads 
\begin{equation}\label{eq:qtrit_Psi_0}
   |\Psi_0\rangle = \mathsf{A}_{21} \mathsf{A}_{10} |0\rangle 
   = (|0\rangle+i|1\rangle-|2\rangle)/\sqrt{3}.
\end{equation}
We call the combination $\mathsf{U}_p \equiv \mathsf{A}_{21} \mathsf{A}_{10}$
the preparation operator.  The particular construction of the state
$|\Psi_0\rangle$ does not produce the lowest Fourier harmonic $|\Psi_h\rangle
= [|0\rangle + |1\rangle + |2\rangle]/\sqrt{3}$; if we insist to work with the
canonical expression for the Fourier transformation, we have to properly
redefine the phases of the computational basis states, $|0\rangle \to
|0\rangle$, $|1\rangle \to -i|1\rangle$, and $|2\rangle \to -|2\rangle$.
Otherwise, we can actually start our counting basis with any Fourier harmonic
of the computational basis or even with any other balanced state as it is
produced in a convenient physical preparation step (as we have done above).
Also note that our preparation operator $\mathsf{U}_p$ is not acting as a
Fourier transformation on the other basis states but is a simpler operator
that only transforms the state $|0\rangle$ into a properly balanced state.

Next, we find the wave functions $|\Psi_1\rangle$ and $|\Psi_2\rangle$ after
{\it passage of 1 and 2 particles}: Upon passage of one particle, the
semi-classical qutrit states $|\nu\rangle$ pick up a phase $\exp (2\pi i \nu
/3)$. The unitary operator
\begin{equation}\label{eq:qtrit_C}
   \mathsf{C}_1 = \exp[(2\pi i/3) (0\,|0\rangle \langle 0|
   +1\,|1\rangle \langle 1|+2\,|2\rangle \langle 2|)]
\end{equation}
generates the additional counting states 
\begin{eqnarray}\label{eq:qtrit_Psi_12}
  &&\!\!\!\!|\Psi_1\rangle = \mathsf{C}_1 |\Psi_0\rangle
   =\frac{|0\rangle+ie^{2\pi i/3}|1\rangle-e^{4\pi i/3}|2\rangle}{\sqrt{3}}, \\
  &&\!\!\!\!|\Psi_2\rangle = \mathsf{C}_1 |\Psi_1\rangle
   = \frac{|0\rangle+ie^{4\pi i/3}|1\rangle-e^{8\pi i/3}|2\rangle}{\sqrt{3}}.
\end{eqnarray}
The further application of $\mathsf{C}_1$ brings us back to $|\Psi_0\rangle$,
i.e., $\mathsf{C}_1$ is cyclic, $\mathsf{C}_1^3 = 1$ (i.e., $\Theta = 0$).  It
is easy to check that the set $|\Psi_0\rangle$, $|\Psi_1\rangle$,
$|\Psi_2\rangle$ forms a new orthonormalized basis in the qutrit's Hilbert
space spanned by the semi-classical states $|0\rangle$, $|1\rangle$,
$|2\rangle$.

We come to the {\it readout} step. We define the inverse preparation operator
$\mathsf{U}_p^{-1} \equiv [\mathsf{A}_{21}\mathsf{A}_{10}]^{-1} =
\mathsf{A}_{10}^\dagger \mathsf{A}_{21}^\dagger$; using the explicit forms in
Eqs.\ (\ref{eq:qtrit_prep_10_ex}) and (\ref{eq:qtrit_prep_21_ex}) we obtain
the expression
\begin{equation}\label{eq:qtrit_prep_p_ex}
   \mathsf{U}_{p}^{-1}
          = \frac{1}{\sqrt{6}}\left(
          \begin{array}{ccc}
          \sqrt{2} & -i\sqrt{2} & -\sqrt{2}\\
          -2i & 1 & -i \\
          0 & -i\sqrt{3} & \sqrt{3}
          \end{array}
          \right).
\end{equation}
The application of $\mathsf{U}_p^{-1}$ takes the state $|\Psi_0\rangle$ back
to $|0\rangle$, while $\mathsf{U}_p^{-1} |\Psi_{1,2}\rangle$ are still 
superpositions of the semi-classical states $|1\rangle$ and $|2\rangle$, 
\begin{eqnarray}\label{eq:qtrit_U_p}
   \mathsf{U}_p^{-1} |\Psi_0\rangle &=& |0\rangle, \\
   \mathsf{U}_p^{-1} |\Psi_1\rangle &=& -i(|1\rangle -|2\rangle)/\sqrt{2}, \\
   \mathsf{U}_p^{-1} |\Psi_2\rangle &=& -i(|1\rangle + |2\rangle)/\sqrt{2}.
\end{eqnarray}
The {\it divisibility test} by 3 then involves the application of
$\mathsf{U}_p^{-1}$ to the state $\mathsf{C}_1^n |\Psi_0\rangle$ obtained
after passage of the $n$ particles and a measurement of the qutrit state; if
the qutrit resides in the state $|0\rangle$, then $n$ is divisible by 3.

Finding the {\it modulus of $n$} to the base 3 is slightly more involved. We
have to rotate the states $\mathsf{U}_p^{-1} |\Psi_{1,2}\rangle$ such as to
recover the original states $|1\rangle$ and $|2\rangle$ of the computational
basis.  Using the spin language in the two-dimensional space spanned by
$|1\rangle$ and $|2\rangle$, this operation is achieved by a `rotation' by
$\pi/2$ around the $z$-axis, followed by a rotation by $-\pi/2$ around the
$x$-axis. The former operation is executed by the phase operator
$\mathsf{P}_2 \equiv \exp(i(\pi/2) |2\rangle \langle 2|)$, while the
latter is noting but the operation $\mathsf{A}_{21}$. The combination
\begin{eqnarray}\label{eq:qtrit_mod_3}
   \mathsf{M} &=& \mathsf{A}_{21}\mathsf{P}_2 \mathsf{U}_p^{-1} \\
              &=& \frac{1}{\sqrt{3}}\left(
              \begin{array}{ccc}
              1 & -i & -1\\
              -i & -e^{4\pi i/3} & ie^{2\pi i/3} \\
              1 & -ie^{2\pi i/3} & -e^{4\pi i/3}
              \end{array}
              \right)
   \nonumber
\end{eqnarray}
then determines the modulus of $n$, since 
\begin{eqnarray}
   \nonumber
   \mathsf{M} |\Psi_0\rangle &=& |0\rangle, \\ 
   \nonumber
   \mathsf{M} |\Psi_1\rangle &=& -i|1\rangle, \\
   \nonumber 
   \mathsf{M} |\Psi_2\rangle &=& |2\rangle.
\end{eqnarray}
Depending on the outcome $|0\rangle$, $|1\rangle$, or $|2\rangle$ after the
measurement of the qutrit, the number $n$ is divisible by 3 modulo 0, 1, or 2.
The modulus can be found with a sequential measurement scheme: after
application of $\mathsf{U}_p^{-1}$ and measurement of the state $|0\rangle$,
the outcome tells that the modulus is 0 (if the particle is found in state
$|0\rangle$) or 1 or 2 (if the particle is not detected); in the latter case
the operator $\mathsf{A}_{21}\mathsf{P}_2$ is applied and the measurement of
well $|1\rangle$ will provide the final result, with a modulus 1 if the
particle is detected in state $|1\rangle$ and a modulus 2 if it is not found
(then the particle resides in state $|2\rangle$).

In the above derivation, we have adopted those phases which naturally appear
in the simplest manipulation of the three-well system---as a result, the
matrix Eq.\ (\ref{eq:qtrit_mod_3}) is not the canonical (inverse) Fourier
transform. The following steps (for the general case with $N$ counting states)
then relate the obtained matrix $\mathsf{M}$ with the canonical form of
$\mathrm{F}^{-1}$. Chosing phases $\alpha_k$ in the definition of eigenstates
$|k\rangle$ of the counting operator $\mathsf{C}_1$, we interrelate the
counting and computational bases through
\begin{equation}\label{eq:Psi_n}
   |\Psi_n\rangle = \frac{1}{\sqrt{N}}
   \sum_k e^{2\pi i\, kn/N} e^{-i\alpha_k} |k\rangle.
\end{equation}
Second, let us assume that our physical manipulations have produced a
measurement operator $\mathsf{M}$ such that
\begin{equation}\label{eq:MPsi_j_j}
   \mathsf{M}|\Psi_n\rangle = e^{i\beta_n} |n\rangle.
\end{equation}
Then the measurement operator $\mathsf{M}$ and the canonical inverse
Fourier transform $\mathrm{F}^{-1}$ are related via
\begin{equation}\label{eq:M_PFP}
   \mathsf{M} = \mathsf{P}_{[\beta]} \mathrm{F}^{-1} 
   \mathsf{P}_{[\alpha]}
\end{equation}
with the unitary (phase) operators
\begin{equation}\label{eq:phase_chi}
   \mathsf{P}_{[\chi] }
              = \left(
              \begin{array}{cccc}
              e^{i\chi_0} & 0 & \dots & 0\\
              0& e^{i\chi_2} & \dots & 0\\
              \vdots & \vdots & \vdots & \vdots\\
              0 & 0 & \dots & e^{i\chi_{N-1}}
              \end{array}
              \right)
\end{equation}
with phases $[\chi] = [\alpha], [\beta]$.  For the particular situation of the
qutrit ($N=3$), the phases $[\alpha]$ and $[\beta]$ are defined by the
operators $\mathsf{U}_p$, Eq.\ (\ref{eq:qtrit_Psi_0}), and $\mathsf{M}$, Eq.\
(\ref{eq:qtrit_mod_3}). Using these phases in Eq.\ (\ref{eq:M_PFP}), one
easily verifies that the expression Eq.\ (\ref{eq:qtrit_mod_3}) indeed
corresponds to the inverse Fourier transformation $\mathsf{F}^{-1}$.

The generalization of the algorithm to $K$-qutrit registers follows the same
steps as above.  We assume that all elements in the $K$-qutrit register
initially reside in the state $|0\rangle_j$, i.e., the state of the register
encodes the state $|0\rangle_{\scriptscriptstyle Q}$ of the computational
basis.  The qutrits $j>1$ serve the counting of particle clusters: the qutrit
$j$ counts groups of $3^{(j-1)}$ particles, hence the elementary phase shift
in $\mathsf{C}_1$ is $\exp(2\pi i/3^j)$.  Correspondingly, subsequent qutrits
are each coupled to the wire a factor 3 less than the previous. Let us go
through the algorithm: the {\it preparation} of the $j$-th qutrit is identical
to the first one. Upon {\it passage of a particle}, the semi-classical states
$|\nu\rangle$ in the $j$-th qutrit pick up phases $\exp(2\pi i \nu/3^j)$.

The readout for the {\it divisibility} check involves the application of
$\mathsf{U}_p^{-1}$ to all $K$ qutrits and subsequent test for the
semi-classical state $|0\rangle_j$ in each qutrit: if all qutrits $j<k+1$
reside in $|0\rangle_j$ and the qutrit $j=k+1$ does not, then $n$ is divisible
by $3^k$.

In order to recover the {\it number $n$ in the base 3 representation}, the
qutrit states need to be corrected for the passage of incomplete groups of
particles before application of $\mathsf{M}$, Eq.\ (\ref{eq:qtrit_mod_3}),
except for the first one, which is directly measured after application of
$\mathsf{M}$. The result $0$, $1$, or $2$ of the measurement is stored in
$n_K$.  Before measuring the next qutrit $j=2$, all counters $J>1$ need to be
corrected for the modulus of $n$: The states $|\nu\rangle$ of the $J$-th
qutrit are given the additional phase shifts $-2\pi n_K \nu \, 1/3^J$.
Subsequently, the operator $\mathsf{M}$ is applied to the qutrit $j=2$, its
state is measured, and the result $0$, $1$, or $2$ is stored in $n_{K-1}$.
Iterating the process, the result $n_{K-j+1}$ of the measurement of the $j$-th
qutrit is used to correct for the passage of incomplete groups of $3^j$
particles by adding phases $-2\pi n_{K-j+1} \nu \, 3^{j-1}/3^J$ to the states
$|\nu\rangle$ of the qutrits $J>j$. After application of $\mathsf{M}$ to the
qutrit $j+1$, its state is measured and the result is stored in $n_{K-j}$.

Let us briefly analyze the three operators $\mathsf{U}_p$,
$\mathsf{U}_p^{-1}$, and $\mathsf{M}$: since $\mathsf{U}_p$ merely takes the
state $|0\rangle$ into the balanced state $|\Psi_0\rangle$ (but does not act
as a Fourier transformation on the others), its inverse $\mathsf{U}_p^{-1}$
does not describe an inverse quantum Fourier transformation. Only after
augmentation of $\mathsf{U}_p^{-1}$ to the measuring operator $\mathsf{M}$,
cf.\ Eq.\ (\ref{eq:qtrit_mod_3}), we arrive at the required inverse quantum
Fourier transformation allowing for the readout of the individual qutrit
states.  The subsequent readout of the qutrit register does not require a
fully quantum inverse transformation, rather, the semi-classical version using
sequential measurements and manipulations (executing the compensation for the
passage of incomplete groups of particles) is sufficient. Below, we will
encounter other implementations, where the preparation operator $\mathsf{U}_p$
already acts as the complete quantum Fourier transformation on the
semi-classical computational states $|n\rangle$; in this case the measurement
operator is trivially given by $\mathsf{M} = \mathsf{U}_p^{-1}$. The reason
for sticking to two different operators $\mathsf{U}_p$ and $\mathsf{M}^{-1}$
in the above discussion is due to the simplicity of the preparation step when
using $\mathsf{U}_p$. This is particularly advantageous in the case where one
is interested in the divisibility of $n$ by a power $3^k$, as the readout only
involves the inverse operator $\mathsf{U}_p^{-1}$. Using the full Fourier
transformation $\mathsf{M}^{-1}$ for the preparation instead leads to a much
more difficult hardware implementation of the preparation, cf.\ the following
section \ref{sec:qudits}.

\subsection{Generalization to qudits}\label{sec:qudits}

The further generalization to base-$d$ counting with qudits follows the same
ideas as those developed for the base-2 and base-3 counting with qubits and
qutrits.  In order to set up the algorithm, we have to define the three steps
`preparation through $\mathsf{U}_p$', `counting with $\mathsf{C}_1$', and
`measurement' with the inverse Fourier transformation $\mathsf{M}$. The first
two steps are clear: the preparation of the initial balanced counting state
$|\Psi_0\rangle$ starts from the computational state $|0\rangle$ and proceeds
with the subsequent shift of weight in the wave function to the neighboring
well, always leaving behind an amplitude with weight $1/\sqrt{d}$; the
individual steps involve the operators
\begin{equation}\label{eq:A}
   \mathsf{A}^{\chi}_{k,k-1}
   =           \left(
              \begin{array}{cccccc}
              1 & 0  & \dots & \dots & 0 & 0\\
              \vdots & \vdots & \vdots & \vdots & \vdots & \vdots\\
              0 & \dots & \cos\chi & i\sin\chi& \dots & 0 \\
              0 & \dots & i\sin\chi & \cos\chi & \dots& 0\\
              \vdots & \vdots & \vdots & \vdots & \vdots & \vdots\\
              0 & 0 & \dots & \dots & 0 & 1\\
              \end{array}
              \right)
\end{equation}
with the $(k-1,k)$ non-trivial $2\times 2$ block shifting the amplitude
between the wells $k-1$ and $k$, cf.\ Eq.\ (\ref{eq:qtrit_prep_10}). This
operation is physically implemented through lowering the barrier between two
neighboring wells $k-1$ and $k$. Similarly, the counting operator
$\mathsf{C}_1$ is given by the straightforward generalization of Eq.\
(\ref{eq:qtrit_C}) involving the rotation- or phase-operators
\begin{equation}\label{eq:P}
   \mathsf{P}^{\phi}_{k}
   =          \left(
              \begin{array}{cccccc}
              1 & 0  & \dots & \dots & 0 & 0\\
              \vdots & \vdots & \vdots & \vdots & \vdots & \vdots\\
              0 & \dots & 1 & 0 & \dots & 0 \\
              0 & \dots & 0 & e^{i\phi} & \dots & 0\\
              \vdots & \vdots & \vdots & \vdots & \vdots & \vdots\\
              0 & 0 & \dots & \dots & 0 & 1\\
              \end{array}
              \right)
\end{equation}
with $\phi = 2\pi k/d$ and implemented through changing the gating bias of the
well $k$.

The implementation of the inverse Fourier transformation $\mathsf{M}$ with the
help of the physical operators Eqs.\ (\ref{eq:A}) and (\ref{eq:P}) is more
difficult but still possible, since this set of operations (gates) is
universal, i.e., given a unitary $d\times d$ matrix, it can be constructed
from a product of operators made from $\mathsf{A}^\chi_{k,k-1}$ and
$\mathsf{P}^\phi_k$.  The proof of this statement is similar to the proof of
the universality of two-level unitary gates, cf.\ Ref.\
\onlinecite{NielsonChuang}: Given an unitary $d\times d$ matrix (or operator)
$\mathsf{U}$, the idea is to reduce $\mathsf{U}$ in an iterative procedure to
the unit operator by right-multiplication with amplitude and phase operators.
E.g., consider the entries $\mathsf{U}_{0,k-1} = \beta = |\beta|
e^{i\varphi_\beta}$ and $\mathsf{U}_{0,k} = \alpha= |\alpha|
e^{i\varphi_\alpha}$ in $\mathsf{U}$ (we number rows and columns with indices
from $0$ to $d-1$). The product $\mathsf{U}\, \mathsf{P}^\phi_{k-1}\,
\mathsf{A}^\chi_{k,k-1}$ generates the new entries $\beta\, e^{i\phi}
\cos\chi+i\alpha\, \sin\chi$ and $\alpha\, \cos\chi + i\beta\, e^{i\phi}
\sin\chi$ and we can replace the new entry at the position $(0,k)$ by 0 if we
choose the angles $\chi = \arctan(|\alpha/\beta|)$ and $\phi = \pi/2 +
\varphi_\alpha - \varphi_\beta$ (if $\beta = 0$ then $\chi=\pi/2, \phi=0$ and
$\alpha=0$ implies zero angles). The new $(0,k-1)$ entry reads $\beta' = i
e^{i\varphi_\alpha} \sqrt{|\alpha|^2+|\beta|^2}$. Repeating this step $d-1$
times we replace the top row by zeros, except for the $(0,0)$ entry, which we
bring to unity with an additional phase operation. Hence $2d-1$ elementary
amplitude and phase operations take the first row to the vector
$(1,0,0,\dots,0)$ and a total of $\sum_1^d(2k-1) = d^2$ elementary operations
take the unitary operator $\mathsf{U}$ to unity; the desired operator then is
obtained by a simple inversion.

In order to illustrate the procedure, we derive the Fourier transform
$\mathrm{M}^{-1}$,
\begin{equation}\label{eq:Min}
   \mathsf{M}^{-1}
   = \frac{1}{\sqrt{3}}
              \left(
              \begin{array}{ccc}
              1 & 1  & 1 \\
              1 & e^{2\pi i/3} & e^{4\pi i/3} \\
              1 & e^{4\pi i/3} & e^{8\pi i/3} \\
              \end{array}
              \right)
\end{equation}
for the qutrit. In the implementation of the first step we find the angles
$\chi = \arctan(1) = \pi/4$ and $\phi = \pi/2$ and the product
$\mathsf{M}^{-1}_1= \mathsf{M}^{-1}\, \mathsf{P}^{\pi/2}_1\,
\mathsf{A}^{\pi/4}_{2,1}$ produces the matrix
\begin{equation}\label{eq:M1in}
   \mathsf{M}^{-1}_1
   =\frac{1}{\sqrt{3}}
              \left(
              \begin{array}{ccc}
              1 & i\sqrt{2}  & 0 \\
              1 & -i/\sqrt{2} & -i\sqrt{3/2} \\
              1 &  -i/\sqrt{2} & i\sqrt{3/2} \\
              \end{array}
              \right).
\end{equation}
The angles for the second step read $\chi = \arctan(\sqrt{2})$ and $\phi = \pi$
and we obtain the matrix $\mathsf{M}^{-1}_2= \mathsf{M}_1^{-1}\,
\mathsf{P}^{\pi}_0\, \mathsf{A}^{\chi}_{1,0}\, \mathsf{P}^{\pi}_0$
\begin{equation}\label{eq:M2in}
   \mathsf{M}_2^{-1}
   =\frac{1}{\sqrt{2}}
              \left(
              \begin{array}{ccc}
              \sqrt{2} & 0  & 0 \\
              0 & -i & -i \\
              0 & -i & i \\
              \end{array}
              \right).
\end{equation}
Next, the angles for the third step read $\chi = \pi/4$ and $\phi = \pi/2$ and
the product $\mathsf{M}^{-1}_3= \mathsf{M}_2^{-1}\, \mathsf{P}^{\pi/2}_1\,
\mathsf{A}^{\pi/4}_{2,1}$ generates the second row in the form $(0,1,0)$
already---no further phase rotation is needed in this step. Finally, we have
to compensate for the phase of the $(2,2)$ matrix element, which is done by
the additional rotation $\mathsf{M}^{-1}_4 = \mathsf{M}^{-1}_3
\mathsf{P}^{3\pi/2}_2$, and we arrive at the unit matrix.  Collecting all
factors, we obtain the measurement operator $\mathsf{M}$ expressed through
elementary shift and phase operators,
\begin{eqnarray}\label{eq:M_qutrit}
   \mathsf{M}= [\mathsf{P}^{\pi/2}_1\, \mathsf{A}^{\pi/4}_{2,1}]
   [\mathsf{P}^{\pi}_0\,\mathsf{A}^{\chi}_{1,0}\,\mathsf{P}^{\pi}_0]
   [\mathsf{P}^{\pi/2}_1\, \mathsf{A}^{\pi/4}_{2,1}][\mathsf{P}^{3\pi/2}_2].
\end{eqnarray}

The result Eq.\ (\ref{eq:M_qutrit}) corresponds to the expression Eq.\
(\ref{eq:qtrit_mod_3}) up to phases. Indeed, above, we have presented a
minimal algorithm with a preparation step $\mathrm{U}_p$ generating the first
computational state with phases $(1,i,-1)$ instead of the canonical ones [the
trivial phases $(1,1,1)$]. As a result, the final measurement operator Eq.\
(\ref{eq:qtrit_mod_3}) corresponds to the inverse Fourier transform up to
phases. Of course one could easily introduce additional phase operations
$\mathsf{P}^{\phi}_k$ and remove the non-canonical phases (with
$\mathsf{U}_p=\mathsf{P}^{3\pi/2}_1\,\mathsf{P}^{\pi}_2\,
\mathsf{A}^{\pi/4}_{2,1}\, \mathsf{A}^{\chi}_{1,0}$ with $\chi =
\arctan\sqrt{2}$), however, such additional gates only render the algorithm
more involved. Also, it is important to note that the divisibility check {\it
does not} require the implementation of the inverse Fourier transformation and
hence it is worth while to know how to implement a minimal preparation
operator $\mathrm{U}_p$.

\section{Implementations of Qutrits and Qudits}\label{sec:implementations}

The base $d$ counting and factorization algorithm obviously requires a set of
suitable and well operating elementary quantum devices, qubits, qutrits, or
qudits. Starting out from qubits and their analogy with a spin-1/2 system, the
most natural generalization is to try a spin-1 system for the implementation
of qutrits and possibly a spin-$d$ system for the qudits; although this idea
can be realized in principle for the case of a spin-1 system, the preparation
and measurement algorithm is rather complex (see below) and should be viewed
as a Gedanken experiment rather than a realistic proposal.  The next idea then
is to generalize the concept of the double-dot charge qubit---this road has
been pursued above and works fine in theory, however, the implementation of
multi-dot charge qubits may turn out difficult. As an alternative, one may try
to emulate a $d$-spin qudit ($d$-level system) by a system of spin-1/2 qubits
(two-level systems). Provided we admit two-qubit
interactions in our manipulation scheme, we then are able to define all the
necessary operations required by the algorithm.  Hence, we can offer a
scalable route for the implementation of qudits using qubits as elementary
units.  Below, we will discuss the various issues related to the
implementation of qutrits and qudits in more detail, putting our main emphasis
on the understanding of the qutrits, their implementation and manipulation.

\subsection{Spin-1 qutrit} \label{sec:imp_spin_1}

The most straightforward attempt to generalize the base 2 counting with qubits
(spin-1/2 two-level systems) to the base 3 counting with qutrits makes use of
a spin-1 three-level system with the orthogonal (computational) basis
$|l,m\rangle_z = |1,1\rangle_z = |0\rangle$, $|1,0\rangle_z = |1\rangle$, and
$|1,-1\rangle_z = |2\rangle$, where $l$ and $m$ denote the angular momentum
and magnetic quantum numbers. As in the previous cases (two-level qubit,
three-level qutrit), we have to prepare the system in the initial counting
state $\Psi_0$, e.g., the lowest harmonic of the computational basis or
another balanced state. The logic taking us to a valid balanced state is the
following: {\it i)} we first note, that experimentally we can produce either
axial states (polarized along a direction  ${\bf n}$) such as
$|1,1\rangle_{\bf n}$, or planar polarized states $|1,0\rangle_{\bf n}$,
hence we will focus on these types of states.  {\it ii)} As 
one cannot construct axial states with equal weights for all basis states
$|0\rangle, |1\rangle, |2\rangle$, we concentrate on the planar states.
{\it iii)} the most general planar state (with a director parallel to ${\bf
n}$ defined by the direction angles $\varphi$ and $\theta$) assumes the form
(in the computational basis)
\begin{equation}\label{eq:planar_n}
  |1,0\rangle_{\bf n} = \frac{1}{\sqrt{2}}
              \left(
              \begin{array}{ccc}
              -\sin\theta \,e^{-i\varphi}\\
              \noalign{\vskip 3 pt}
              \sqrt{2}\,\cos\theta \\
              \noalign{\vskip 3 pt}
              \sin\theta \,e^{i\varphi} 
              \end{array}
              \right).
\end{equation}
We demand that all components have equal weights, hence $\theta = \arctan
\sqrt{2}$; choosing $\varphi = \pi/4$, we obtain the balanced state
\begin{equation}
   |\Psi_0\rangle = \frac{-e^{-i\pi/4}|0\rangle+|1\rangle
                   +e^{i\pi/4}|2\rangle}{\sqrt{3}},
\end{equation}
the planar state $|1,0\rangle_{{\bf e}_c}$ with a director ${\bf e}_c =
(1,1,1)/\sqrt{3}$ pointing along the body diagonal; other values of $\varphi$
correspond to another direction ${\bf e}_c$ rotated around the $z$-axis.

The other counting states then are obtained by rotating $|\Psi_0\rangle$ by
the angle $2\pi/3$ (anti-clockwise) around the $z$-axis, $|\Psi_1\rangle =
U_z(2\pi/3) |\Psi_0\rangle$ and $|\Psi_2\rangle = U_z(2\pi/3) |\Psi_1\rangle$;
physically, this rotation is achieved by ensuring that passing electrons
create a local magnetic field pulse along the $z$-axis.  To simplify our
further discussion, we exchange the two axes, the one for the preparation and
the one defining the counting field: our task then is to generate an initial
planar polarized state $|1,0\rangle_z$ and implement the counting step through
rotation by an angle $2\pi/3$ (anti-clockwise) around the axis ${\bf e}_c$.

While the creation of axially polarized states is rather straightforward,
creating a planar polarized state $|1,0\rangle_{\bf n}$ is non-trivial.
Indeed, while we can make use of a simple deterministic procedure to prepare
an axially polarized state (by switching on a magnetic field and relaxing the
spin through coupling to a bath), the planar polarized state required here is
more difficult to obtain.  A simple preparation can be implemented with the
help of a Stern-Gerlach apparatus, however, this procedure is a statistical
one, with a one-half probability for a positive outcome. In fact, preparing an
initial state polarized along $x$, $|1,1\rangle_x = (|1,1\rangle_z
+|1,-1\rangle_z)/2+ |1,0\rangle_z/\sqrt{2}$, using a Stern-Gerlach setup
directed along $z$, and selecting particles with a straight trajectory (i.e.,
selecting the `middle spot' in the Stern-Gerlach setup) we obtain the desired
spin state $|1,0\rangle_z$ (a procedure to generate a proper initial state
with unit probability is discussed later).
\begin{figure}[ht]
 \includegraphics[width=5.0cm]{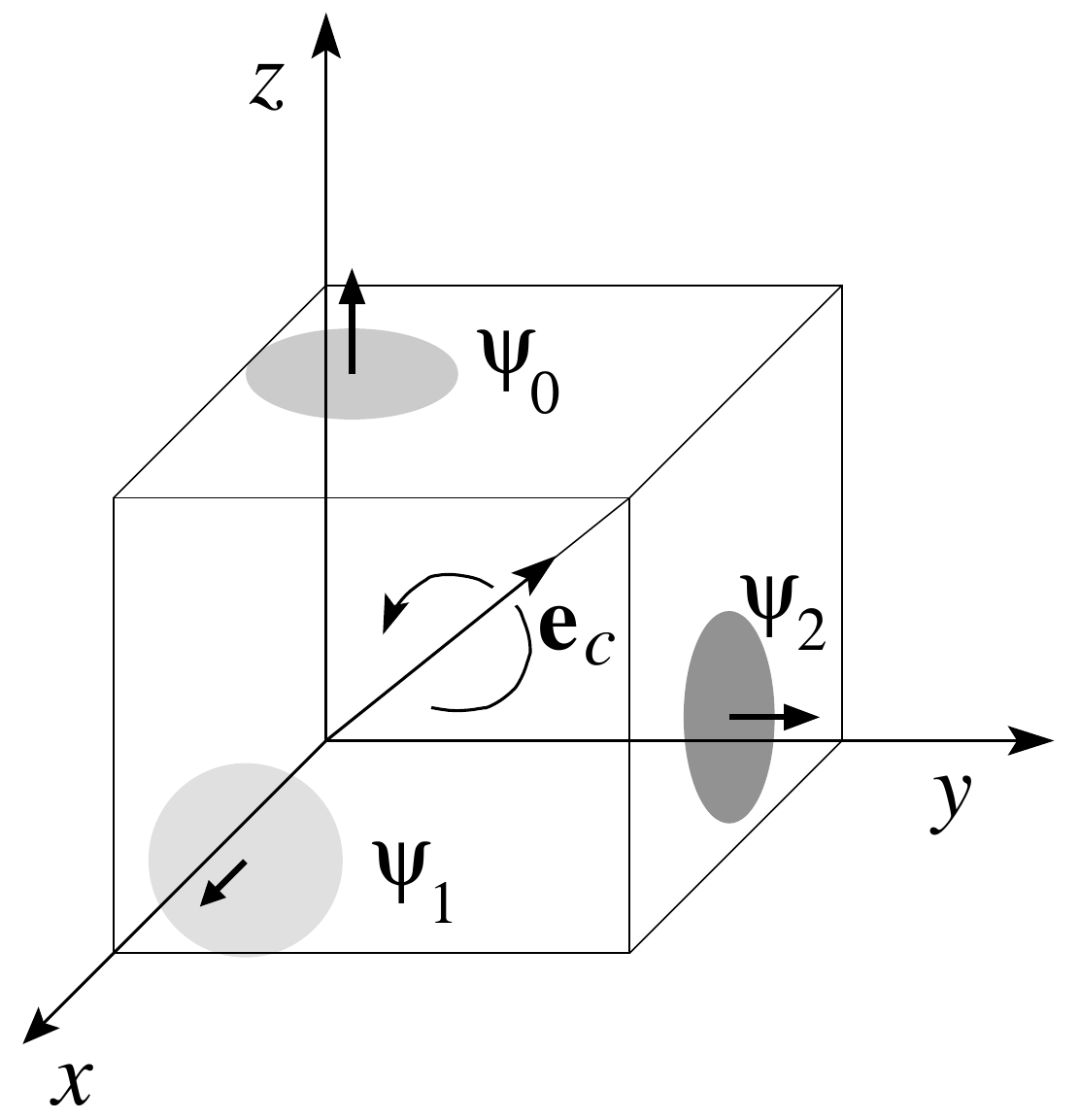}
  \caption[]{\label{fig:spin1} Spin-1 qutrit. After preparation in the state
  $|1,0\rangle_z$, the system is in a state of planar ($xy$) polarization,
  i.e., the spin has no component along $z$. Upon passage of a particle, the
  polarization plane is rotated (by the angle $2\pi/3$ around ${\bf e}_c$)
  from $xy$ to $yz$, for the second particle to $zx$, and for the third back
  to $xy$. The divisibility test involves a measurement with a Stern-Gerlach
  apparatus with the field directed along the $z$-axis.}
\end{figure}

With the counting state $|\Psi_0\rangle=|1,0\rangle_z$ properly prepared, we
define the counting step through a rotation by the angle $2\pi/3$ around the
axis ${\bf e}_c = (1,1,1)/\sqrt{3}$, cf.\ Fig.\ \ref{fig:spin1}, 
\begin{eqnarray}\label{eq:counting_rot}
   \mathsf{C}_1 &=& \exp[-i(2\pi/3){\bf e}_c\cdot{\bf L}/\hbar] \\
   \nonumber
   &=& \exp\left[-\frac{2\pi i}{3\sqrt{3}}
              \left(
              \begin{array}{ccc}
              1 & e^{-i\pi/4} & 0\\
              e^{i\pi/4} & 0 & e^{-i\pi/4} \\
              0 & e^{i\pi/4} & -1
              \end{array}
              \right)\right].
\end{eqnarray}
As the three-fold iteration takes us back to the original state
$|1,0\rangle_z$, our counting operator is cyclic and generates the complete
orthogonal counting basis
\begin{eqnarray}\label{eq:spin1_count}
   |\Psi_0\rangle &=& |1,0\rangle_z, \\
   \nonumber
   |\Psi_1\rangle &=& |1,0\rangle_x, \\
   \nonumber
   |\Psi_2\rangle &=& |1,0\rangle_y.
\end{eqnarray}

\begin{figure}[ht]
 \includegraphics[width=8.0cm]{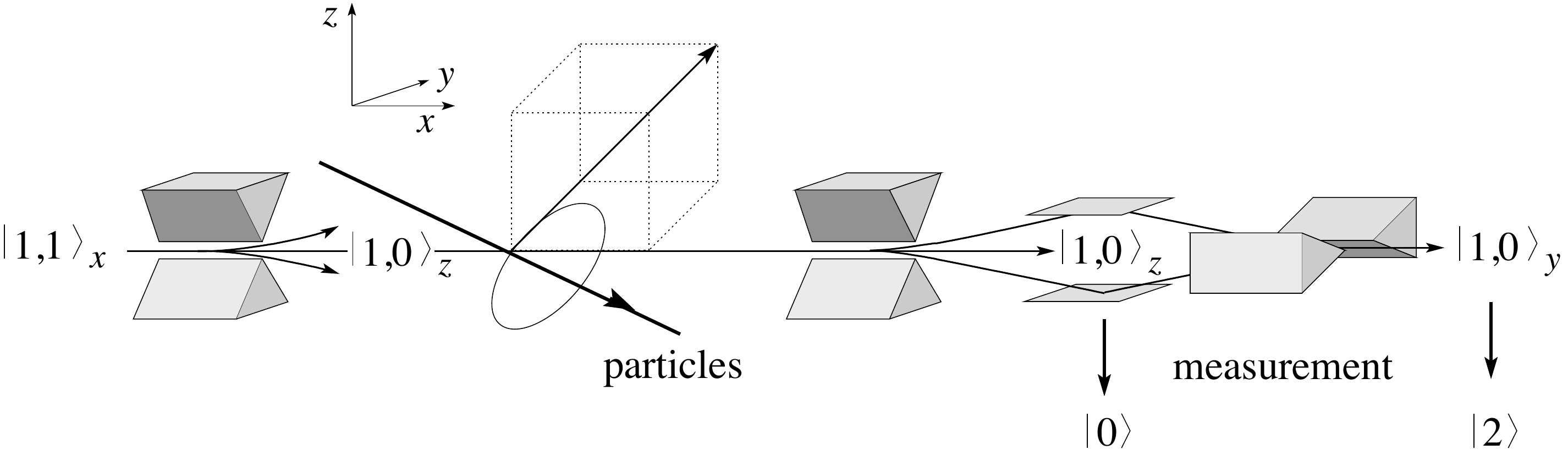}
  \caption[]{\label{fig:SG} Setup for the counting with a spin-1 qutrit.
  Sending the initial polarized state $|1,1\rangle_x$ into a first
  Stern-Gerlach (SG) apparatus with axis $z$ and selecting the middle
  (undeflected) trajectory, we obtain the planar state $|1,0\rangle_z$ with
  probability 1/2.  The counted particles passing by rotate this state around
  the axis ${\bf e}_c= (1,1,1)/\sqrt{3}$ and generate the planar counting
  states $|1,0\rangle_z$, $|1,0\rangle_x$, and $|1,0\rangle_y$ (note that the
  counter particle has to be stopped and trapped during the time of
  interaction with the particles passing by). These are analyzed in a second
  SG apparatus with axis $z$---if the spin is not deflected, the number $n$ of
  counted particles is divisible by 3. Combining the two deflected beams and
  analyzing the superposition in a further SG apparatus directed along the
  $y$-axis, we find the mudulus of $n$, which is 2 if the spin is undeflected
  and 1 else.}
\end{figure}

The measurement step for the divisibility check after the passage of the
particles involves a second Stern-Gerlach experiment directed along the
$z$-axis---if the particle moves again on the straight trajectory, its
polarization was unchanged by the passage of the $n$ particles and hence $n$
is divisible by 3. The other states $|1,0\rangle_x = (|1,-1\rangle_z -
|1,1\rangle_z)/ \sqrt{2}$ and $|1,0\rangle_y = i(|1,1\rangle_z +
|1,-1\rangle_z)/ \sqrt{2}$, cf.\ Eq.\ (\ref{eq:planar_n}), give no
contribution to the signal on the straight trajectory, cf.\ Fig. \ref{fig:SG}.

As usual, the measurement of the number's modulus (the counting measurement)
is more involved. We then make use of the other outcomes $|1,1\rangle_z$ and
$|1,-1\rangle_z$ of the second Stern-Gerlach experiment and bring them to
interference further down their trajectories, cf.\ Fig.\ \ref{fig:SG}. Testing
the resulting state, e.g., the state $|1,0\rangle_y = i (|1,1\rangle_z +
|1,-1\rangle_z) /\sqrt{2}$ (we choose symmetric trajectories with equal
phases) in a third Stern-Gerlach apparatus polarized along the $y$-axis then
identifies the counting state $|1,0\rangle_y$ on the straight trajectory (the
middle spot, the number's modulus is 2); if no spin is measured, the counting
state $|1,0\rangle_x$ has been realized and the number's modulus is 1. 

Finally, we look for a preparation step with unit efficiency. This can be
achieved by making further use of the states $|1,\pm 1 \rangle_z$ behind the
first Stern-Gerlach device (we still use the input state $|1,1\rangle_x$ to
the Stern-Gerlach apparatus): superimposing these two states, we obtain the
component $i(|1,1\rangle_z+|1,-1\rangle_z)/2 = |1,0\rangle_y/\sqrt{2}$ and
rotating this planar state back to the $z$-axis, we have managed to transform
the initial spin $|1,1\rangle_x$ into two subsequent wave-packets with weight
1/2, each describing the same spin state $|1,0\rangle_z$. Hence, although we
cannot generate a planar state out of an axial state by simple rotation, the
use of a Stern Gerlach apparatus and proper manipulation of all three
amplitudes of the split wave function allows one to generate the desired
planar state, though split into two wave-packets with weight 1/2 each.

Obviously, the above scheme is a very complex one and should be regarded as a
`Gedanken' experiment rather than a realistic setup.  Nevertheless, it is
interesting to see, that a spin-1 qubit can, at least in principle, be used
for the implementation of the base-three counting algorithm.

\subsection{Triple-dot qutrit} \label{sec:imp_triple_dot}

The triple-dot qutrit has been discussed above in section \ref{sec:qutrits}.
Preparation, counting, and readout can be properly implemented via voltage
pulses acting on the semi-classical states (phase shifts) or on the barriers
in between (amplitude shifts). A drawback is the need to design a new device
when going from base 2 to base 3 counting and, more generally, each time a new
prime factor is to be tested. The emulation of qutrits (and qudits) through
qubits described below allows one to stay with only one computational unit for
all base-$d$ counting tasks and factorizations.

\subsection{Emulation of spin-1 qutrit} \label{sec:imp_emulation}

The basic idea we pursue here is to emulate the qudits needed in the base-$d$
counting and factorization through simpler qubits. We start by combining 2
qubits into a qutrit. 

\subsubsection{Emulation using a spin triplet}\label{sec:triplet}

An obvious way to choose three appropriate states in the product Hilbert space
${\cal H}_{1/2}\otimes{\cal H}_{1/2}$ of the two qubits is to make use of the
decomposition into singlet and triplet sectors,
\begin{equation}\label{eq:H_x_H}
   {\cal H}_{1/2}\otimes{\cal H}_{1/2} = {\cal H}_{0}\oplus{\cal H}_{1},
\end{equation}
and use the three-dimensional triplet space ${\cal H}_{1}$. Contrary to the
simple spin-1 qutrit, the emulated version using two qubits provides us with
the necessary degrees of freedom to perform all of the required steps
(preparation with $\mathsf{U}_p$ and readout with $\mathsf{M}$) in the
counting and factorization algorithm with the help of one- ($\sigma_x^{(i)},
\sigma_z^{(i)}$) and two-qubit ($\sigma_z^{\rm\scriptscriptstyle
(1)}\sigma_z^{\rm\scriptscriptstyle (2)}$) operations.  Referring to the
previous paragraph, we start from the computational basis $|0\rangle=
|\uparrow\uparrow\rangle$, $|1\rangle = (|\uparrow\downarrow \rangle
+|\downarrow\uparrow\rangle)/\sqrt{2}$, and $|2\rangle = |\downarrow
\downarrow\rangle$ and seek for those manipulations which provide us with the
counting states in the form of three orthogonal planar polarized states with
balanced weights. The latter take the form Eq.\ (\ref{eq:planar_n}), whereas
the axial states can be written as (again in the computational basis)
\begin{equation}\label{eq:axial}
  |1,1\rangle_{\bf n} = \frac{1}{2}
              \left(
              \begin{array}{ccc}
              (1+\cos\theta) \,e^{-i\varphi}\\
              \noalign{\vskip 3 pt}
              \sqrt{2}\,\sin\theta \\
              \noalign{\vskip 3 pt}
              (1-\cos\theta) \,e^{i\varphi}
              \end{array}
              \right),
\end{equation}
with $\bf n$ the direction of the spin axis. Note that once these states are
chosen, the operators $\sigma_z^{(i)}$ and $\sigma_z^{\rm\scriptscriptstyle
(1)}\sigma_z^{\rm\scriptscriptstyle (2)}$ only change the relative phases in
$|1,0\rangle_{\bf n}$ and $|1,1\rangle_{\bf n}$ but not the relative weights
in the amplitudes. It is then our task to construct three orthogonal planar
states out of the computational basis, which consists of two axial and one
planar state. In this way, we can find a preparation operator $\mathsf{U}_p$
which is congruent to the Fourier transformation, i.e., $\mathsf{U}_p =
\mathsf{M}^{-1}$ transforms the computational basis $\{|k\rangle\}|_{k=0}^{2}$
into the counting basis $\{|\Psi_n\rangle\} |_{n=0}^{2}$.

\begin{figure}[ht]
 \includegraphics[width=7.0cm]{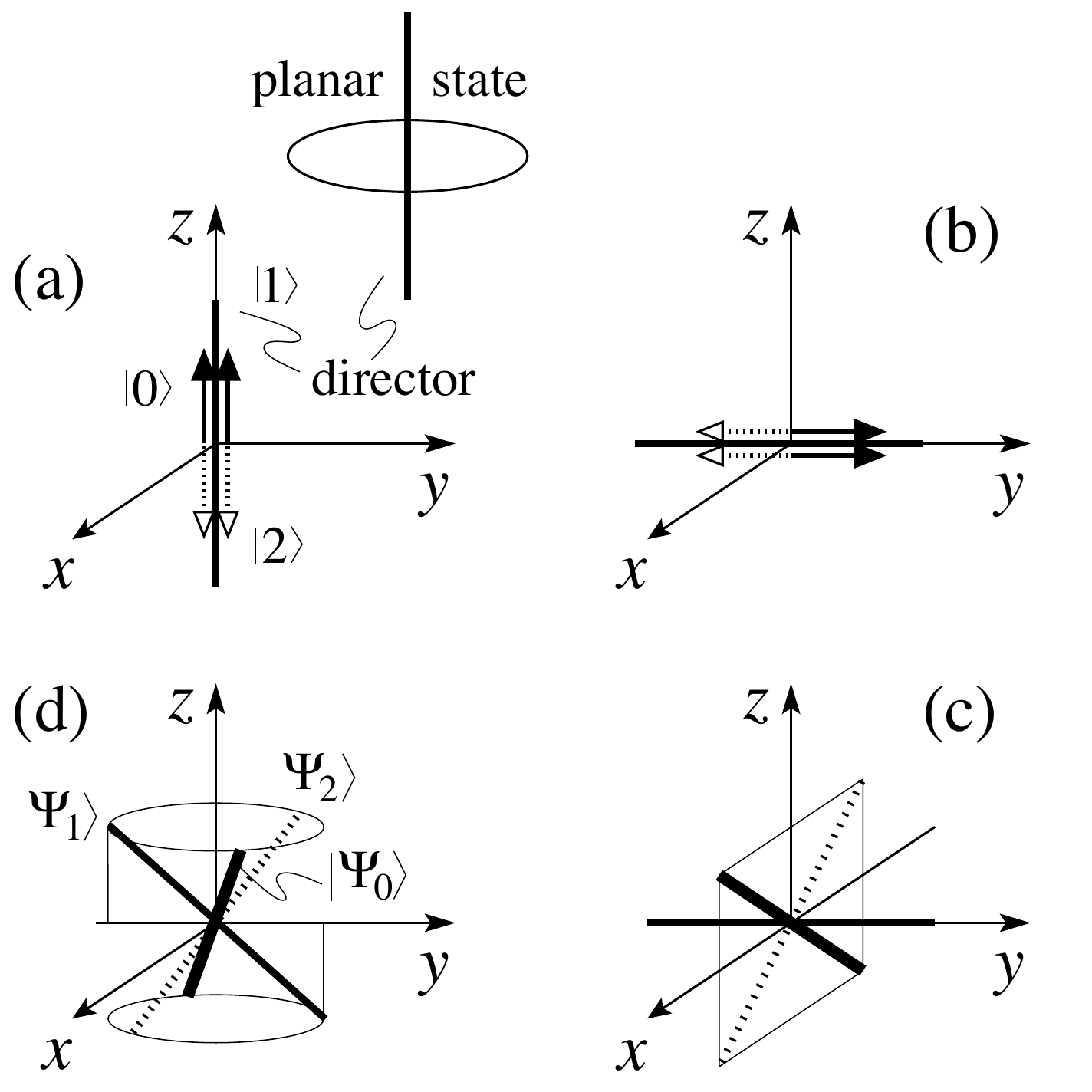}
  \caption[]{\label{fig:F_rot} Rotations generating the quantum Fourier
  transformation on the triplet sector of a two-qubit system; we denote the
  axial states $|0\rangle = |\uparrow\uparrow\rangle$ and $|2\rangle =
  |\downarrow\downarrow\rangle$ by double arrows and planar states (e.g.,
  $|1\rangle = [|\uparrow\downarrow\rangle + |\downarrow\uparrow\rangle]
  /\sqrt{2}$) by a director. (a)$\to$(b) Rotation by $-\pi/2$ around the
  $x$-axis, (b)$\to$(c) conditional rotation transforming the two axial states
  into planar ones with directors along the diagonals in the $xz$-plane,
  (c)$\to$(d) rotation by $\varphi= -\arctan(1/\sqrt{2})$ around the $x$-axis
  of three planar states to position them symmetrically around the $z$-axis.
  The last step defines the counting states which transform among one another
  through a rotation by the angle $-2\pi/3$ around the $z$-axis.}
\end{figure}

We first note, that the planar states Eq.\ (\ref{eq:planar_n}) always have the
same weight in the components $|0\rangle$ and $|2\rangle$ and hence we first
have to rotate the axial states of the computational bases into the
$xy$-plane.  This is done with the operator $e^{i\pi \sigma_x/2}$ where
$\sigma_x = (\sigma_x^{\rm\scriptscriptstyle (1)} +
\sigma_x^{\rm\scriptscriptstyle (2)})/2$ and results (up to a phase) in the
states [$\theta = \pi/2$ in Eq.\ (\ref{eq:axial})]
\begin{equation}\label{eq:axial_y}
  |1,1\rangle_{{\bf e}_y} = \frac{1}{2}
              \left(
              \begin{array}{ccc}
              1\\
              \noalign{\vskip 3 pt}
              \sqrt{2}\,i \\
              \noalign{\vskip 3 pt}
              -1
              \end{array}
              \right), \quad
  |1,1\rangle_{-{\bf e}_y} = \frac{1}{2}
              \left(
              \begin{array}{ccc}
              -1\\
              \noalign{\vskip 3 pt}
              \sqrt{2}\,i \\
              \noalign{\vskip 3 pt}
              1
              \end{array}
              \right),
\end{equation}
cf.\ Fig.\ \ref{fig:F_rot}. Note that the simple rotation takes the planar
state $|1\rangle$ into a planar state (with a director along $y$). In order to
map the axial states in Eq.\ (\ref{eq:axial_y}) to planar states we choose
$\theta = \pi/4$ in Eq.\ (\ref{eq:planar_n}) and obtain the candidate states
\begin{equation}\label{eq:planar_45}
  |1,0\rangle_{{\bf n}_{\pi/4}} = \frac{1}{2}
              \left(
              \begin{array}{ccc}
              -e^{-i\varphi}\\
              \noalign{\vskip 3 pt}
              \sqrt{2} \\
              \noalign{\vskip 3 pt}
              e^{i\varphi}
              \end{array}
              \right);
\end{equation}
the planar and axial states in Eqs.\ (\ref{eq:planar_45}) and
(\ref{eq:axial_y}) now have equal amplitudes. In order to select the
appropriate phases $\varphi$ in the planar states Eqs.\ (\ref{eq:planar_45}),
we note that the phase differences between the  $|0\rangle$ and $|2\rangle$
components of the axial states are equal to $\pm \pi$; these have to be
matched with the phase differences $\pi-2\varphi$ in the planar states and
hence we choose planar states with $\varphi = 0, \pi$,
\begin{equation}\label{eq:planar_45_diag}
  |1,0\rangle_{x,-z} = \frac{1}{2}
              \left(
              \begin{array}{ccc}
              1\\
              \noalign{\vskip 3 pt}
              \sqrt{2} \\
              \noalign{\vskip 3 pt}
              -1 
              \end{array}
              \right), \quad
  |1,0\rangle_{x,z} = \frac{1}{2}
              \left(
              \begin{array}{ccc}
              -1\\
              \noalign{\vskip 3 pt}
              \sqrt{2} \\
              \noalign{\vskip 3 pt}
              1
              \end{array}
              \right).
\end{equation}
The two states are characterized by directors pointing along the
$xz$-diagonals, cf.\ Fig.\ \ref{fig:F_rot}. In order to map the axial states
Eq.\ (\ref{eq:axial_y}) to the planar states Eq.\ (\ref{eq:planar_45_diag}) we
make use of the two-qubit operator
\begin{equation}\label{eq:U_chi}
   \mathsf{U}_\chi = e^{i\chi\sigma_z^{\rm\scriptscriptstyle
   (1)}\sigma_z^{\rm\scriptscriptstyle (2)}}.
\end{equation}
The latter adds the phases $\chi$ ($-\chi$) to the components $|0\rangle$,
$|2\rangle$ ($|1\rangle$) and hence leaves the relative phase between the
components $|0\rangle$, $|2\rangle$ unchanged while adding a relative shift
$-2\chi$ to the middle one. Hence choosing $\chi = -\pi/4$, we can map the
axial state $|1,1\rangle_{{\bf e}_y}$ to the planar state $|1,0\rangle_{x,z}$
and $|1,1\rangle_{-{\bf e}_y}$ to the planar state $|1,0\rangle_{x,-z}$, cf.\
Fig.\ \ref{fig:F_rot}. Since the component $|1\rangle$ in the planar state
$|1,0\rangle_{{\bf e}_y}$ has weight 0, the operator $\mathsf{U}_{-\pi/4}$
leaves it unchanged.

Hence we have arrived at three orthogonal planar states directed along $y$ and
along the $xz$-diagonals; a final rotation $e^{i\varphi\sigma_x}$ with
$\sigma_x = (\sigma_x^{\rm\scriptscriptstyle (1)} +
\sigma_x^{\rm\scriptscriptstyle (2)})/2$ by $\varphi = \arctan(1/\sqrt{2})$
around the $x$-axis then arranges the three states symmetrically around the
$z$-axis. Combining the three operations, we obtain the preparation and
inverse measurement operator
\begin{equation}\label{eq:U_p}
   \mathsf{U}_p = \mathsf{M}^{-1} = e^{i\varphi\sigma_x} e^{-i\pi
   \sigma_z^{\rm\scriptscriptstyle (1)} \sigma_z^{\rm\scriptscriptstyle
   (2)}/4} e^{i\pi \sigma_x/2}.
\end{equation}
The actions of the three operations in Eq.\ (\ref{eq:U_p}) on the
computational basis states $\{|n\rangle\}|_{n=0}^{2}$ are illustrated in Fig.\
\ref{fig:F_rot}; the resulting counting basis is given by the three (planar)
states $|\Psi_n\rangle = \mathsf{U}_p |n\rangle$, 
\begin{eqnarray}\nonumber
   |\Psi_0\rangle 
   &=& \frac{e^{-5\pi i/12}}{\sqrt{3}} |0\rangle
   + \frac{e^{3\pi i/4}}{\sqrt{3}}|1\rangle
   + \frac{e^{11\pi i/12}}{\sqrt{3}}|2\rangle, \\ \label{eq:U_p_psi_0}
   |\Psi_1\rangle
   &=& \frac{e^{i\pi/4}}{\sqrt{3}} |0\rangle
   + \frac{e^{i3\pi/4}}{\sqrt{3}}|1\rangle
   + \frac{e^{i\pi/4}}{\sqrt{3}}|2\rangle,
   \\ \nonumber
   |\Psi_2\rangle
   &=& \frac{e^{i 11\pi/12}}{\sqrt{3}} |0\rangle
   + \frac{e^{i3\pi/4}}{\sqrt{3}}|1\rangle
   + \frac{e^{-i5\pi/12}}{\sqrt{3}}|2\rangle.
\end{eqnarray}

One can confirm, that the {\it counting} step involving the application of the
operator $\mathsf{C}_1 = \exp(2\pi i\, \sigma_z/3)$ with $\sigma_z =
(\sigma_z^{\rm\scriptscriptstyle (1)} + \sigma_z^{\rm\scriptscriptstyle
(2)})/2$, indeed transforms the three counting states in Eq.\
(\ref{eq:U_p_psi_0}) into each other, i.e., $|\Psi_n\rangle = \mathsf{C}_1^n
|\Psi_0\rangle = \mathsf{C}_n |\Psi_0\rangle$ (note the symmetric definition
of $\sigma_z$ used here and compare with the corresponding definition of
$\mathsf{C}_1$ in the following paragraph, section \ref{sec:effemul}).

\subsubsection{Efficient emulation}\label{sec:effemul}

Although the emulation of the qutrit by the triplet sector of two qubits can
be done consistently, this scheme is inefficient in scaling up to larger
primes. Indeed, emulating a spin-2 system within this scheme requires 4
qubits, whereby most of the Hilbert space (an eleven dimensional sector) is
not used since only the component ${\cal H}_{2}$ out of the decomposition
\begin{equation}\label{eq:H^4}
   {\cal H}_{1/2}^{\otimes 4} = {\cal H}_{2}\oplus 3{\cal H}_{1}
  \oplus 2 {\cal H}_{0}
\end{equation}
is needed. A suitably scalable case should make maximal use of the emulating
qubits, i.e., use a large fraction of the Hilbert space. In choosing the
appropriate subset of states in the multi-qubit Hilbert space, we have to
select states with an equidistant spectrum. E.g., for the case $d=3$ and 2
qubits, we choose three consecutive states from $|0\rangle = |\uparrow
\,\uparrow\rangle$, $|1\rangle = |\uparrow\,\downarrow\rangle$, $|2\rangle =
|\downarrow\,\uparrow\rangle$, $|3\rangle = |\downarrow\,\downarrow\rangle$
and an elementary phase shift operator $\mathsf{C}_1 = \exp[-i\lambda
(\sigma_z^{\rm\scriptscriptstyle (1)} + \sigma_z^{\rm\scriptscriptstyle
(2)}/2)]$; upon passage of a particle, the four states then pick up the phases
$\exp(-3i\lambda/2)$, $\exp(-i\lambda/2)$, $\exp(i\lambda/2)$, and
$\exp(3i\lambda/2)$. Choosing $\lambda = 2\pi/3$ and the first three states
$|0\rangle$, $|1\rangle$, and $|2\rangle$, the (relevant sector of the)
operator $\mathsf{C}_1$,
\begin{equation}\label{eq:C12qubits}
   \mathsf{C}_1 = \left(
          \begin{array}{cccc}
             e^{-i\pi} & 0 & 0 & 0\\
             0 & e^{-i\pi/3} & 0 & 0 \\
             0 & 0 & e^{i\pi/3} & 0 \\
             0 & 0 & 0 & e^{i\pi}
           \end{array}
        \right),
\end{equation}
coincides with the expression Eq.\ (\ref{eq:qtrit_C}), up to an overall phase
$\exp(3i\lambda/2) = \exp(i\pi)$ [to be added to (\ref{eq:C12qubits})]. The
generalization of this scheme to an $E$-qubit emulation of a qudit is
straightforward,
\begin{equation}\label{eq:gen_P_1}
   \mathsf{C}_1 = \exp\Bigl[-i\lambda\sum_{l=1}^E
   \sigma_z^{\rm\scriptscriptstyle (l)}/2^{(l-1)}\Bigr].
\end{equation}

In addition to the shift operator $\mathsf{C}_1$, we need to know the form
of the preparation operator $\mathsf{U}_p$. For the above qutrit emulation,
this operator takes the form (see Appendix \ref{app:Up} for the derivation)
\begin{eqnarray}\label{eq:em_U_p}
   \mathsf{U}_p &=&
   e^{-i\pi\sigma^{\rm\scriptscriptstyle (2)}_{x}/4}
   e^{i\pi\sigma^{\rm\scriptscriptstyle (1)}_{z}
          \sigma^{\rm\scriptscriptstyle (2)}_{z}/8}
   e^{i\pi\sigma^{\rm\scriptscriptstyle (2)}_{z}/8} \\ \nonumber
   &&\qquad
   e^{i\pi\sigma^{\rm\scriptscriptstyle (2)}_{x}/4}
   e^{i\pi\sigma^{\rm\scriptscriptstyle (1)}_{z}/4}
   e^{i\theta\sigma^{\rm\scriptscriptstyle (1)}_{x}/2}
\end{eqnarray}
with the angle $\theta = 2\arctan(1/\sqrt{2})$. As this is not yet the full
Fourier transformation on the computational basis $|k\rangle$, $k = 0,1,2$, we
still have to find the measurement operator $\mathsf{M}$. In order to
accomplish this task, one notes that the three counting states
$|\Psi_n\rangle$, $n=0,1,2$ are entangled, while the three computational
states $|k\rangle$, $k=0,1,2$ are not. The inverse Fourier transformation
$\mathsf{M}$ then has to disentangle the counting states, a criterion
which helps us in finding its explicit form. In the end, the measurement
operator $\mathsf{M}$ is obtained in the form of a product of three unitary
operations $\mathsf{U}_i$, $i=0,1,2$,
\begin{equation}\label{eq:MUUU}
   \mathsf{M} = \mathsf{U}_0 \mathsf{U}_1 \mathsf{U}_2,
\end{equation}
where $\mathsf{U}_2$ and $\mathsf{U}_1$ serve to disentangle the three
counting states $|\Psi_n\rangle$, $n=0,1,2$, and $\mathsf{U}_0$ is a
conditional Hadamard operation, turning the spin of qubit 2 into the $z$-axis
if the qubit 1 is in the state $|\uparrow\rangle$.  The detailed derivation of
Eqs.\ (\ref{eq:em_U_p}) and (\ref{eq:MUUU}) and the form of the operators
$\mathsf{U}_i$, $i=0,1,2$, is given in Appendix \ref{app:Up}.

\section{Relation to phase estimation algorithm and its application}
\label{sec:PEA}

It turns out, that our counting algorithm has much in common with the phase
estimation algorithm (PEA); the following discussion of the PEA is formulated
in a way as to make this connection apparent. The phase estimation algorithm
first appeared as a part of Shor's factorization algorithm \cite{Shor_94}; an
extended separate algorithm was presented by Kitaev \cite{Kitaev} and later by
Cleve et al.\ \cite{Cleve}. The  phase estimation algorithm attempts to find
the `phase' $0 \leq \varphi < 1$ in the eigenvalue $\exp(2\pi i \varphi)$ of a
unitary operator $\mathsf{U}$ associated with a given eigenvector $|u\rangle$.
In the version of Refs.\ \onlinecite{NielsonChuang,Cleve}, this is achieved
with the help of two qubit-registers, one of which (the second) is storing the
vector $|u\rangle$ and acts on it with the operators $\mathsf{U}^{2^{j-1}}$,
$j = 1, \dots, K$, to generate the phases $\exp(2\pi i\, 2^{j-1} \varphi)$.
The other (first) register consists of $K$ qubits and produces the desired
phase estimate in the following manner: with all qubits initialized in the
state $|0\rangle_j$, a Hadamard operation generates the balanced states
$(|0\rangle_j+|1\rangle_j) /\sqrt{2}$ for all qubits in the first register. A
controlled $\mathsf{U}^{2^{K-j}}$ operation between the second register and
the $j$-th qubit in the first register then puts the qubit into the state
$(|0\rangle_j+\exp(2\pi i\, 2^{K-j} \varphi)|1\rangle_j) /\sqrt{2}$, hence
generating the quantum Fourier transform (we express the product state through
the computational basis and assume that the phase $\varphi$ can be represented
by $K$ binary digits)
\begin{equation}\label{eq:PE}
   \mathsf{F}(|2^K \varphi\rangle) = 
   \frac{1}{\sqrt{2^K}}\sum_{k} e^{2 \pi i k \varphi} 
   |k\rangle_{\scriptscriptstyle Q}
\end{equation}
in the first register.  A final inverse Fourier transformation then generates
the state $|2^K \varphi\rangle_{\scriptscriptstyle Q} |u\rangle$ and the
projective measurement of the $K$-qubit register in the computational basis
provides us with the phase $\varphi$; for an arbitrary phase $0< \varphi < 1$
we obtain a $K$-binary-digit estimate $\varphi_\mathrm{dig}$ of the phase
$\varphi$.

Comparing this algorithm with our counting setup, we identify the action of
the $n$ particles traversing the quantum wire with the action of the second
register in the PEA, with the correspondence $|n\rangle_{\scriptscriptstyle
\Phi} \leftrightarrow |u\rangle$.  The controlled $\mathsf{U}^{2^j}$ operation
in the PEA is replaced by the coupling of the wire to the qubits: the
interaction of the $n$ particles with the last ($K$-th) qubit has to be
identified with the action of the controlled $\mathsf{U}$ operator in the PEA,
hence $\varphi = n/2^K$.  The qubits $j < K$ are more strongly coupled to the
wire, that corresponds to higher powers of the operator $\mathsf{U}$ in the
PEA; in fact, the $j$-th qubit coupling is enhanced by the factor $2^{K-j}$
and its interaction with the particles in the wire corresponds to the action
of the controlled $\mathsf{U}^{2^{K-j}}$ operator. Finally, the intermediate
states in Eqs.\ (\ref{eq:qFt}) and (\ref{eq:PE}) agree with one another with
the identification $n/N = n/2^K \leftrightarrow \varphi$. The final states
exhibit the correspondence $|n\rangle_{\scriptscriptstyle Q}
|n\rangle_{\scriptscriptstyle \Phi} \leftrightarrow |2^K\varphi_\mathrm{dig}
\rangle_{\scriptscriptstyle Q} |u\rangle$. Note that our divisibility
algorithm has no counterpart in the PEA.

This analogy immediately allows us to profit from the performance analysis
\cite{NielsonChuang,Cleve} of the PEA: Assume that we wish to measure the
phase $\varphi$ in the PE problem to an accuracy of $1/2^A$ (i.e., we want to
encode $\varphi$ with $A$ bits) and be sure of our measurement result with a
probability $P = 1-\epsilon$ at least, then the setup must involve $K = A +
\lceil \log_2(2+1/2\epsilon) \rceil$ qubits. 

This result can be applied to our counting algorithm.  Consider the case,
where a non-integer number $x=n + \delta n$ has passed the counter, with $n$
an integer and $0 < \delta n < 1$ a real number. Such a situation may occur
when the interaction between the particles and the counting qubits is still
finite at the moment when the readout procedure starts, corresponding to the
passage of a fraction of a full charge.  Concentrating first on the
implementation with a fully-quantum inverse Fourier transformation, the
performance analysis of the PEA tells us that we still can measure the number
to any desired precision.  E.g., if we want to be able to measure the number
$n < N = 2^K$ such that $|n_\mathrm{meas}-x| < 1/2$ with a probability $P =
1-2^{-r}$, we need to be able to resolve fractional charges $\delta n \sim
2^{-r} \ll 1$, i.e., we have to add additional qubits which measure
half-charges (turning by $2\pi$ on the passage of one particle), quarter
charges (rotating by $4\pi$), etc. The entire setup then has to involve
$\approx K+\log_2(1/2^{-r}) = K+r$ qubits. This result can be extended to
qudits: requiring a precision $P = 1-d^{-r}$, we need $\approx
K+\log_d(1/d^{-r}) = K+r$ qudits. Hence, we can trade additional qudits in the
counting process against a higher probability to obtain a correct integer
result.

Next, we consider the semi-classical inverse Fourier transformation; we
demonstrate below that this semi-classical scheme exhibits the same stability
as the fully-quantum version, although the passage of a fractional charge
strongly affects the conditional measurement of subsequent qubits, e.g., for
$\delta n = 1/2$ the measurement of the first qubit gives a random input for
the direction of measurement of the second qubit.  However, this error does
not propagate through the entire measurement scheme.  Instead, the
measurements of higher qubits recover from false results measured for lower
qubits. Formally, this can be proven by comparing the two probabilities
$P_\mathrm{qF}(n;x)$ (using a full quantum Fourier transformation) and
$P_\mathrm{scF}(n;x)$ (using a semi-classical Fourier transformation) to find
the integer number $n$ when a non-integer number $x$ has passed by the
qubit register.

Indeed, let us calculate the probability $P_\mathrm{scF}(n,x)$ to find the
integer result $n = \sum_{j=1}^K n_j 2^{K-j}$ upon measurement of a
non-integer signal $x$ by the qubit register.  The probability $p_1(n_K,x)$
that the first qubit provides the value $n_K$ is given by the matrix element
between the qubit state $(|0\rangle_1 + e^{2\pi i x/2}|1\rangle_1)/\sqrt{2}$ and
the state $(|0\rangle_1 + e^{\pi i n_K}|1\rangle_1)/\sqrt{2}$ to be measured
(i.e., the projection $|+y\rangle_1$ for $n_K=0$ or $|-y\rangle_1$ for $n_K=1$,
cf.\ Sec.\ \ref{sec:qubits})
\begin{equation}\label{eq:p1}
   p_1(n_K;x)= \frac{|(_1\langle 0| + e^{-\pi i n_K}{_1\langle 1|})
                    (|0\rangle_1 + e^{2\pi i x/2}|1\rangle_1)|^2}{4}.
\end{equation}
Next, the conditional probability to measure the value $n_{K-1}$ for the
second qubit is given by the product $p_1(n_K;x) p_2(n_{K-1},n_K;x)$ with
\begin{eqnarray}\nonumber
   p_2(n_{K-1},n_K;x)&=&
                |(_2\langle 0|+e^{-\pi i(n_{K-1}+n_K/2)}{_2\langle 1|})
                     \\ \label{eq:p2} && ~~
                 (|0\rangle_2 + e^{2\pi i x/4}|1\rangle_2)|^2/4.
\end{eqnarray}
Using Eq.\ (\ref{eq:n_rel}), we can rewrite the first factor in
$p_2(n_{K-1},n_K;x)$ in the simpler (and equivalent to the second factor) form
\begin{equation}\label{eq:p2n}
   p_2(n;x) = \frac{|(_2\langle 0| + e^{-2\pi i n/4}{_2\langle 1|})
                          (|0\rangle_2 + e^{2\pi i x/4}|1\rangle_2)|^2}{4}.
\end{equation}
The straightforward iteration of this scheme then produces the final
result for $P_\mathrm{scF}(n,x)$ in the product form
\begin{equation}\label{eq:P_scF}
   P_\mathrm{scF}(n,x) = \prod_{j=1}^{K} p_j(n;x),
\end{equation}
where 
\begin{equation}\label{eq:pjn}
    p_j(n,x) = \frac{|(_j\langle 0| + e^{-2\pi i n/2^j}{_j\langle 1|})
			  (|0\rangle_j + e^{2\pi i x/2^j}|1\rangle_j)|^2}{4}.
\end{equation}
Evaluating the product, the result Eq.\ (\ref{eq:P_scF}) is easily rewritten
in the form $P_\mathrm{scF}(n,x) = |_{\scriptscriptstyle Q}\langle \Psi_n|
\Psi_x\rangle_{\scriptscriptstyle Q}|^2$ with the generalized counting state
(cf.\ Eq.\ (\ref{eq:fi}))
\begin{equation}\label{eq:Fou}
    |\Psi_x\rangle_{\scriptscriptstyle Q} = \frac{1}{\sqrt{2^K}} 
    \sum_{k=0}^{2^K-1} 
    \exp(2\pi i \,x k/2^K) |k\rangle_{\scriptscriptstyle Q}.
\end{equation}
Expressing the counting state $|\Psi_n\rangle_{\scriptscriptstyle Q}$ as a
Fourier transform of the computational state $|n\rangle_{\scriptscriptstyle
Q})$, $|\Psi_n \rangle_{\scriptscriptstyle Q} = \mathsf{F}
(|n\rangle_{\scriptscriptstyle Q})$, we arrive at the result $P_\mathrm{scF}
(n,x) = |_{\scriptscriptstyle Q}\langle n| x\rangle_{\scriptscriptstyle
Q}|^2$, where $|x\rangle_{\scriptscriptstyle Q}= \mathsf{F}^{-1}(|\Psi_x
\rangle_{\scriptscriptstyle Q})$ is defined in terms of the back transformed
counting state. But this is nothing else than the probability
$P_\mathrm{qF}(n,x)$ to find the number $n$ in a one-shot measurement of the
qubit register after application of an inverse quantum Fourier transformation
on the detected state $|\Psi_x \rangle_{\scriptscriptstyle Q}$, hence
$P_\mathrm{scF}(n,x)= P_\mathrm{qF}(n,x)$. We conclude that the semi-classical
and the fully-quantum Fourier transformation exhibit the same stability to
systematic errors introduced by incomplete (non-integer) counting.
Furthermore, we note that the semi-classical algorithm is rather robust with
respect to random errors; the latter can be handled with a classical
multi-qubit error correction scheme combined with a simple majority rule, cf.\
Ref.\ \onlinecite{LSB_09}.
 
\subsection{Quantum metrology: voltage measurement}\label{sec:ADC}

\begin{figure}[ht]
 \includegraphics[width=7.0cm]{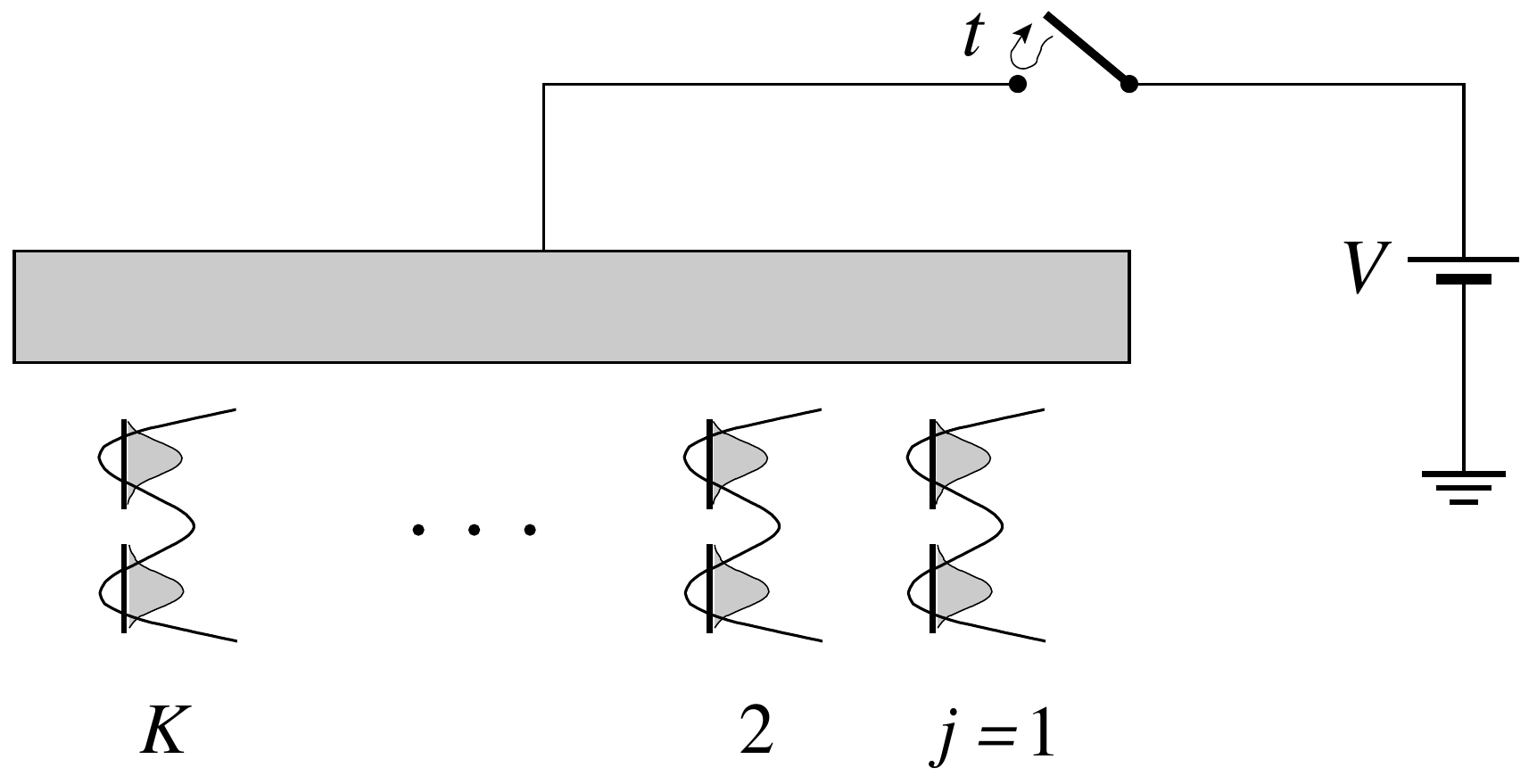}
  \caption[]{\label{fig:ADC} Quantum voltage-detector made from $K$ counter
  qubits.  A voltage $V$ applied over a time interval $t$ generates a
  continuous phase $\varphi = e V t/\hbar$, which can be measured by the
  $K$-qubit register and translated into a binary digital output signal.}
\end{figure}

The above discussion allows us to use our counting device to measure
continuous variables.  The insight paves the way for its use as a quantum
voltage-detector, a particular form of an analog to digital converter (ADC).
Consider a setup similar to the one in Fig.\ \ref{fig:counting} but with the
wire replaced by a finite metallic object, see Fig.\ \ref{fig:ADC}.

Assuming that the applied voltage translates into a relative shift $\delta_K =
\alpha_K eV$ ($\delta_1 = \alpha_1 eV$) of the most weakly (strongly) coupled
qubit [which is the qubit $j=K$ ($j=1$)], the reading of the qubit array after
detaching the voltage provides us with a binary digital number $n$ which
translates into the phase $\varphi$ according to
\begin{equation}\label{eq:ADC}
   \varphi = \frac{e V t}{\hbar} = \frac{2\pi}{\alpha_K}\frac{n}{2^K}
   = \frac{2\pi}{\alpha_1}\frac{n}{2},
\end{equation}
{\it provided that} the quantity $2\alpha_1 e V t/h$ can be represented by an
integer number.  In order to handle successfully the general case with an
arbitrary drive $Vt$, we have to add additional qubits in order to arrive at a
digital estimate of $\varphi$: for a result with a relative precision $1/2^K$
and with a probability better than $1-\epsilon$, we need a measuring device
with $K + \lceil \log_2(2+1/2\epsilon) \rceil$ qubits. 

The sensitivity of our quantum ADC improves as $1/t$, where $t$ is the time of
observation; this is because the accumulated phase increases linearly in time,
while the most sensitive qubit always resolves phases of $\pi$. The precision
then is by a factor $1/\sqrt{t}$ better than the usual classical sensitivity,
which scales as $1/\sqrt{t}$, and agrees with standard expectations
\cite{Lloyd}. In particular, using the straightforward algorithm described in
Sec.\ \ref{sec:efficiency} to measure a voltage $V < V_\mathrm{max} = \pi\,
\hbar/e\tau$ with an accuracy $\delta V$, we either need $M = (\pi/\phi)^2 =
(V_\mathrm{max}/\delta V)^2$ qubits or have to run repeated experiments over a
time $t = M\tau$, where $\tau$ is the measuring time of the individual
experiment. Solving for the desired accuracy $\delta V$, we obtain the scaling
$\delta V \sim \sqrt{\hbar V_\mathrm{max}/ e t}$, i.e., the precision improves
only with the square-root of the overall measuring time $t$, the same as for
the usual classical case.

\section{Multi-particle entanglement}\label{sec:mpentanglement}

\begin{figure}[ht]
 \includegraphics[width=7.0cm]{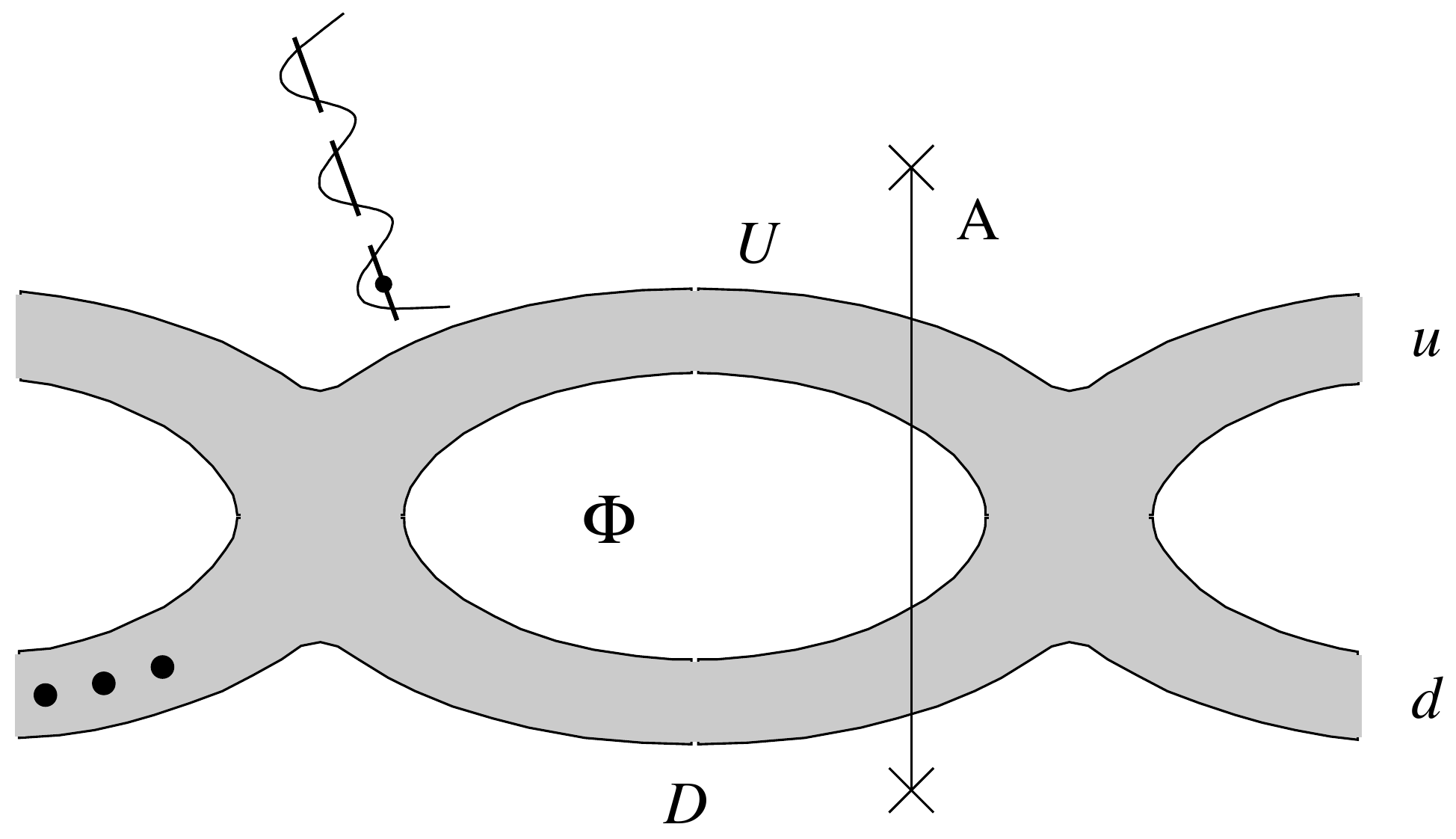}
  \caption[]{\label{fig:MZ} Mach-Zehnder interferometer with qutrit counter.
  Particles enter the interferometer through the left leads (here the bottom
  lead) and are measured on the right. The qutrit counter in the upper arm $U$
  detects the passage of particles. The magnetic flux $\Phi$ through the loop
  allows to tune the phase difference when propagating along different arms.
  }
\end{figure}

Another application of our counting device is the generation of multi-particle
entangled states with the help of a Mach-Zehnder interferometer, see Fig.\
\ref{fig:MZ}. Injecting particles into the device through the lower left arm,
the splitter generates a superposition of number states in the two arms of the
interferometer; measuring the counter placed near the upper arm and selecting
a particular reading projects the system to the desired entangled state.  The
functionality of the device has been described in detailed in Ref.\
\onlinecite{LSB_09} before. Here, we use the device with our qutrit counter to
generate the original GHZ (Greenberger-Horne-Zeilinger) state \cite{GHZ} and
to unveil in more detail the entanglement between the counter states and the
physical number states in the quantum wire; the generalization to other cases
with more particles and counters follows the previous discussion in Ref.\
\onlinecite{LSB_09}.

Consider a particle entering the Mach-Zehnder interferometer from the
lower-left lead and propagating along one of the two leads $U$ or $D$, see
Fig.\ \ref{fig:MZ}.  The wave function can propagate along two trajectories,
the upper arm $U$ where the particle picks up a phase $\varphi_U$ and the
counter is activated, or the lower arm $D$ accumulating a phase $\varphi_D$
and leaving the counter state unchanged. The total wave function evaluated at
the position A then assumes the form
\begin{equation}
   \label{eq:Psi1}
   \Psi_{1A} = t\, e^{i\varphi_U} |\Uparrow\rangle \otimes |\Psi_1\rangle
            + r\, e^{i\varphi_D} |\Downarrow\rangle\otimes|\Psi_0\rangle,
\end{equation}
where $t$ and $r$ denote the transmission and reflection coefficients of the
beam splitter and we have introduced a pseudo-spin notation to describe the
propagation of the particles along the two arms: a pseudo-spin $\Uparrow$
($\Downarrow$) refers to the particle propagating in the upper (lower) arm.
The qutrit state depends on the particle's trajectory and reads either
$|\Psi_1\rangle$ if the particle has passed in the nearby upper arm or
$|\Psi_0\rangle$ if the particle passed through the lower arm of the
interferometer.

Next, we inject three particles (from the bottom left) into the Mach-Zehnder
(MZ) loop.  We assume the three individual wave functions describing the
initial state to be well separated in space, allowing us to ignore exchange
effects in our (MZ) geometry. The wave function at the position A then reads
\begin{eqnarray}
   \nonumber
   &&\Psi_{3A}
     =\bigl[t^3\,e^{3i\varphi_U}|\!\Uparrow, \Uparrow, \Uparrow\rangle
     +r^3\,e^{3i\varphi_D}|\!\Downarrow,\Downarrow,\Downarrow\rangle\bigr]
        \otimes |\Psi_0\rangle
          \\ \label{eq:Psi3A}
      &&\qquad\quad
      +t r^2 \,e^{i(\varphi_U+2\varphi_D)}
      \bigl[|\!\Uparrow,\Downarrow,\Downarrow \rangle
           +|\!\Downarrow,\Uparrow,\Downarrow \rangle
       \\ \nonumber
      &&\qquad\qquad\qquad
           +|\!\Downarrow,\Downarrow,\Uparrow \rangle \bigr]
               \otimes |\Psi_1\rangle 
            \\ \nonumber
      &&\qquad\quad
      +t^2 r \,e^{i(2\varphi_U+\varphi_D)}
      \bigl[|\!\Uparrow,\Uparrow,\Downarrow \rangle
           +|\!\Uparrow,\Downarrow,\Uparrow \rangle
            \\ \nonumber
      &&\qquad\qquad\qquad
           +|\!\Downarrow,\Uparrow,\Uparrow \rangle \bigr]
               \otimes |\Psi_2\rangle.
\end{eqnarray}
Assuming scattering coefficients for a symmetric beam splitter, e.g., $tt^* =
1/2$, $r^2 = (-1)/2$, and $r\,t^*=\pm i\, e^{-i\varphi_t}/2$ (with $\varphi_t$
the transmission phase), the projection to the counter state $|\Psi_0\rangle$
provides one with the GHZ-like state $|\Psi_{\rm\scriptscriptstyle GHZ}\rangle
= (|\!\Uparrow, \Uparrow, \Uparrow\rangle \mp i \,e^{3i(\varphi_D-\varphi_U-
\varphi_t)} |\!\Downarrow,\Downarrow,\Downarrow\rangle)/\sqrt{2}$;
manipulation of the flux $\Phi$ in the Mach-Zehnder loop then allows one to
implement the desired entangled state.  Furthermore, the wave function Eq.\
(\ref{eq:Psi3A}) unveils the entanglement between the (split) number states
and the counter.  Note that the indistinguishability of particles exploited in
the above entanglement process is an `artificial' one defined by the qutrit
detector, rather than the `fundamental' one of identical particles.

\section{Discussion and Conclusion}\label{sec:conclusion}

Summarizing, we have generalized the binary (base 2) quantum counting
algorithm and divisibility test by $2^k$ to ternary (base 3) and higher
counting systems. This extension is quite non-trivial in several respects: On
the device level, the qubits used in the base 2 algorithm have to be replaced
by qutrits for a ternary counting system and to qudits for a base $d$
algorithm. Since the algorithm is based on two subsequent quantum Fourier
transformations, suitable manipulation schemes have to be defined in order to
implement a quantum Fourier transformation on the level of individual qubits,
qutrits, and qudits. Furthermore, rather than developing new hardware for
every new counting base, we have discussed how to use qubits in order to
emulate qudits, with particular emphasis on the qutrits.

Starting from binary counting with qubits, it seems not immediately clear how
to generalize the concept. It turns out, that defining the quantum counting
task on an elementary level through a one to one correspondence between the
counting objects and distinguishable states in a Hilbert space provides us
with a constructive scheme how this task can be achieved. Also, the analysis
of the unary counting scheme naturally introduces the quantum Fourier
transformation as the basic operation in a non-demolitian counting process.
Indeed, the task of counting naturally introduces a shift operator
$\mathsf{C}_1$ in the counting space, taking one counting state
$|\Psi_n\rangle_{\scriptscriptstyle Q}$ into the next
$|\Psi_{n+1}\rangle_{\scriptscriptstyle Q}$. When expressing these counting
states through the eigenstates $|n\rangle_{\scriptscriptstyle Q}$ of
$\mathsf{C}_1$, then the counting operation only adds a phase $\exp(2\pi i
n/N)$ to each of these states. Hence using these eigenstates
$|n\rangle_{\scriptscriptstyle Q}$, which are nothing but the Fourier
transforms of the counting states $|\Psi_n\rangle_{\scriptscriptstyle Q}$, as
our computational basis provides for us a `soft' non-demolitian counting
scheme.  Choosing another computational basis involves an energy exchange
between the counted object and the counting system and introduces a much more
severe perturbation.

The aforementioned basic understanding of quantum counting has provided us
with a constructive scheme for the counting algorithm: starting out with a set
of measureable states (the computational basis
$\{|n\rangle_{\scriptscriptstyle Q}\}$) which evolve with a prescribed phase
accumulation upon the passage of particles, we have to prepare out of them a
balanced state $|\Psi_0\rangle_{\scriptscriptstyle Q}$ which is used as the
first counting state. With the appropriate phase increment picked up during
counting, this first counting state evolves to the next and returns to the
first one after an $N$-cycle. As we have seen, we do not have to enter the
cycle in the lowest harmonic, any harmonic will do, and even arbitrary phases
in a balanced state (an equal-weight superposition of the computational basis
states) are acceptable.  In the latter case, the Fourier transformation is
modified with additional phases, which are not harmful to the algorithm,
however.

Equipped with this general scheme, we have been able to generalize the binary
algorithm to a ternary and to base-$d$ counting. We have seen, that one
possible straightforward extension of the hardware, going from a spin-1/2 to a
spin-1 system, poses severe problems due to the lack of suitable operators in
the state preparation.  Indeed, although any superposition of two states are
allowed in quantum mechanics, it might be difficult to prepare this
superposition in practice.  E.g., if we want to superpose two different
eigenstates of the same Hamiltonian, one has to act on the states with an
operator which might not be available in the given physical system; this
situation is actually encountered in the spin-1 system. However,
emulating a spin-1 system through the triplet sector of two qubits provides
a viable alternative: the two-qubit operation combined with single-qubit
rotations provide sufficient degrees of freedom to carry out all required
operations for counting, the state preparation and the inverse Fourier
transformation. Further emulation of qudits with qubits, however, should not
be done in the spin-$d$ sector, as this is a waist of resources. Instead, a
straightforward sequence of neighboring energy states will do.  An important
element to realize is that the Fourier transformation, which we can handle
semi-classically `between' the qudits, has to be fully quantum `within' the
qudits. Hence the larger the chosen counting basis $d$ is, the larger is the
part of the inverse Fourier transformation which is done fully quantum.

We have demonstrated, that the substitution of the semi-classical inverse
Fourier transformation for the fully-quantum does not entail any disadvantage
for the counting algorithm; this is an important result, both with respect to
the stability of the counting algorithm against systematic errors (non-integer
counting) as well as its application in metrology. It thus appears that the
conditional operation in the quantum Fourier transform can be fully
substituted by a measurement combined with a conditional operation in the
semi-classical scheme.  In both cases, a non-integer reading may affect some
of the last digits of the seeked number, but the leading digits are not
compromized.  Furthermore, adding additional qubits (digits) allows to trade
an extension of the hardware for a better precision in the output.

In terms of applications of our counter, we have generalized the scheme
producing multi-particle entanglement in a Mach-Zehnder interferometer and
have proposed its use as a quantum voltage-detector, a particular example of
an analog-digital converter. We may speculate that our quantum counting scheme
can be generalized to other broadband measurement algorithms and thus
contribute to other applications in quantum metrology\cite{Lloyd}. 

\bigskip

We thank Klaus Ensslin, Alexei Kitaev, Gerald Milburn, Renato Renner, and
Peter Shor for discussions and acknowledge financial support by the
Pauli-Center at ETH Zurich and the Russian Foundation for Basic Research under
grant No.\ 08-02-00767-a.

\appendix
\section{Derivation of $\mathsf{U}_p$ and $\mathsf{M}$ for the
qubit-emulated qutrit} \label{app:Up}

Motivated by our general discussion in section \ref{sec:fourier}, we start out
with the counting basis in the form [we choose a balanced state
$|\Psi_0\rangle$ and apply the counting operator $\mathsf{C}_1$ in Eq.\
(\ref{eq:C12qubits}) to obtain, up to an overal phase]
\begin{eqnarray}\label{eq:ent_cs}
   |\Psi_0\rangle &=& \frac{1}{\sqrt{3}} \bigl( |\uparrow\uparrow\rangle 
   + |\uparrow\downarrow\rangle + |\downarrow\uparrow\rangle \bigr)
   \\ \nonumber
   |\Psi_1\rangle &=& \frac{1}{\sqrt{3}} \bigl( |\uparrow\uparrow\rangle 
   + e^{2\pi i/3} |\uparrow\downarrow\rangle 
   + e^{4\pi i/3} |\downarrow\uparrow\rangle\bigr)
   \\ \nonumber
   |\Psi_2\rangle &=& \frac{1}{\sqrt{3}} \bigl( |\uparrow\uparrow\rangle
   + e^{4\pi i/3} |\uparrow\downarrow\rangle 
   + e^{2\pi i/3} |\downarrow\uparrow\rangle\bigr)
   \\ \nonumber
   |\Psi_3\rangle &=& |\downarrow \downarrow\rangle.
\end{eqnarray}
These counting states are entangled, whereas those defining the
computational basis, $|0\rangle = |\!\uparrow \uparrow\rangle$, $|1\rangle =
|\!\uparrow \downarrow \rangle$, and $|2\rangle = |\!\downarrow \uparrow
\rangle$, are not. This feature can be conveniently exploited in finding the operators
$\mathsf{U}_p$ and $\mathsf{M}$ for preparation and readout.

\begin{figure}[ht]
 \includegraphics[width=7.0cm]{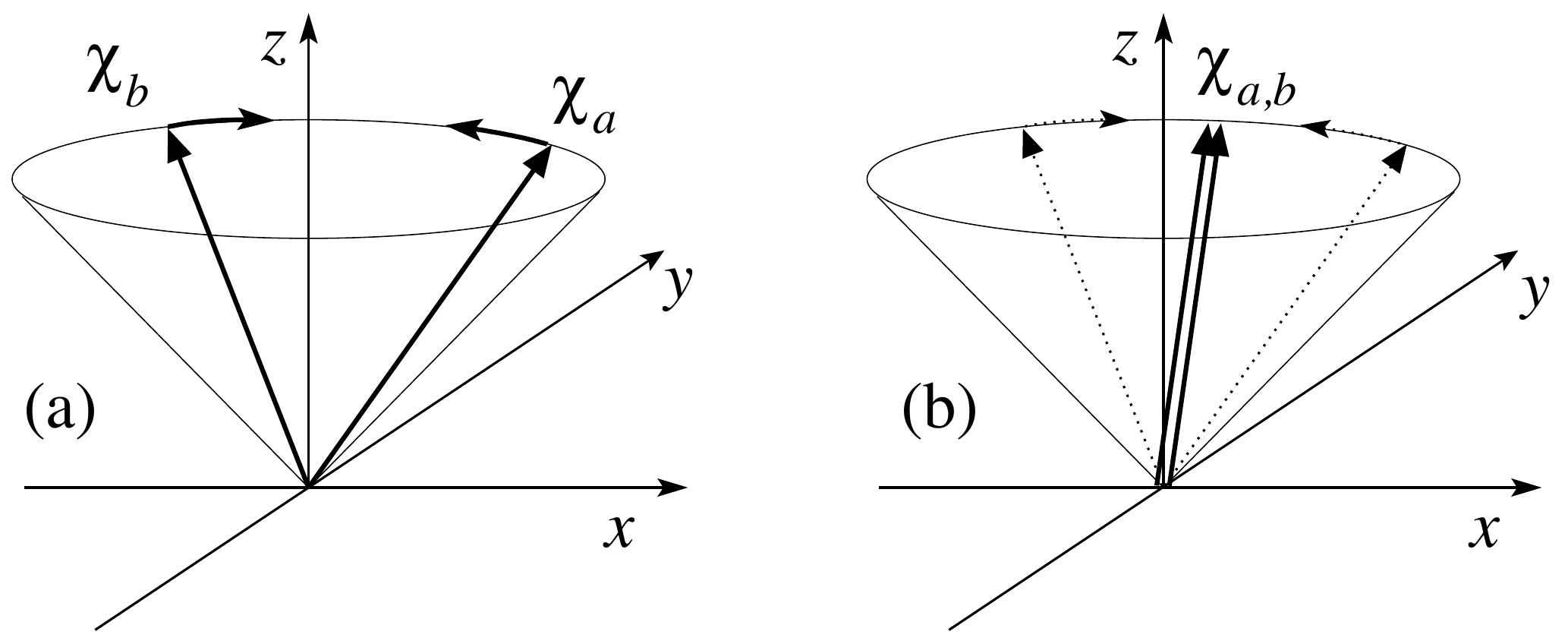}
  \caption[]{\label{fig:U_dis} Action of $\mathsf{U}_\varphi$ in disentangling
  a state. The two components $|\chi_{a,b}\rangle$ to be aligned point along
  the same polar angle $\theta$. The conditional rotation by $\varphi$ aligns
  the two components if $\varphi$ matches the azimuthal angle difference,
  $\varphi = \varphi_b - \varphi_a$. }
\end{figure}

We begin with the preparation step: rather then finding $\mathsf{U}_p$, we
search for the inverse operator $\mathsf{U}_p^{-1}$ which disentangles the
state $|\Psi_0\rangle$. In order to accomplish this task, we have the
operators $\sigma^{\rm\scriptscriptstyle (1)}_{x}$, $\sigma^{\rm
\scriptscriptstyle (2)}_{x}$, $\sigma^{\rm\scriptscriptstyle (1)}_{z}$,
$\sigma^{\rm\scriptscriptstyle (2)}_{z}$, and the two-qubit operator
$\sigma^{\rm\scriptscriptstyle (1)}_{z} \sigma^{\rm\scriptscriptstyle
(2)}_{z}$ at our disposal. In our construction below, we will make heavy use
of the conditional rotation
\begin{equation}\label{eq:U_varphi}
   \mathsf{U}_\varphi \equiv 
   \exp(-i\varphi \, \sigma^{\rm\scriptscriptstyle (1)}_{z} 
                  \sigma^{\rm\scriptscriptstyle (2)}_{z}/4).
\end{equation}
This operator will generate the decisive step in the disentanglement of the
qubit states. Let us consider a general (entangled) two-qubit state
\begin{equation}\label{eq:chi_ab}
   |\phi\rangle = |\uparrow\rangle_1|\chi_a \rangle_2
   + |\downarrow\rangle_1|\chi_b\rangle_2
\end{equation}
with normalization $\langle\phi|\phi\rangle = \langle\chi_a|\chi_a\rangle +
\langle\chi_b| \chi_b\rangle = 1$. The state $|\phi\rangle$ is not entangled
if and only if $|\chi_a \rangle = \alpha |\chi_b \rangle$ or one of the states
$|\chi_a \rangle$, $|\chi_b \rangle$ vanishes. Acting with
$\mathsf{U}_\varphi$ on $|\phi\rangle$,
\begin{eqnarray}\label{eq:cond_rot}
   \mathsf{U}_\varphi |\phi\rangle
   &=& |\uparrow\rangle_1 
   e^{-i\varphi\sigma^{\rm\scriptscriptstyle (2)}_{z}/4} |\chi_a\rangle_2
   \\ \nonumber
   &&\qquad\qquad
   + |\downarrow\rangle_1 
   e^{i\varphi\sigma^{\rm\scriptscriptstyle (2)}_{z}/4}|\chi_b\rangle_2,
\end{eqnarray}
we find that $\mathsf{U}_\varphi$ disentangles $|\phi\rangle$ if $\theta_a =
\theta_b = \theta$ and $\varphi = \varphi_b - \varphi_a$, see Fig.\
\ref{fig:U_dis} (here, the angles $\theta_{a,b}$ and $\varphi_{a,b}$ denote
the directions of the second spin described with the states
$|\chi_{a,b}\rangle$).  Furthermore, we need the property of
$\mathsf{U}_\varphi$ that it transforms a product state $|\psi\rangle_1
|\chi\rangle_2$ into a product state if either $|\psi\rangle_1$ or
$|\chi\rangle_2$ is directed along the $z$-axis.

\begin{figure}[ht]
 \includegraphics[width=6.0cm]{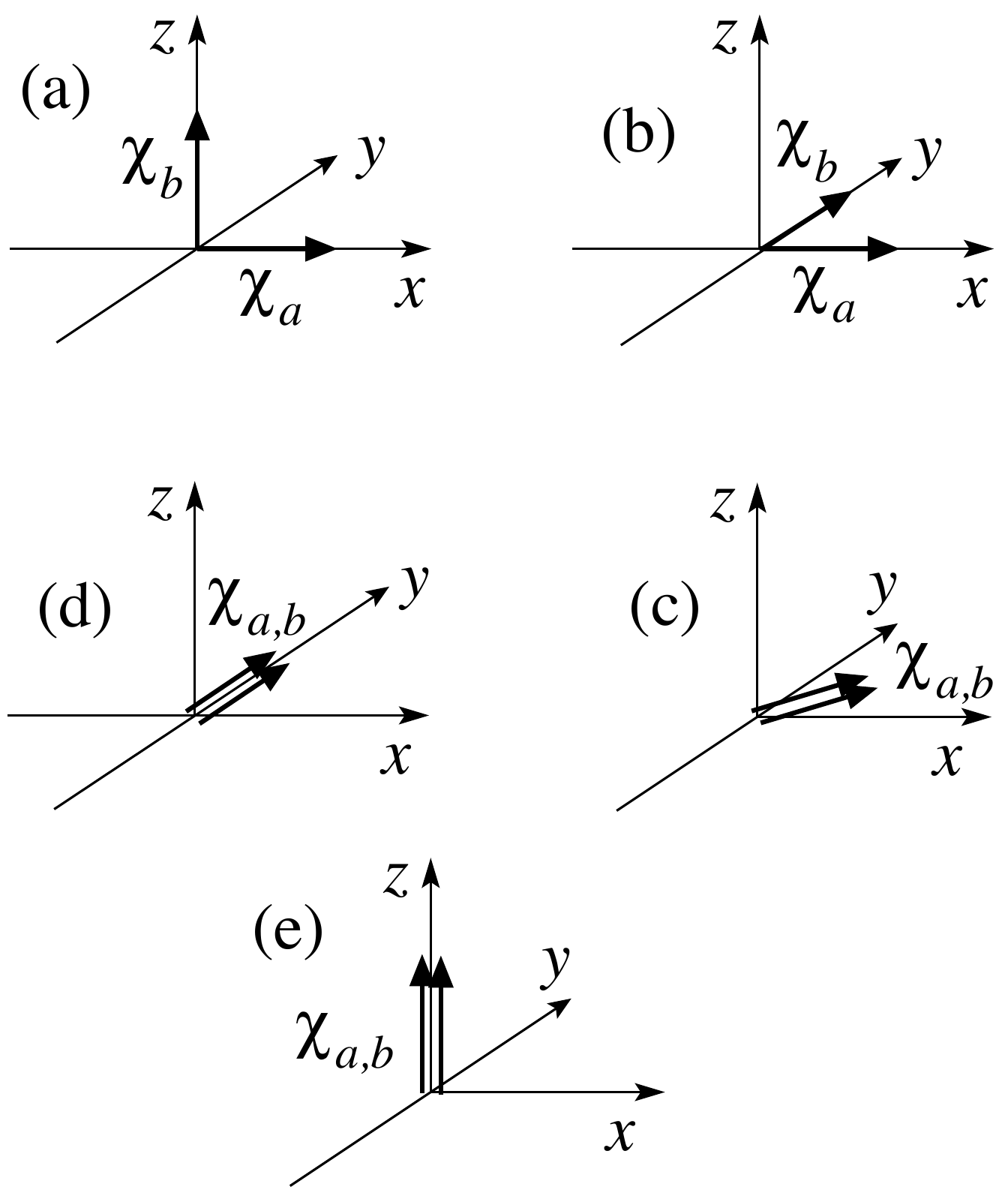}
  \caption[]{\label{fig:Up}  Rotations generating the operator
  $\mathsf{U}_p^{-1}$; shown are the rotations acting on the components
  $|\chi_a\rangle_2 \propto {(|\uparrow \rangle_2}+{|\downarrow\rangle_2)}$
  and $|\chi_b\rangle_2 \propto |\uparrow\rangle_2$ in $|\Psi_0\rangle$, cf.\
  Eq.\ (\ref{eq:psi_0_chi}) (the arrows reflect the polarization angles of the
  spin states). The first rotation (a) $\to$ (b) by $-\pi/2$ around
  the $x$-axis rotates $\chi_b$ into the direction along the $y$-axis. The
  subsequent two-qubit rotation $\mathsf{U}_{\pi/2}$, cf.\ (b) $\to$ (c),
  aligns both components along the $xy$-diagonal; this step disentangles the
  state.  The next rotation by $\pi/4$ around the $z$-axis makes the two
  states $\chi_{a,b}$ point along the $y$-axis and the subsequent rotation by
  $\pi/2$ around the $x$-axis aligns them parallel to the $z$-axis.  The final
  two rotations of the qubit 1 (not shown) transforms the state
  $|\psi\rangle_1 |\uparrow\rangle_2$ into the state $e^{-i\pi/4}
  |\uparrow\rangle_1 |\uparrow\rangle_2$, which coincides (up to a phase) with
  the computational state $|0\rangle = |\uparrow\uparrow \rangle$.}
\end{figure}

In order to find the operator $\mathsf{U}_p^{-1}$ disentangling
$|\Psi_0\rangle$, we write the latter in the form
\begin{equation} \label{eq:psi_0_chi}
   |\Psi_0\rangle = |\uparrow\rangle_1
   \frac{|\uparrow\rangle_2+|\downarrow\rangle_2}{\sqrt{3}}
   + \frac{|\downarrow\rangle_1 |\uparrow\rangle_2}{\sqrt{3}},
\end{equation}
hence $|\chi_a\rangle_2 = (|\uparrow\rangle_2+|\downarrow\rangle_2)/\sqrt{3}$
and $|\chi_b \rangle_2 = |\uparrow\rangle_2/\sqrt{3}$. The sequence of
operations shown in Fig.\ \ref{fig:Up} disentangles the state $|\Psi_0\rangle$
by aligning $|\chi_a\rangle_2$ and $|\chi_b\rangle_2$ along the diagonal in
the $xy$-plane and produces the state $|\psi\rangle_1 |\uparrow\rangle_2$ with
\begin{equation} \label{eq:psi_up}
   |\psi\rangle_1 = \sqrt{2/3}\, |\uparrow\rangle_1
   + \sqrt{1/3}\,|\downarrow\rangle_1.
\end{equation}
The remaining one-qubit operations $\exp(-i\pi\sigma^{\rm\scriptscriptstyle
(1)}_{z}/4)$ and $\exp(-i\theta\sigma^{\rm\scriptscriptstyle (1)}_{x}/2)$ with
$\theta = 2\arctan(1/\sqrt{2})$ then produces the computational state
$|\uparrow\uparrow\rangle$ up to a phase, $\mathsf{U}_p |\uparrow\uparrow
\rangle = \exp(i\pi/4) |\Psi_0\rangle$. One easily verifies that the above
sequence of operations produces the operator $\mathsf{U}_p$, Eq.\
(\ref{eq:em_U_p}).

The construction of the inverse quantum Fourier transformation $\mathsf{M}$
for the readout follows the same scheme, i.e., we look for the operator
$\mathsf{M}$ which disentangles the counting states $|\Psi_j\rangle$ and maps
$|\Psi_2\rangle$ to $|2\rangle = |\!\downarrow \uparrow \rangle$,
$|\Psi_1\rangle$ to $|1\rangle = |\!\uparrow \downarrow \rangle$, and
$|\Psi_0\rangle$ to $|0\rangle = |\!\uparrow \uparrow\rangle$; this will be
done in three consecutive steps,
\begin{equation}\label{eq:MUUUA}
   \mathsf{M} = \mathsf{U}_0 \mathsf{U}_1 \mathsf{U}_2,
\end{equation}
with $\mathsf{U}_i$, $i = 0,1,2$, appropriate unitary operators serving to
disentangle the three counting states $|\Psi_j\rangle$, $j=0,1,2$, and
producing the simple computational states $|j\rangle$, $j=0,1,2$.
\begin{figure}[ht]
 \includegraphics[width=6.0cm]{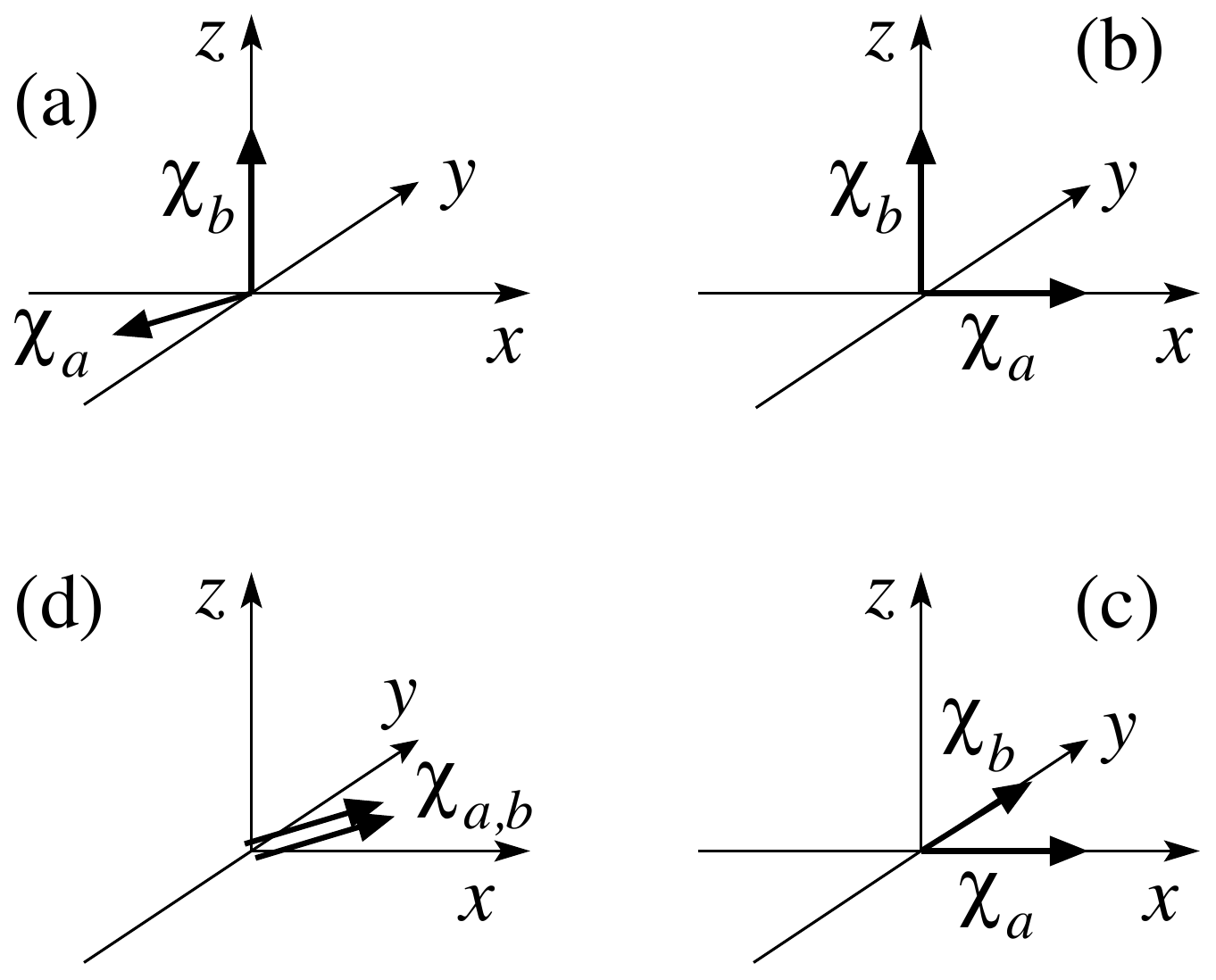}
  \caption[]{\label{fig:U2} Rotations generating the operator $\mathsf{U}_2$;
  shown are the operations acting on the components $|\chi_a\rangle_2 \propto
  {(|\uparrow \rangle_2} + e^{i4\pi/3}{|\downarrow\rangle_2)}$ and
  $|\chi_b\rangle_2 \propto |\uparrow\rangle_2$ in $|\Psi_2\rangle$ of qubit 2
  in Eq.\ (\ref{eq:dis_psi_2}) (vectors mark the polarizations of the spin
  states).  The first rotation (a) $\to$ (b) by $2\pi/3$ around the $z$-axis
  rotates the in-plane component $\chi_a$ into the $x$-axis. The second
  rotation by $-\pi/2$ around the $x$-axis makes $\chi_b$ point along the
  $y$-axis, cf.\ (b) $\to$ (c). The subsequent two-qubit rotation
  $\mathsf{U}_{\pi/2}$, cf.\ (c) $\to$ (d), aligns both components along the
  $xy$-diagonal; this step disentangles the state. The remaining rotations of
  qubit 2 by $\pi/4$ around the $z$-axis and by $\pi/2$ around the $x$-axis
  serve to prepare the operation $\mathsf{U}_1$ and take the spin 2 back to
  the $z$-axis. The state $|\Psi_3\rangle$ merely picks up a phase, while the
  states $|\Psi_0\rangle$ and $|\Psi_1\rangle$ remain entangled, cf.\ Eqs.\
  (\ref{eq:U2P0}). }
\end{figure}

We start with the disentanglement of $|\Psi_2\rangle$, which we write in the
form
\begin{equation}\label{eq:dis_psi_2}
   |\Psi_2\rangle = |\uparrow\rangle_1
   \frac{|\uparrow\rangle_2+e^{i4\pi/3}|\downarrow\rangle_2}{\sqrt{3}}
   + \frac{e^{i2\pi/3} |\downarrow\rangle_1 |\uparrow\rangle_2}{\sqrt{3}}
\end{equation}
and require $\mathsf{U}_2$ to disentangle $|\Psi_2\rangle$ while leaving the
fourth state $|\Psi_3\rangle$ unchanged (up to a phase). This task is
accomplished by the operator
\begin{eqnarray}\label{eq:U_2}
   \mathsf{U}_2
   &=&
   e^{-i\pi\sigma^{\rm\scriptscriptstyle (2)}_{x}/4}
   e^{-i\pi\sigma^{\rm\scriptscriptstyle (2)}_{z}/8}
   e^{-i\pi\sigma^{\rm\scriptscriptstyle (1)}_{z}
           \sigma^{\rm\scriptscriptstyle (2)}_{z}/8} \\ \nonumber
   &&\qquad\qquad\qquad\qquad
   \times e^{i\pi\sigma^{\rm\scriptscriptstyle (2)}_{x}/4}
   e^{-i\pi\sigma^{\rm\scriptscriptstyle (2)}_{z}/3}
   \\ \nonumber
   &=& \frac{1}{\sqrt{2}} \left(
          \begin{array}{cccc}
             e^{-i\pi/3} & e^{i\pi/3} & 0 & 0\\
             e^{i2\pi/3} & e^{i\pi/3} & 0 & 0 \\
             0 & 0 & \sqrt{2} e^{-i\pi/3} & 0 \\
             0 & 0 & 0 & \sqrt{2} e^{i\pi/3}
           \end{array}
        \right),
\end{eqnarray}
where the first expression is used in an implementation through single and
two-qubit gates and the second provides the simpler overall unitary matrix,
The first two rotations prepare the states $|\chi_a\rangle_2$ and
$|\chi_b\rangle_2$ to allow the two-qubit operator $\mathsf{U}_{\pi/2}$ to
align them and hence disentangle the state, see Fig.\ \ref{fig:U2}; the
remaining one-qubit operators acting on qubit 2 serve to prepare the state for
the action of $\mathsf{U}_1$.  The operator $\mathsf{U}_2$ leaves the state
$|\Psi_3\rangle = |\!\downarrow \downarrow \rangle$ parallel to itself,
$\mathsf{U}_2 |\!\downarrow \downarrow \rangle = \exp(i\pi/3)
|\!\downarrow\downarrow \rangle$, and transforms $|\Psi_2\rangle$ into the
product state
\begin{equation}\label{eq:U2}
   \mathsf{U}_2|\Psi_2 \rangle = \frac{\bigl(\sqrt{2}e^{-i\pi/3}
   |\uparrow\rangle_1
   +e^{i\pi/3}|\downarrow\rangle_1\bigr) |\uparrow\rangle_2}{\sqrt{3}}.
\end{equation}
The other two states remain entangled,
\begin{eqnarray}\label{eq:U2P0}
   \mathsf{U}_2|\Psi_0 \rangle &=& \Bigl(\frac{|\uparrow\rangle_1}{\sqrt{6}}
   +\frac{e^{-i\pi/3}|\downarrow\rangle_1}{\sqrt{3}}\Bigr) |\uparrow
   \rangle_2,
   \\ \nonumber
   &&\qquad\qquad\qquad\qquad
   + \frac{e^{i\pi/2}|\uparrow\rangle_1 |\downarrow\rangle_2}{\sqrt{2}}
   \\ \nonumber
   \mathsf{U}_2|\Psi_1 \rangle \!&=&\!
   \Bigl(\frac{e^{-i2\pi/3}|\uparrow\rangle_1}{\sqrt{6}}
   \!-\!\frac{|\downarrow\rangle_1}{\sqrt{3}}\Bigr) |\uparrow \rangle_2 \\
   \nonumber
   &&\qquad\qquad\qquad\qquad
   + \frac{e^{i5\pi/6} |\uparrow\rangle_1 |\downarrow\rangle_2}{\sqrt{2}}.
\end{eqnarray}
Note that the state of qubit 1 is identical in the above two wave functions
(pull out a factor $e^{-2\pi i/3}$ in $\mathsf{U}_2|\Psi_1 \rangle$).
\begin{figure}[h]
 \includegraphics[width=6.0cm]{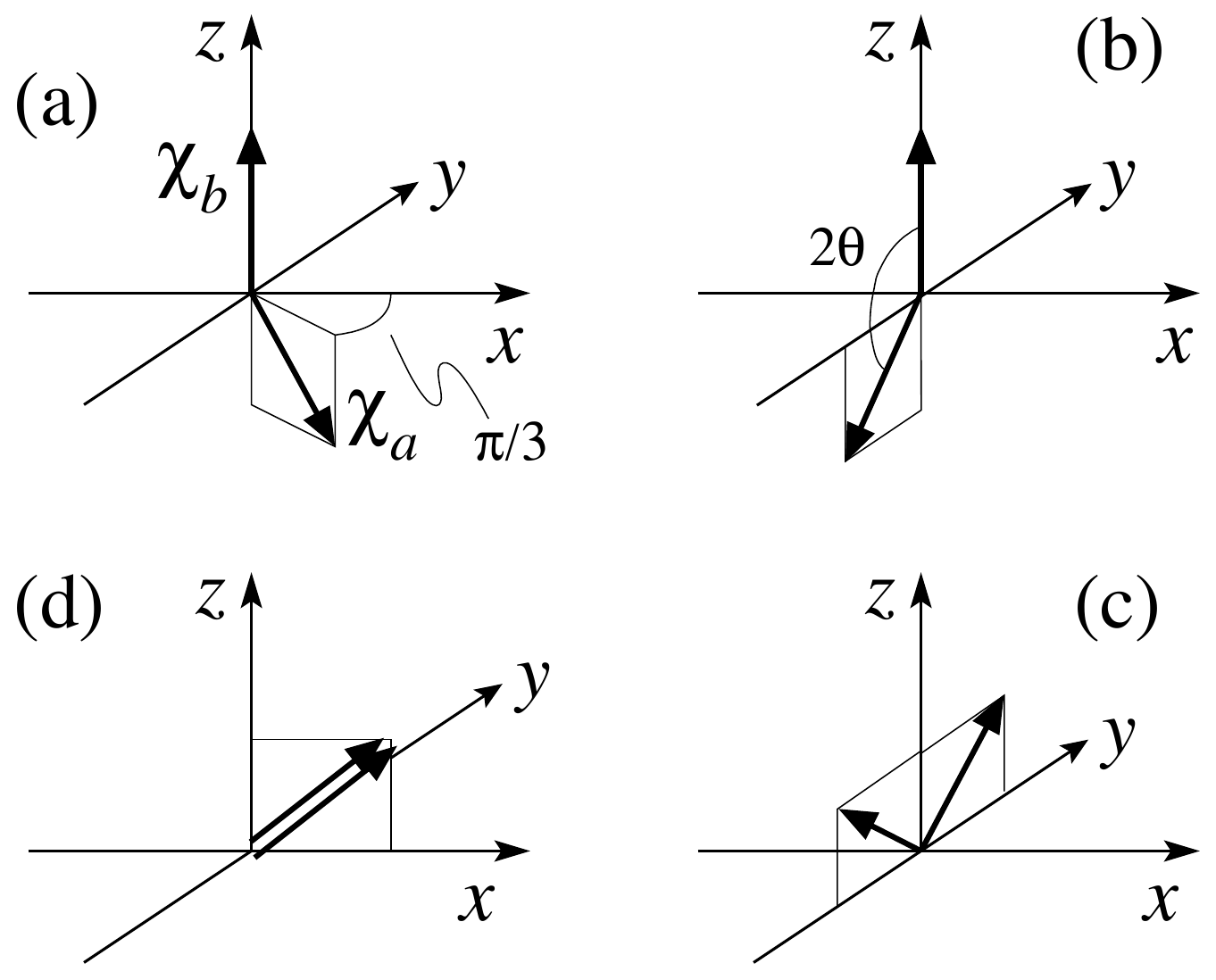}
  \caption[]{\label{fig:U1} Rotations generating the operator $\mathsf{U}_1$;
  shown are the rotations acting on the components $|\chi_a\rangle_1 \propto
  {(|\uparrow\rangle_1/\sqrt{2}} + e^{-i\pi/3}{|\downarrow\rangle_1)}$ and
  $|\chi_b\rangle_1 \propto |\uparrow\rangle_1$ in
  $\mathsf{U}_2|\Psi_1\rangle$ and $\mathsf{U}_2|\Psi_0\rangle$ of qubit 1,
  cf.\ Eq.\ (\ref{eq:U2P0})  (vectors mark the polarizations of the spin
  states). The first rotation (a) $\to$ (b) by $-\pi/6$ around the $z$-axis
  rotates the component $|\chi_a\rangle$ into the $yz$-plane. The second
  rotation by $-\theta$ around the $x$-axis takes the two states symmetrically
  around the $z$-axis, cf.\ (b) $\to$ (c).  The subsequent two-qubit rotation
  $\mathsf{U}_{\pi}$, cf.\ (c) $\to$ (d), aligns both components in the
  $xz$-plane; this step disentangles the state.  The remaining rotations of
  qubit 1 serve to align its state along the $z$-axis.}
\end{figure}

Next, the operator $\mathsf{U}_1$,
\begin{eqnarray}\label{eq:U_1}
   \mathsf{U}_1
   &=&
   e^{-i\theta\sigma^{\rm\scriptscriptstyle (1)}_{x}/2}
   e^{-i\pi\sigma^{\rm\scriptscriptstyle (1)}_{z}/4}
   e^{-i\pi\sigma^{\rm\scriptscriptstyle (1)}_{z}
           \sigma^{\rm\scriptscriptstyle (2)}_{z}/4} \\ \nonumber
   &&\qquad\qquad\qquad\qquad
   \times e^{i\theta\sigma^{\rm\scriptscriptstyle (1)}_{x}/2}
   e^{i\pi\sigma^{\rm\scriptscriptstyle (1)}_{z}/12}
   \\ \nonumber
   &=& \!\! \left(
          \begin{array}{cccc}
             \sqrt{\frac{1}{3}} e^{-i5\pi/12} & 0 & 
             \sqrt{\frac{2}{3}}e^{-i\pi/12} & 0\\
             0 & e^{i\pi/12} & 0 & 0 \\
             \sqrt{\frac{2}{3}} e^{-i11\pi/12} & 0 & 
             \sqrt{\frac{1}{3}}e^{i5\pi/12} & 0 \\
             0 & 0 & 0 & e^{-i\pi/12}
           \end{array}
        \right),
\end{eqnarray}
where $\theta = \arctan(\sqrt{2})$, takes the remaining two entangled states
$\mathsf{U}_2|\Psi_0\rangle$ and $\mathsf{U}_2|\Psi_1\rangle$ into product
states,
\begin{eqnarray}\label{eq:U12P0}
   \mathsf{U}_1\mathsf{U}_2|\Psi_0 \rangle
   &=& e^{-i5\pi/12} \frac{|\uparrow\rangle_1 (|\uparrow\rangle_2
   -|\downarrow\rangle_2)}{\sqrt{2}},
   \\ \nonumber
   \mathsf{U}_1\mathsf{U}_2|\Psi_1 \rangle
   &=& e^{i11\pi/12} \frac{|\uparrow\rangle_1 (|\uparrow\rangle_2
   +|\downarrow\rangle_2)}{\sqrt{2}},
\end{eqnarray}
while leaving the product states in product states,
\begin{eqnarray}\label{eq:U12P2}
   \mathsf{U}_1 \mathsf{U}_2|\Psi_2 \rangle
   &=& e^{i3\pi/4}|\downarrow\rangle_1|\uparrow\rangle_2,
   \\ \nonumber
   \mathsf{U}_1 \mathsf{U}_2|\Psi_3 \rangle
   &=& e^{i\pi/4}|\downarrow\rangle_1|\downarrow\rangle_2.
\end{eqnarray}
The action of the (conditional) rotations in $\mathsf{U}_1$ leading to the
disentanglement of $|\Psi_0\rangle$ and $|\Psi_1\rangle$ is shown in Fig.\
\ref{fig:U1}.

Finally, the operator $\mathsf{U}_0$,
\begin{eqnarray}\label{eq:U_0}
   \mathsf{U}_0
   &=&
   e^{-i\pi\sigma^{\rm\scriptscriptstyle (2)}_{x}/4}
   e^{i\pi\sigma^{\rm\scriptscriptstyle (2)}_{z}/8}
   e^{i\pi\sigma^{\rm\scriptscriptstyle (1)}_{z}
          \sigma^{\rm\scriptscriptstyle (2)}_{z}/8} \\ \nonumber
   &&\qquad\qquad\qquad\qquad
   \times e^{i\pi\sigma^{\rm\scriptscriptstyle (2)}_{x}/4}
   \\ \nonumber
   &=& \frac{1}{\sqrt{2}} \left(
          \begin{array}{cccc}
             1 & -1 & 0 & 0\\
             1 & 1 & 0 & 0 \\
             0 & 0 & \sqrt{2} & 0 \\
             0 & 0 & 0 & \sqrt{2}
           \end{array}
        \right).
\end{eqnarray}
acts as a Hadamard operation on the qubit 2 if the qubit 1 is in the
$|\uparrow\rangle_1$ state and leaves the qubit 2 unchanged if the qubit 1 is
in the state $|\downarrow\rangle_1$, hence $\mathsf{U}_0$ is a controlled
Hadamard.  The combined action $\mathsf{M}$ of the operators $\mathsf{U}_2$,
$\mathsf{U}_1$, and $\mathsf{U}_0$ finally take the counting states into the
computational basis states,
\begin{eqnarray}\label{eq:qtrit_M_0123}
   \mathsf{M} |\Psi_0\rangle &=& e^{-i\,5\pi/12} |0\rangle, \\ \nonumber
   \mathsf{M} |\Psi_1\rangle &=& e^{i\,11\pi/12} |1\rangle, \\ \nonumber
   \mathsf{M} |\Psi_2\rangle &=& e^{i\,3\pi/4} |2\rangle,\\ \nonumber
   \mathsf{M} |\Psi_3\rangle &=& e^{i\,\pi/4} |3\rangle,
\end{eqnarray}
which is nothing but the desired inverse Fourier transformation (up to phases).


\begin{thebibliography}{99}
\bibitem{repulsive} Assuming a repulsive interaction, a similar setup with
   well energies decreasing to the left can be drawn, where the passing
   particle brings two adjacent levels into resonance, allowing the counter
   particle to tunnel between two wells; with an appropriate fine tuning of
   the interaction strength and passage time, the counter particle will
   undergo complete tunneling between two wells. Such a device is required to
   evolve coherently during tunneling only, but may loose its phase after each
   completion of the tunneling step.
\end{thebibliography}
\end{document}